\documentclass[document]{elsarticle}
\usepackage[margin=4cm]{geometry}
\usepackage[utf8]{inputenc}
\usepackage{hyperref}
\usepackage{graphicx}
\usepackage{epstopdf, epsfig}
\usepackage{commath}
\usepackage{amsmath,amssymb,amsthm,tabstackengine}
\usepackage{relsize}
\usepackage{fancybox}

\usepackage{float}
\usepackage{caption}
\usepackage{subcaption}
\usepackage{mathtools}
\usepackage{multicol}
\usepackage{dblfloatfix}
\usepackage{array}
\usepackage[title]{appendix}
\usepackage{hyperref}
\usepackage{float}
\usepackage{enumitem}
\usepackage{kbordermatrix}

\usepackage{pdflscape}
\usepackage{relsize}

\usepackage{etoolbox}
\usepackage{tikz}
\usetikzlibrary{tikzmark}
\usetikzlibrary{calc}

\usepackage{pdflscape}
\usepackage{relsize}
\usepackage{tikz}
\usepackage{rotating}                   
\usetikzlibrary{shapes}
\usetikzlibrary{intersections}
\usetikzlibrary{positioning}
\usetikzlibrary{arrows}
\usetikzlibrary{arrows.meta}
\usetikzlibrary{matrix}

\usepackage{colortbl}

\usepackage{tkz-euclide}
\usetkzobj{all}

\usepackage{makecell}
\usepackage{algorithm}
\usepackage{algpseudocode}

\algnewcommand{\LineComment}[1]{ \Statex \( \hspace{0.6cm} \triangleright\) #1}

\newcommand{\ALGtikzmarkcolor}{black}
\newcommand{\ALGtikzmarkextraindent}{4pt}
\newcommand{\ALGtikzmarkverticaloffsetstart}{-.5ex}
\newcommand{\ALGtikzmarkverticaloffsetend}{-.5ex}
\makeatletter
\newcounter{ALG@tikzmark@tempcnta}

\newcommand\ALG@tikzmark@start{%
    \global\let\ALG@tikzmark@last\ALG@tikzmark@starttext%
    \expandafter\edef\csname ALG@tikzmark@\theALG@nested\endcsname{\theALG@tikzmark@tempcnta}%
    \tikzmark{ALG@tikzmark@start@\csname ALG@tikzmark@\theALG@nested\endcsname}%
    \addtocounter{ALG@tikzmark@tempcnta}{1}%
}

\def\ALG@tikzmark@starttext{start}
\newcommand\ALG@tikzmark@end{%
    \ifx\ALG@tikzmark@last\ALG@tikzmark@starttext
    \else
        \tikzmark{ALG@tikzmark@end@\csname ALG@tikzmark@\theALG@nested\endcsname}%
        \tikz[overlay,remember picture] \draw[\ALGtikzmarkcolor] let \p{S}=($(pic cs:ALG@tikzmark@start@\csname ALG@tikzmark@\theALG@nested\endcsname)+(\ALGtikzmarkextraindent,\ALGtikzmarkverticaloffsetstart)$), \p{E}=($(pic cs:ALG@tikzmark@end@\csname ALG@tikzmark@\theALG@nested\endcsname)+(\ALGtikzmarkextraindent,\ALGtikzmarkverticaloffsetend)$) in (\x{S},\y{S})--(\x{S},\y{E});%
    \fi
    \gdef\ALG@tikzmark@last{end}%
}

\apptocmd{\ALG@beginblock}{\ALG@tikzmark@start}{}{\errmessage{failed to patch}}
\pretocmd{\ALG@endblock}{\ALG@tikzmark@end}{}{\errmessage{failed to patch}}
\makeatother

\DeclareMathOperator{\DIAG}{diag}

\DeclarePairedDelimiter{\ceil}{\lceil}{\rceil}
\newcolumntype{P}[1]{>{\centering\arraybackslash}p{#1}}
\newcommand{\txt}[1]{\mathsmaller{\rm{#1}}}
\newcommand{\sign}[1]{\text{sign}\left( #1\right)}
\newcommand{\defeq}{\vcentcolon=}
\newcommand{\eqdef}{=\vcentcolon}
\newcommand{\diag}[1]{\,\text{diag}\left( #1 \right)}
\newcommand{\rank}[1]{\rm{rank}\left( #1 \right)}

\newcommand{\bm}[1]{\boldsymbol{#1}}
\newcommand{\SIGN}[1]{\rm{sign}(}

\newtagform{Alph}[\renewcommand{\theequation}{\Alph{equation}}]()
\newtagform{roman}[\renewcommand{\theequation}{\textit{\roman{equation}}}]()
\newtagform{scroman}[\renewcommand{\theequation}{\scshape\roman{equation}}][]

\newtheorem*{myproof*}{Proof}
\newtheorem{lemma}{Lemma}
\newtheorem{remark}{Remark}
\newtheorem{corollary}{Corollary}
\newtheorem{assumption}{Assumption}
\newtheorem{proposition}{Proposition}
\newtheorem{property}{Property}
\newtheorem{theorem}{Theorem}
\newcommand{\bemph}[1]{{\upshape#1}} 
\newcommand{\ep}[1]{\bemph{(}#1\bemph{)}} 

\makeatletter
\DeclareRobustCommand*\uell{\mathpalette\@uell\relax}
\newcommand*\@uell[2]{
  \setbox0=\hbox{$#1\ell$}
  \setbox1=\hbox{\rotatebox{10}{$#1\ell$}}
  \dimen0=\wd0 \advance\dimen0 by -\wd1 \divide\dimen0 by 2
  \mathord{\lower 0.1ex \hbox{\kern\dimen0\unhbox1\kern\dimen0}}
}

\makeatletter
\newcommand{\vast}{\bBigg@{4}}
\newcommand{\Vast}{\bBigg@{5}}
\makeatother



\usepackage[leqno]{empheq}
\usepackage[skins,theorems]{tcolorbox}

\tcbset{colback=gray!20!white,highlight math style={enhanced, colframe=black,colback=gray!20!white,arc=6pt,boxrule=0pt}}

\newtcbox{\otherbox}[1][]{nobeforeafter,math upper,tcbox raise base,
enhanced,frame hidden,boxrule=0pt,interior style={top color=gray!10!white,
bottom color=gray!10!white,middle color=gray!50!white},
fuzzy halo=1pt with black,#1}

\newtcbox{\mybox}[1][]{nobeforeafter,math upper,tcbox raise base, enhanced,boxrule=0.5pt,drop lifted shadow,#1,colframe=gray!50!black,colback=white}

\definecolor{bluegray}{rgb}{0.4, 0.6, 0.8}
\definecolor{myblue}{rgb}{.8, .8, 1}
\newcommand*\mybluebox[1]{%
\colorbox{bluegray}{\hspace{1em}#1\hspace{1em}}}

{\endEmphEqMainEnv}

\newcommand\encircle[1]{%
  \tikz[baseline=(X.base)] 
    \node (X) [draw, shape=circle, inner sep=-2] {\strut #1};}

\makeatletter
\newcommand{\leqnos}{\tagsleft@true\let\veqno\@@leqno}
\newcommand{\reqnos}{\tagsleft@false\let\veqno\@@eqno}
\reqnos
\makeatother

\makeatletter
\def\@author#1{\g@addto@macro\elsauthors{\normalsize%
    \def\baselinestretch{1}%
    \upshape\authorsep#1\unskip\textsuperscript{%
      \ifx\@fnmark\@empty\else\unskip\sep\@fnmark\let\sep=,\fi
      \ifx\@corref\@empty\else\unskip\sep\@corref\let\sep=,\fi
      }%
    \def\authorsep{\unskip,\space}%
    \global\let\@fnmark\@empty
    \global\let\@corref\@empty  
    \global\let\sep\@empty}%
    \@eadauthor={#1}
}
\makeatother

\journal{arxiv.org}
\begin{document}
\begin{frontmatter}

\title{Pipe Roughness Identification of Water Distribution Networks: A Tensor Method}

\author{Stefan Kaltenbacher\corref{cor1}\fnref{EMAIL}}
\author{Martin Steinberger}
\author{Martin Horn}
\address{Institute of Automation and Control\\ Graz University of Technology}
\cortext[cor1]{Address all correspondence to this author.}
\fntext[EMAIL]{\texttt{Email address:} \href{mailto:s.kaltenbacher@tugraz.at}{s.kaltenbacher@tugraz.at} }

\address{Inffeldgasse 21b, 8010 Graz, Austria}

\begin{abstract}

The identification of pipe roughnesses in a water distribution network is formulated as nonlinear system of algebraic equations which turns out to be demanding to solve under real-world circumstances. 
This paper proposes an enhanced technique to numerically solve this identification problem, extending the conventional \textit{Newton-Raphson} approach with second-order derivatives in the determination of the search direction. 
Enabled through some interesting mathematical findings, the resulting formulation can be represented compactly and thus facilitates the development of an efficient and more robust solving-technique. Algorithms on the basis of this more enhanced solving method are then compared to a customized \textit{Newton-Raphson} approach in simulation examples.

\end{abstract}

\begin{keyword}
Tensor Method; Numerical Root Finding; Roughness Calibration; Water Distribution Networks
\end{keyword}

\end{frontmatter}

\section{Introduction}

For the model-based reconstruction of the flow and pressure distribution in a drinking water network
lots of parameters need to be determined.
In this context and given a limited amount of distributed sensors, the unique identification of individual hydraulic friction parameters per pipe in a water distribution network is particularly challenging. 
For instance, their knowledge is essential for the model-based detection and localization of leakages which are mathematically hardly distinguishable from increased friction parameters as far as the steady-state network is concerned.

Effectively, this paper enhances the numerical solving of the problem formulation for the model-based identification of individual roughness values per pipe in the network as proposed in \citep{PipeRoughness_arxiv}. For a more in-depth and comprehensive explanation of the original problem formulation as well as a more detailed literature overview, the reader is referred to \citep{PipeRoughness_arxiv}. 
Also, \citep{PipeRoughness_arxiv} primarily focuses on the deduction of the concrete circumstances which allow a unique solution to this identification problem. Thereby, a set of assumptions \citep[table 1]{PipeRoughness_arxiv} has been introduced which do also apply for this paper. These assumptions and algorithms (originally stated in paper \citep{PipeRoughness_arxiv}) do also have the same reference number in the present manuscript.  

\subsection{Notation}
Generally, vectors and matrices are highlighted bold and italic and are consistently assigned to variables featuring lower- and upper-case letters respectively.
%
 %
%
\noindent Bold $\bm{1}_{x}$ and  $\bm{0}_{x}$ with size $x$ characterize a matrix or vector filled with ones or zeros, whereas size $x$ is only provided if it is unclear from the context. For instance, $\bm{1}_3 = [\begin{matrix} 1 &1 &1\end{matrix}]^T$ or
\[
\bm{0}_{2\times3} = 
\left[
\begin{matrix}
0 &0&0\\
0&0&0
\end{matrix}
\right] .
\]
%
Block matrix entries left blank can be filled with $\bm{0}_x$ of corresponding size $x$.
\noindent The bracket-operator $[\bm{A}]_{ij} = A_{ij}$ applied on matrix $\bm{A} \in \mathbb{K}^{n\times m}$ of a number field $\mathbb{K}$, e.g. $\mathbb{K}=\mathbb{R}$ or $\mathbb{K}=\mathbb{C}$, selects element $A_{ij}$ of matrix $\bm{A}$ in row  $i\in \{1,2,\ldots,n\}$ and column $j \in \{1,2,\ldots,m\}$.
%
\noindent \sloppy Bold letter $\bm{e}_i$ utilizing index $i \in \mathbb{N}$ characterizes a unity vector $\bm{e}_i = [\begin{matrix}0 &\ldots &0 &1 &0 &\ldots &0 \end{matrix}]^T$ with appropriate size where $[\bm{e}_i]_{j} = 0 \,\, \forall i\ne j$ but $[\bm{e}_i]_i=1$.
\noindent In the set of integers $\mathbb{Z}_{\{-1,0,1\}}$, the  subscript  $\{-1,0,1\}$  highlights that only subset $\{-1,0,1\}$ instead of all integers is used.
The ceil operator $\ceil{.}$ is applied to denote the rounding to the next higher integer. The equality symbol supplemented with double dots, as in $\text{ex}_1 \eqdef \text{ex}_2$, denotes an explicit definition which assigns the expression at the equality symbol, i.e.  $\text{ex}_1$, to the expression at the double dots, i.e. $\text{ex}_2$. Please also consider Appendix \ref{app:derivatives} for the definition of derivatives of scalar, vector or matrix functions with respect to vectors.

\newtheorem*{Hadamard}{Definition: Hadamard Product}
\begin{Hadamard}
Let $\bm{A},\bm{B} \in \mathbb{C}^{n\times m}$. The \textit{Hadamard} product of $\bm{A}$ and $\bm{B}$ is defined by $[\bm{A} \odot \bm{B}]_{ij} = A_{ij} B_{ij}$ for all $1\le i \le n,1\le j \le m$.
\end{Hadamard}

This \textit{Hadamard} operator \citep{cite:Hadamard} is also utilized to display element-wise exponentiations as well as inversions in a more compact manner, for instance
\begin{equation*}
\begin{aligned}
[\bm{A}^{\odot^{2}}]_{ij} &= A^2_{ij}\\
[\bm{A}^{\odot^{1/2}}]_{ij} &= A^{1/2}_{ij}\\
[\bm{A}^{\odot^{-1}}]_{ij} &= A^{-1}_{ij}
\end{aligned}
\qquad \qquad \forall 1\le i \le n,1\le j \le m 
\end{equation*}
which certainly provides that  $[\bm{A}]_{ij}=A_{ij} \ne 0 \,\, \forall i,j$ in the context of the inversion $\bm{A}^{\odot^{-1}}$. 
Also, suppose $\beta \in \mathbb{C}$ and $\bm{A, B, C} \in \mathbb{C}^{n \times m}$ the operator is commutative $\bm{A} \odot \bm{B} = \bm{B} \odot \bm{A}$ as well as linear, thus additive $\bm{C} \odot (\bm{A} + \bm{B}) = \bm{C} \odot \bm{A} + \bm{C} \odot \bm{B}$ and homogeneous $\beta (\bm{A} \odot \bm{B}) = (\beta \bm{A}) \odot \bm{B} = \bm{A} \odot (\beta \bm{B})$. 

\section{Preliminaries}

\paragraph{Network Hydraulics}
Represented by a single connected graph \citep[Assumption 1]{PipeRoughness_arxiv}, one distinguishes between  $\mathfrak{I} = \{1,2,\ldots, n_{\rm{j}} \}$ \textit{inner} nodes/vertices and $\mathfrak{S}=\{n_{\rm{j}}+1,\ldots,n_{\rm{s}}+n_{\rm{s}}\}$ source nodes of the network such that their intersection is empty $\mathfrak{I} \cap \mathfrak{S} = \{ \}$ whereas their combination $\mathfrak{I} \cup \mathfrak{S} = \mathfrak{N}=\{1,2,\ldots,n_{\rm{j}}+n_{\rm{s}}\}$ yields the complete set  $\mathfrak{N}$ of all nodes. Thereby, source nodes characterize those where nodal pressure heads $[\bm{h}_s]_i = h_{(i+n_{\rm{j}})}$ for $i=1,\ldots,n_s$ (in m) are known, which is the case at pumps or reservoirs for instance. In analogy, $[\bm{h}]_i = h_{i\in \mathfrak{I}}$ (in m) denotes the vector of nodal pressure heads at the inner nodes which are separated in a part where $n_{\rm{p}}$ out of $n_{\rm{j}}$ are measured  
\begin{subequations}
\begin{equation}
\label{eq:Ch}
\bm{y}_h = \bm{C}_h \bm{h}, \qquad \text{with} \qquad
\bm{C}_h = \left[
\begin{matrix}
 \bm{e}_{p_1} & \bm{e}_{p_2} &\ldots &\bm{e}_{p_{n_{{\rm{p}}}}} 
\end{matrix}
\right]^T
\in \mathbb{Z}^{n_{{\rm{p}}} \times n_{{\rm{j}}}}_{\{0,1\}}
\end{equation}
and its complementary part
\begin{equation} \label{eq:Chb}
\bm{h}_N = \bar{\bm{C}}_h \bm{h} \qquad \text{with} \qquad \bar{\bm{C}}_h = 
\left[
\begin{matrix}
 \bm{e}_{\bar{p}_1} & \bm{e}_{\bar{p}_2} & \ldots & \bm{e}_{\bar{p}_{n_{{\rm{j}}}-n_{{\rm{p}}}}} 
\end{matrix}
\right]^T \in \mathbb{Z}^{(n_{\rm{j}}-n_{\rm{p}}) \times n_{\rm{j}}}_{\{0,1\}}
\end{equation}
\end{subequations}
which is not measured. Matrix $\bm{C}_h$ and its complementary part $\bar{\bm{C}}_h$ are comprised of unity vectors $\bm{e}_i \in \mathbb{Z}_{\{0,1\}}^{n_{j}}$ with the indices $i\in \mathcal{P}=\{p_1,p_2,\ldots,p_{n_{\rm{p}}}\} \subseteq \mathfrak{I}$ and $i \in \bar{\mathcal{P}}=\{\bar{p}_1,\bar{p}_2,\ldots,\bar{p}_{n_{{\rm{j}}}-n_{{\rm{p}}}}\}\subseteq \mathfrak{I}$ respectively such that $\mathcal{P} \cap \bar{\mathcal{P}} = \{\}$ and $\mathcal{P} \cup \bar{\mathcal{P}} = \mathfrak{I}$. This means, for instance, that $\bm{C}_h^T\bm{C}_h + \bar{\bm{C}}_h^T \bar{\bm{C}}_h = \bm{I}_{n_{\rm{j}}}$ (see \citep[Lemma 1]{cite:ModelingHydraulicNetworks}). One also has to consider the geographical elevation at each \textit{inner} node denoted by $[\bm{z}]_i = z_{i\in\mathfrak{I}}$ (in m), whereas it is assumed that source (pressure) heads $\bm{h}_s$ are already increased by the nodal elevation at respective source nodes $\mathfrak{S}$.

Apart from nodal heads, the nodal consumption are denoted by $[\bar{\bm{q}}]_i=q_{i\in\mathfrak{I}}$ which characterize a volumetric flow rate (in m$^3$/s). These nodal consumption are thereby a linear combination of $\mathfrak{P}=\{1,2,\ldots,n_{\rm{{\uell}}}\}$ edge/pipe flows (i.e. volumetric flow rates) denoted by $[\bm{x}_Q]_i = Q_{i \in \mathfrak{P}}$ such that
\begin{equation} \label{eq:ssEQ_1}
\bm{A} \bm{x}_Q = \bar{\bm{q}} 
\end{equation}
utilizing the graph's incidence matrix $\bm{A} \in \mathbb{Z}^{n_{\rm{j}} \times n_{\uell} }_{\{-1,0,1\}}$ of the \textit{inner} nodes $\mathfrak{I}$. The complete incidence matrix $[\begin{matrix}-\bm{A}^T &\tilde{\bm{C}}_s\end{matrix}]^T \in \mathbb{Z}_{\{-1,0,1\}}^{(n_{\rm{j}}+n_{\rm{s}}) \times n_{\uell}}$ also considers source nodes $\mathfrak{S}$ regarding $\tilde{\bm{C}}_s$. The directed graph thereby counts pipe flows influent to \textit{inner} nodes $\mathfrak{I}$ positively and flows effluent of \textit{inner} nodes negatively as far as $\bm{A}$ (e.g. in  \eqref{eq:ssEQ_1}) is concerned. However, the opposite is the case for $\tilde{\bm{C}}_s$ where flows effluent of source nodes $\mathfrak{S}$ (as they should be) are counted positively which is per definition influent to the $\mathfrak{I}$ \textit{inner} nodes. A concrete example (same as in \citep{PipeRoughness_arxiv}) of these matrices is provided in section \ref{sec:sim_example}.

The matrix orthogonal to $\bm{A}^T$, the transposed incidence matrix, (given \citep[Assumption 1]{PipeRoughness_arxiv}) is the so-called cycle matrix $\bm{S} \in \mathbb{Z}_{\{-1,0,1\}}^{(n_{\uell}-n_{\rm{j}}) \times n_{\uell}}$ which accounts for $n_{\uell}-n_{\rm{j}}$ linear independent cycles in the network, i.e. $\rank{\bm{S}} = n_{\uell}-n_{\rm{j}}$ such that $\bm{S}\bm{A}^T = \bm{0}$ (an example for $\bm{S}$ is provided in \citep{PipeRoughness_arxiv}). 

\paragraph{Colebrook-White Flow}

The Darcy-Weisbach formula together with Colebrook-White's implicit friction factor is applied to express pipe friction (see \citep{PipeRoughness_arxiv}), whereas it is assumed that each pipe flow is in the turbulent flow regime \citep[Assumption 8]{PipeRoughness_arxiv} while minor losses are neglected \citep[Assumption 3]{PipeRoughness_arxiv}. For the inversion of the steady-state hydraulic network \citep[Assumption 6]{PipeRoughness_arxiv} with respect to the roughnesses $\bm{\epsilon} \in \mathbb{R}_{\ge 0}^{n_{\uell}}$, this implicit relation is inverted yielding \citep[Eq. (9)]{PipeRoughness_arxiv}
\begin{equation}
Q = f_t(\epsilon,\Delta h) = -\sign{\Delta h}\frac{2}{\ln(10)} \sqrt{\frac{\abs{\Delta h}}{k}} \ln\Bigg(\underbrace{\frac{\epsilon}{3.7 d} + 2.51  \frac{\eta A}{\rho d} \sqrt{\frac{k}{\abs{\Delta h}}}}_{\eqdef \ell(\epsilon, \Delta h)} \Bigg)
\label{eq:ft}
\end{equation}
which expresses the turbulent flow in each pipe $\mathfrak{P}$ as function on the respective roughness $\epsilon$ and the head loss $\Delta h$ of that pipe. Mind that $\epsilon, \Delta h$ as well as the pipes diameter $d$, its cross section area $A$ and $k=\frac{l}{2dg A^2}$, including the pipe's length $l$, would also need a pipe index concerning $\mathfrak{P}$. This index was omitted in \eqref{eq:ft} for readability. Abbreviation $g\approx9.81$ m/s$^2$ thereby denotes the gravitational acceleration. Besides, $\rho$ denotes the water density and $\eta$ the dynamic water viscosity which are treated as constants, although they do vary with temperature.  

As function \eqref{eq:ft} will be central for ongoing considerations the reader shall be reminded of \citep[Remark 3]{PipeRoughness_arxiv} which says that $f_t(\epsilon, \Delta h)$ has a point of discontinuity at $\Delta h =0$ and $\forall \epsilon$, but this is not of concern since $\Delta h=0$ implies $Q=0$, meaning that the pipe flow is motionless and thus certainly not turbulent.


\paragraph{Network and Sensor Configuration}

In order to accommodate for a large number of unknowns, namely the roughnesses $\bm{\epsilon}\in \mathbb{R}_{\ge0}^{n_{\uell}}$ and the not-measured pressure heads $\bm{h}_N^{(i)} \in \mathbb{R}_{\ge0}^{n_{\rm{j}} - n_{\rm{p}}}$ for all $i \in \mathfrak{M}=\{1,2,\ldots,n_{\rm{m}}\}$, several sets of linear independent measurements \citep[Assumption 5]{PipeRoughness_arxiv} denoted by $\mathfrak{M}$, are required. The minimal number of measurements to obtain as many \textit{nodal} equations as unknowns is $n_{\rm{m,min}} = \ceil{n_{\uell}/n_{\rm{p}}}$ \citep[Eq. (14)]{PipeRoughness_arxiv}. Thereby, linear independency can be achieved by the variation of the nodal consumption $\bar{\bm{q}}^{(i)}$ and/or the source heads $\bm{h}_s^{(i)}$ along the measurement-sets $i \in \mathfrak{M}$. Please mind that the network has to be in steady-state in each of these measurement-sets \citep[Assumption 6]{PipeRoughness_arxiv}.
In sum, the roughness identification problem formulation is founded on the assumptions summarized in table \ref{tab:calibrationAssumptions}.
\begin{table}[H]
\begin{center}
\begin{tabular} { P{70pt} | P{290pt}  } 
		Assumption                        & Context                   \\ \hline\hline
		 1             &  connected graph with no self-loops   	  	      \\ \hline 
		2                  &  continuous,  strictly monotonically increasing hydraulic friction 		\\ \hline 
		3  & negligible minor losses\\  \hline 
	       4             & known pipe dimensions, source pressure, consumption	\\ \hline 
5     & number and independency of measurements\\  \hline
6   & measurements in steady-state\\  \hline
7            & bounded measurement noise \\  \hline 
\end{tabular}
\caption{Summary of assumptions relevant for roughness identification (see \citep[table 1]{PipeRoughness_arxiv}).}
\label{tab:calibrationAssumptions}
\end{center}
\end{table}

\vspace{-0.6cm}
Note that the assumption numbering is consistent to the one in \citep{PipeRoughness_arxiv}. Table \ref{tab:calibrationAssumptions} summarizes the same assumptions as \citep[table 1]{PipeRoughness_arxiv}, although the context has been adapted slightly to facilitate comprehensibility. Nonetheless, details about the assumptions in table \ref{tab:calibrationAssumptions} can be found in \citep{PipeRoughness_arxiv}.

\setcounter{assumption}{7}
\begin{assumption} \label{ass:FullTurbulent}
Apart from the assumptions in table \ref{tab:calibrationAssumptions}, it is assumed that each pipe flow $j\in\mathfrak{P}$ in each measurement-set $i\in\mathfrak{M}$ is in the turbulent regime, i.e. 
\begin{equation} \label{eq:Re_bla}
\frac{|Q_j^{(i)}| d_j \rho}{A_j \eta} \ge 4000 \qquad \forall j\in \mathfrak{P} \land \forall i \in \mathfrak{M}.
\end{equation}
\end{assumption}
Assumption \ref{ass:FullTurbulent} is consistent with \citep[Assumption 8]{PipeRoughness_arxiv} and is introduced for the Colebrook-White flow \eqref{eq:ft} to be generally applicable.
%

\section{Problem Statement} \label{sec:Newton}
\vspace{-0.2cm}

The roughness identification problem set-up for the full turbulent case is formulated via nodal equations along the $i$-th measurement-set
%
\begin{subequations}
\label{eq:calibrationTur}
\begin{gather}
\bm{A} \bm{x}_Q(\bm{\epsilon}, \bm{h}_N^{(i)}) = \bar{\bm{q}}^{(i)} \label{eq:calibrationTur_1}\\
\Delta\bm{h}^{(i)} = \tilde{\bm{C}}_s \bm{h}_s^{(i)} - \bm{A}^T \bm{C}^T_h \bm{y}_h^{(i)} - \bm{A}^T \bar{\bm{C}}_h^T \bm{h}_N^{(i)} - \bm{A}^T\bm{z} \label{eq:calibrationTur_2}\\
[\bm{x}_Q(\bm{\epsilon}, \Delta\bm{h}^{(i)})]_j =Q^{(i)}_j \stackrel{\eqref{eq:ft}}{=} f_{t,j}([\bm{\epsilon}]_j, [\Delta\bm{h}^{(i)}]_j) \quad  \forall j \in \mathfrak{P} 
\label{eq:calibrationTur_3}
\end{gather}
\end{subequations}
for all $i\in \{1,2,\ldots, n_{{\rm{m}}}\} =\mathfrak{M}$ where the $j$-th flow component \eqref{eq:calibrationTur_3} used for \textit{Kirchhoff} equations \eqref{eq:calibrationTur_1} is calculated via \eqref{eq:ft}, i.e. the flow in the turbulent regime, thereby applying the conservation of energy for the head losses \eqref{eq:calibrationTur_2}.
Mind that apart from the roughnesses $\bm{\epsilon} \in \mathbb{R}^{n_{\uell}}_{\ge 0}$ also the pressure heads at nodes with no sensors $\bm{h}_N^{(i)} \in \mathbb{R}^{n_{{\rm{j}}}-n_{{\rm{p}}}}_{\ge 0}$ in the $i$-th measurement-set are unknown.

In principle, all information needed to determine all the pipes' roughnesses is contained in \eqref{eq:calibrationTur} provided that the assumptions in table \ref{tab:calibrationAssumptions} in addition to Assumption \ref{ass:FullTurbulent} hold. Important to repeatedly emphasize is the requirement of at least $n_{\rm{m,min}} = \ceil{n_{\uell}/n_{\rm{p}}}$ measurement-sets (see \citep{PipeRoughness_arxiv}) which have to be independent from each other. However, set \eqref{eq:calibrationTur} turns out to be particularly difficult to solve even in the unperturbed case when no measurement noise is considered. The reason for that can not only be attributed to the problem's size but to the nonlinear dependency of \eqref{eq:ft} on $\bm{h}_N^{(i)}$.

In sum, the purpose of this paper is the development of techniques which facilitate the solving of \eqref{eq:calibrationTur} whilst computational demand and memory requirements remain manageable.
\vspace{-0.2cm}
\subsection{Introduction to the Tensor Method}
The aim is to solve a considerably large, nonlinear set of equations presented in the form $\bm{f}(\bm{x}) \stackrel{!}{=} \bm{0}$
with a smooth and continuous vector function $\bm{f}: \mathbb{R}^{n} \rightarrow \mathbb{R}^{n}$ by means of an iterative scheme $\bm{x}_k = \bm{x}_{k-1}+ \mu_k \Delta \bm{x}_k$  with step length $\mu_k$ and search direction $\Delta \bm{x}_k$ along iterations denoted by $k$. The equivalence of this equation set to \eqref{eq:calibrationTur} becomes apparent when considering
\begin{subequations}
\label{eq:fx_xQTur_1}
\begin{align}
\bm{f}(\bm{x}) &= \left[
\begin{matrix}
\bm{A} & &\\
 &  \ddots&\\
& &  \bm{A}
\end{matrix}
\right] \left[
\begin{matrix}
\bm{x}_Q(\bm{\epsilon}, \bm{h}_N^{(1)})\\
\bm{x}_Q(\bm{\epsilon}, \bm{h}_N^{(2)})\\
\vdots\\
\bm{x}_Q(\bm{\epsilon}, \bm{h}_N^{(n_{{\rm{m}}})})
\end{matrix}
\right] - \left[
\begin{matrix}
\bar{\bm{q}}^{(1)}\\
\bar{\bm{q}}^{(2)}\\
\vdots\\
\bar{\bm{q}}^{(n_{{\rm{m}}})}\\
\end{matrix}
\right] \quad 
\text{with} \\
 \bm{x}^T &= \left[
\begin{matrix}
\bm{\epsilon}^T & \bm{h}_N^{(1)^T} & \ldots & \bm{h}_N^{(n_{{\rm{m}}})^T}
\end{matrix}
\right] ,
\raisetag{10pt}
\end{align}
\end{subequations}
where nonlinearities appear in the turbulent flow $\bm{x}_Q(\bm{\epsilon},\bm{h}_N^{(i)})$ expressed via \eqref{eq:ft}.
 This should provide convergence to the real root $\bm{x}^* \Rightarrow \bm{f}(\bm{x}^*) = \bm{0}$ of \eqref{eq:calibrationTur}.
However, function $\bm{f} : \mathbb{R}^{n_{{\rm{m}}} n_{{\rm{j}}}} \rightarrow \mathbb{R}^{n_{\uell} + n_{{\rm{m}}} (n_{{\rm{j}}} - n_{{\rm{p}}})}$ has, in general, not the same number of components as the number of variables, i.e. $n_{{\rm{m}}} n_{{\rm{j}}} \ne n_{\uell} + n_{{\rm{m}}} (n_{{\rm{j}}} - n_{{\rm{p}}})$. Its \textit{Jacobian}
\begin{equation}
\bm{J}(\bm{x}) =  \left[
\begin{matrix}
\bm{A} & &\\
 &  \ddots&\\
& &  \bm{A}
\end{matrix}
\right] \left[
\begin{matrix}
\frac{\partial \bm{x}_Q(\bm{\epsilon},\bm{h}_N^{(1)})}{\partial \bm{\epsilon}} & \frac{\partial \bm{x}_Q(\bm{\epsilon},\bm{h}_N^{(1)})}{\partial \bm{h}_N^{(1)}} &\bm{0} &\ldots &\bm{0} \\
\frac{\partial \bm{x}_Q(\bm{\epsilon},\bm{h}_N^{(2)})}{\partial \bm{\epsilon}} & \bm{0}&  \frac{\partial \bm{x}_Q(\bm{\epsilon},\bm{h}_N^{(2)})}{\partial \bm{h}_N^{(2)}}  &\ldots &\bm{0}\\
\vdots & \vdots & &\ddots\\
\frac{\partial \bm{x}_Q(\bm{\epsilon},\bm{h}_N^{(n_{{\rm{m}}})})}{\partial \bm{\epsilon}} & \bm{0}&  \bm{0}&  \ldots &\frac{\partial \bm{x}_Q(\bm{\epsilon},\bm{h}_N^{(n_{{\rm{m}}})})}{\partial \bm{h}_N^{(n_{{\rm{m}}})}}  
\end{matrix}
\right],
\label{eq:Jacobian}
\end{equation}
with $\bm{J} \in \mathbb{R}^{n_{{\rm{m}}} n_{{\rm{j}}} \times n_{\uell} + n_{{\rm{m}}} (n_{{\rm{j}}} - n_{{\rm{p}}})}$ is supposed to have full $\rank{\bm{J}(\bm{x})} =  n_{\uell} + n_{{\rm{m}}} (n_{{\rm{j}}} - n_{{\rm{p}}})$ in reference to Assumption 5. Abbreviation $\bm{J}_{k-1} \defeq \bm{J}(\bm{x}_{k-1})$ is utilized along solving-iteration $k$. 
In \citep{PipeRoughness_arxiv} a modified version of \textit{Newton-Raphson}'s algorithm was applied which makes use of a customized method to vary the step length. In this paper, however, we will focus on how to obtain a search direction which accounts for additional nonlinearities of $\bm{f}(\bm{x})$ when compared to the \textit{Newton} step $\Delta \bm{x}_k = -\bm{J}^{-1}_{k-1} \bm{f}(\bm{x}_{k-1})$ (suppose $\bm{J}$ is square for the sake of simplicity).

\subsection{First Turbulent Flow Derivatives} \label{sec:FirstDerivatives}
For the determination of the \textit{Jacobian} according to \eqref{eq:Jacobian}, the derivatives of \eqref{eq:ft} with respect to $[\bm\epsilon]_i = \epsilon_i$ $\forall i\in\mathfrak{P}$ (over all pipes) and $[\bm{h}_N]_j = h_{N,j}$ $\forall j\in \bar{\mathcal{P}}$ (over all nodes with no sensors) are required.  In analogy to \citep[section 5.1]{PipeRoughness_arxiv}, the parameter index along each pipe, that is $i$, e.g. used for diameter $d_i$ $\forall i\in \mathfrak{P}$, is omitted in the following derivatives to improve readability.
One obtains
\begin{subequations}
\label{eq:dQdx}
\begin{equation}
\frac{\partial f_{t,i}}{\partial \epsilon_i} \eqdef p_{\epsilon,i} (\epsilon_i, \Delta h_i) \mathrel{\hat=} \frac{\partial f_{t}}{\partial \epsilon} = -\frac{2}{\ln{(10)}} \text{sign}(\Delta h) \sqrt{\frac{\abs{\Delta h}}{k}}   \frac{1}{3.7 d  \, \ell(\epsilon, \Delta h)} \quad \forall i\in \mathfrak{P}
\label{eq:dQde_scalar}
\end{equation}
followed by $\frac{\partial f_{t,i}}{\partial h_{N,j}} = \frac{\partial f_{t,i}}{\partial \Delta h_i} \frac{\partial \Delta h_i}{\partial h_{N,j}} \eqdef p_{\Delta h, i} (\epsilon_i,\Delta h_i) \frac{\partial \Delta h_i}{\partial h_{N,j}} \quad \forall i\in \mathfrak{P} \land j \in \bar{\mathcal{P}} \mathrel{\hat=}$
\begin{equation}
\frac{\partial f_t}{\partial h_N} = - \frac{1}{\ln{(10)}} \left( \sqrt{\frac{1}{k \abs{\Delta h}}} \ln{(\ell(\epsilon,\Delta h))} - 2.51 \frac{\eta A}{\rho d} \frac{\abs{\Delta h}^{-1}}{\ell{(\epsilon, \Delta h})}\right) \frac{\partial \Delta h}{\partial h_N} 
\label{eq:dQdhN_scalar}
\end{equation}
\end{subequations}
where the partial derivative of $\Delta h$ with respect to $h_N$ is constant due to
\begin{equation}
\label{eq:dhdhN}
\frac{\partial \Delta \bm{h}}{\partial \bm{h}_N} \stackrel{\eqref{eq:calibrationTur_2}}{=} -\bm{A}^T \bar{\bm{C}}_h, \,\,\,
\begin{aligned} 
 \quad [\Delta \bm{h}]_i &=[\bm{h}_{\txt{loss}}(\bm{x}_Q)]_i = \Delta h_i \,\,\forall i\in \mathfrak{P} \\
 \quad [\bm{h}_N]_j &= [\bar{\bm{C}}_h \bm{h}]_j = h_{N,\bar{p}_j} \,\, \forall \bar{p}_j\in \bar{\mathcal{P}} \land j \in \{1,\ldots,n_{\rm{j}}-n_{\rm{p}}\}
\end{aligned}
 \end{equation}
when considering vector dependencies. As a remark, note that the authors assumed that $\frac{\partial}{\partial \Delta h} \text{sign}(\Delta h) = 0$ neglecting the \textit{Dirac-Delta} $\delta(\Delta h)$ function. The scalar partial derivatives \eqref{eq:dQdx} can now be summarized in vector notation as follows
\begin{equation}
\label{eq:dxQ_der}
\begin{aligned}
\frac{\partial \bm{x}_Q(\bm{\epsilon},\Delta \bm{h}^{(i)})}{\partial \bm{\epsilon}} &= \diag{\bm{p}_{\epsilon}(\bm{\epsilon},\Delta \bm{h}^{(i)})} \\
 \frac{\partial \bm{x}_Q(\bm{\epsilon},\Delta \bm{h}^{(i)})}{\partial \bm{h}_N^{(i)}} &= -\diag{\bm{p}_{\Delta h}(\bm{\epsilon},\Delta \bm{h}^{(i)})} \bm{A}^T \bar{\bm{C}}_h^T
\end{aligned}
\quad \forall i \in \mathfrak{M}=\{1,2,\ldots n_{{\rm{m}}}\} 
\end{equation}
where $[\bm{p}_{\epsilon}]_j \stackrel{\eqref{eq:dQde_scalar}}{=} p_{\epsilon,j}(\epsilon_j,\Delta h_j)$ and $[\bm{p}_{\Delta h}]_j \stackrel{\eqref{eq:dQdhN_scalar}}{=} p_{\Delta h,j}(\epsilon_j,\Delta h_j)$ for all $j \in \mathfrak{P}$. Actually, one can recognize that the information concerning $\bm{J}$ which varies along the $i$-th measurement-sets can entirely be stored in vectors by means of $\bm{p}_{\epsilon}$ and $\bm{p}_{\Delta h}$.


\subsection{Tensor Extension of the Search Direction} \label{sec:TensorMethod}

Applying a \textit{Newton-Raphson} type algorithm with step length variation \citep[Algorithm 1]{PipeRoughness_arxiv} on a simulation example in \citep{PipeRoughness_arxiv}, the real root $\bm{x}^*$ had a considerable smaller residual $\Vert \bm{f} \Vert_{\mathcal{L}_1} (\bm{x}^*) = \abs{f_1} + \ldots + \abs{f_{n_{\rm{j}} n_{\rm{m}}}}= v(\bm{x}^*)= 0.011 \times 10^{-2}$ l/s when compared to solutions of this  \citep[Algorithm 1]{PipeRoughness_arxiv}. The potential to improve the result in terms of the residual $v(\bm{x})$ is studied when applying a more advanced method for the determination of the search direction.

Generally, the search direction is determined by setting the truncated \textit{Taylor} series to zero, whereas \textit{Newton} proposes to truncate after the linear term. The primary idea of the \textit{Tensor Method} is to also consider the second order term of this \textit{Taylor} series for the calculation of the search direction with the intention to account for additional non-linearities of $\bm{f}(\bm{x})$.
\begin{equation}
\bm{f}(\bm{x}_k) = \bm{f}(\bm{x}_{k-1}) + \left. \frac{\partial \bm{f}}{\partial \bm{x}} \right|_{\bm{x}_{k-1}} \Delta \bm{x}_k + \frac{1}{2} 
\left[
\begin{matrix}
\Delta \bm{x}_k^T \left. \bm{\mathcal{H}}(f_1)\right|_{\bm{x}_{k-1}} \Delta \bm{x}_k\\
\Delta \bm{x}_k^T \left. \bm{\mathcal{H}}(f_2)\right|_{\bm{x}_{k-1}} \Delta \bm{x}_k\\
\vdots\\
\Delta \bm{x}_k^T \left. \bm{\mathcal{H}}(f_{n_{\rm{m}}n_{\rm{j}}})\right|_{\bm{x}_{k-1}} \Delta \bm{x}_k\\
\end{matrix}
\right]
+  \mathcal{O}(\Delta \bm{x}_k^{3})
\label{eq:Taylor_tensor}
\end{equation}
Symbol $\bm{\mathcal{H}}(f_i)$ denotes the \textit{Hessian} matrix 
of the $i$-th component of $\bm{f}(\bm{x})$ (see also Appendix \ref{app:derivatives}). Following this procedure, the search direction $\Delta \bm{x}_k$ as solution of
\begin{equation}
\bm{f}(\bm{x}_{k-1})
+ 
\left. \frac{\partial \bm{f}}{\partial \bm{x}} \right|_{\bm{x}_{k-1}} \Delta \bm{x}_k
+ 
\frac{1}{2} 
\left[
\begin{matrix}
\Delta \bm{x}_k^T \left. \bm{\mathcal{H}}(f_1)\right|_{\bm{x}_{k-1}} \Delta \bm{x}_k\\
\Delta \bm{x}_k^T \left. \bm{\mathcal{H}}(f_2)\right|_{\bm{x}_{k-1}} \Delta \bm{x}_k\\
\vdots\\
\Delta \bm{x}_k^T \left. \bm{\mathcal{H}}(f_{n_{\rm{m}}n_{\rm{j}}})\right|_{\bm{x}_{k-1}} \Delta \bm{x}_k\\
\end{matrix}
\right]
= \bm{0} ,
\label{eq:TensorEquation}
\end{equation}
 referred to as \textit{Tensor Equation}, is called the \textit{Tensor Method} \citep{cite:TensorSparse,cite:NumericalOptimization}. 
In practice, however, there are two particularly good reasons why the \textit{Tensor Equation} is hardly applied for the determination of the search direction, although it is known that once \eqref{eq:TensorEquation} is solved, significantly less iterations are necessary for the iterative scheme to converge to $\bm{x}^*$ for solving $\bm{f}(\bm{x}) = \bm{0}$  \citep{cite:TensorSparse}:
\begin{itemize}
\item assuming the problem has as many unknowns as equations one would need to store $n^3$ entries, where $n=n_{{\rm{m}}} n_{{\rm{j}}}$, evaluating \textit{Hessian} $\bm{\mathcal{H}}(f_1), \bm{\mathcal{H}}(f_2), \ldots$ in each iteration
\item the \textit{Tensor Equation} as in \eqref{eq:TensorEquation} is an actual hard problem to solve, especially for large problems
\end{itemize}

In this paper it is shown that the former issue (first bullet point) can actually be completely resolved for the current application. Concerning the second bullet point, however, the authors present findings which lead to the strong presumption that the second problem may be mitigated substantially for the roughness identification problem.

\subsection{Second Turbulent Flow Derivatives} \label{sec:SecondDerivatives}
This section is supplement to section \ref{sec:FirstDerivatives} and provides the appropriate second derivatives of the turbulent flow \eqref{eq:ft} with respect to the roughness $\bm{\epsilon}$ and the not-measured pressure head $\bm{h}_N$. One receives
\begin{subequations}
\begin{align}
\frac{\partial^2 Q_i}{\partial \epsilon_j \partial \epsilon_k} \stackrel{\eqref{eq:Re_bla}}{=} \frac{\partial^2 f_{t,i}}{\partial \epsilon_j \partial \epsilon_k} &= 0 \quad \text{for} \quad  \forall i \ne j \ne k \in \mathfrak{P}, \\
\frac{\partial^2 f_{t,i}}{\partial \epsilon_j \partial h_{N,k}} &= 0 \quad \text{for} \quad \forall i \ne j \in \mathfrak{P} \land k\in \bar{\mathcal{P}}
\end{align}
%
which holds always for all pipes, whereas one has to expect that
\begin{gather}
\frac{\partial^2 f_{t,i}}{\partial \epsilon_i^2} \ne 0 \quad \text{for} \quad i \in \mathfrak{P}, \qquad 
\frac{\partial^2 f_{t,i}}{\partial \epsilon_i \partial h_{N,j}} \ne 0 \quad \text{for} \quad i \in \mathfrak{P} \land j \in \bar{\mathcal{P}}.
\end{gather}
\end{subequations}
%
\begin{remark}
As the turbulent flow \eqref{eq:ft}, i.e. the flow in the turbulent regime where $Re \ge 4000$, is at least two times continuously differentiable with respect to $\epsilon$ and $\Delta h$, the second derivatives of \eqref{eq:ft} satisfy
\begin{subequations}
\begin{align}
\frac{\partial^2 f_{t,i}}{\partial \epsilon_i \partial h_{N,j}} &= \frac{\partial^2 f_{t,i}}{\partial  h_{N,j} \partial \epsilon_i } \quad \forall  i \in \mathfrak{P} \land j \in \bar{\mathcal{P}}\\
\frac{\partial^2 f_{t,i}}{\partial h_{N,k}\partial h_{N,j}} &= \frac{\partial^2 f_{t,i}}{\partial h_{N,j}\partial h_{N,k}} \quad \forall  i \in \mathfrak{P} \land j,k \in \bar{\mathcal{P}} ,
\end{align}
\end{subequations}
i.e. they are symmetrical.
\end{remark}
%
Suppressing indices for the sake of readability, one obtains the following derivatives while applying the chain rule $\frac{\partial \bm{f}_t}{\partial \bm{h}_N} = \frac{\partial \bm{f}_t}{\partial \Delta \bm{h}} \frac{\partial \Delta \bm{h}}{\partial \bm{h}_N}$ and \eqref{eq:dhdhN}: 
\begingroup
\allowdisplaybreaks
\begin{subequations}
\label{eq:flow_derive_1}
\begin{gather}
\begin{aligned}
\label{eq:Q2_eps2}
\frac{\partial^2 f_{t,i}}{\partial \epsilon_i^2} &\mathrel{\hat=}\frac{\partial^2 f_{t}}{\partial \epsilon^2}  =   \frac{\SIGN(\Delta h) \frac{2}{\ln{(10)}}   \sqrt{\frac{\abs{\Delta h}}{k}} \left( \frac{1}{3.7 d}\right)^2   }{\ell^2(\epsilon,\Delta h)} \\
&\eqdef [\bm{p}_{\epsilon^2}(\bm{\epsilon},\Delta \bm{h})]_i = p_{\epsilon^2,i}(\epsilon_i,\Delta h_i) \qquad \forall i \in \mathfrak{P}
\end{aligned}\\
\begin{aligned}
\label{eq:Q2_epsdh}
\frac{\partial^2 f_{t,i}}{\partial \epsilon_i \partial \Delta h_i} &\mathrel{\hat=} \frac{\partial^2 f_{t}}{\partial \epsilon \partial \Delta h} = -\frac{1}{\ln{(10)}} \left( \sqrt{\frac{1}{k \abs{\Delta h}}} \frac{1}{3.7 d \, \ell(\epsilon,\Delta h)} + \frac{2.51 \eta A}{3.7 \rho d^2} \frac{\abs{\Delta h}^{-1}}{\ell^2(\epsilon,\Delta h)} \right) \\
&\eqdef [\bm{p}_{\epsilon \Delta h}(\bm{\epsilon},\Delta \bm{h})]_i = p_{\epsilon \Delta h,i} (\epsilon_i,\Delta h_i)  \qquad \forall i \in \mathfrak{P} \raisetag{0.5cm}
\end{aligned}
\end{gather}
\end{subequations}
\endgroup
The vector function $\bm{f}_t(\bm{\epsilon},\Delta \bm{h}) = [\begin{matrix} f_{t,1}(\epsilon_1,\Delta h_1) &\ldots &f_{t,n_{\uell}}(\epsilon_{n_{\uell}},\Delta h_{n_{\uell}})\end{matrix}]^T$ thereby denotes function \eqref{eq:ft} evaluated for each pipe flow while applying abbreviation $\ell(\epsilon,\Delta h)$ \eqref{eq:ft}. In this context it is clear to see that, e.g., $\frac{\partial \bm{f}_t}{\partial \Delta \bm{h}} = \diag{\bm{p}_{\Delta h}}$ is diagonal.
As in section \ref{sec:FirstDerivatives}, note that the author assumed that $\frac{\partial}{\partial \Delta h} \SIGN (\Delta h) = 0$, neglecting the \textit{Dirac-Delta} $\delta(\Delta h)$ function.
Applying unity vector $\bm{e}_i \in \mathbb{Z}_{\{0,1\}}^{n_{\uell}}$ when writing \eqref{eq:Q2_epsdh} in  vector notation one obtains
\begin{align}
\frac{\partial^2 f_{t,i}}{\partial \bm{\epsilon} \partial \bm{h}_N} &= \frac{\partial}{\partial \Delta \bm{h}_N} \left( \frac{\partial f_{t,i}}{\partial \epsilon_i} \frac{\partial \epsilon_i}{\partial \bm{\epsilon}} \right)^T = \left( \frac{\partial \epsilon_i}{\partial \bm{\epsilon}} \right)^T \frac{\partial^2 f_{t,i}}{\partial \epsilon_i \partial \Delta h_i } \frac{\partial \Delta h_i}{\partial \bm{h}_N} \nonumber\\
&= - \bm{e}_i  p_{\epsilon \Delta h, i}  \bm{e}_i^T \bm{A}^T \bar{\bm{C}}_h^T \quad \forall i \in \mathfrak{P} .
\label{eq:Qtur_dedhN_vec}
\end{align} 

Regarding the second derivative with respect to $\bm{h}_N$ in the scalar form
\begin{equation}
\frac{\partial^2 f_{t,i}}{\partial h_{N,j} \partial h_{N,k}} = \frac{\partial^2 f_{t,i}}{\partial \Delta h_i^2} \frac{\partial \Delta h_i}{\partial h_{N,j} } \frac{\partial \Delta h_i}{\partial h_{N,k} } \qquad i \in \mathfrak{P} \land j,k \in \bar{\mathcal{P}}
\label{eq:Q_der_ijk}
\end{equation}
one obtains
\begin{subequations}
\label{eq:Q2_mixed}
\begin{align}
\frac{\partial^2 f_{t,i}}{\partial \Delta h_i^2}  &\eqdef [\bm{p}_{\Delta h^2}(\bm{\epsilon},\Delta \bm{h})]_i = p_{\Delta h^2,i}(\epsilon_i,\Delta h_i) \quad \forall i \in \mathfrak{P} \mathrel{\hat=}\\
\frac{\partial^2 f_t}{\partial \Delta h^2} &= \frac{\SIGN(\Delta h)}{2 \ln{(10)}} \left[ \frac{\abs{\Delta h}^{-3/2}}{\sqrt{k}} \ln{(\ell(\epsilon,\Delta h))} - 2.51 \frac{\eta A}{\rho d} \frac{\abs{\Delta h}^{-2}}{\ell(\epsilon,\Delta h)} + \left( 2.51 \frac{\eta A}{\rho d}\right)^2 \frac{\sqrt{k} \abs{\Delta h}^{-5/2}}{\ell^2(\epsilon,\Delta h)}\right]
\end{align}
\end{subequations}
yet again suppressing indices (one would simply add, analogously to \eqref{eq:flow_derive_1}, the identical index for all pipe parameters, $\Delta h$ and $\epsilon$).
Applying vector notation, \eqref{eq:Q_der_ijk} yields
\begin{align}
\frac{\partial^2 f_{t,i}}{\partial \bm{h}_N^2} &= \frac{\partial}{\partial \bm{h}_N} \left( \frac{\partial f_{t,i}}{\partial \Delta h_i} \frac{\partial \Delta h_i}{\partial \bm{h}_N}\right)^T = \left( \frac{\partial \Delta h_i}{\partial \bm{h}_N} \right)^T \frac{\partial^2 f_{t,i}}{\partial \Delta h_i^2} \frac{\partial \Delta h_i}{\partial \bm{h}_N} \nonumber\\
&=\left(  \bar{\bm{C}}_h \bm{A} \bm{e}_i \right)  p_{\Delta h^2,i} \left( \bm{e}_i^T \bm{A}^T \bar{\bm{C}}_h^T \right)
 \quad \forall i \in \mathfrak{P} .
\label{eq:Q2_dhN_split}
\end{align}
Finally, things become structurally more appealing by rewriting the derivatives. Denoting
\begin{gather}
\bm{A} = \left[
\begin{matrix}
\bm{a}_1 & \bm{a}_2 &\ldots &\bm{a}_{n_{\uell}}
\end{matrix}
\right]
= \left[ 
\begin{matrix}
\bm{b}_1 &  \bm{b}_2 &\ldots & \bm{b}_{n_{{\rm{j}}}} 
\end{matrix}
\right]^T 
\label{eq:A_split}
\end{gather}
with $\bm{a}_i \in \mathbb{Z}^{n_{{\rm{j}}}}_{\{-1,0,1\}} \, \forall i \in \mathfrak{P}$ and $\bm{b}_i \in \mathbb{Z}_{\{-1,0,1\}}^{n_{\uell}} \, \forall i \in \mathfrak{I}$ while rewriting  \eqref{eq:Q_der_ijk} and \eqref{eq:flow_derive_1}, the second turbulent flow derivatives 
\begin{subequations}
\label{eq:Qtur_d2}
\begin{align}
\frac{\partial^2 f_{t,i}}{\partial \bm{h}_{N}^2 } &\stackrel{\eqref{eq:Q2_mixed} \land \eqref{eq:Q2_dhN_split}}{=} p_{\Delta h^2,i}   \bar{\bm{C}}_h \bm{a}_i \bm{a}_i^T \bar{\bm{C}}_h^T  \label{eq:Qtur_dhN}\\
\frac{\partial^2 f_{t,i}}{\partial \bm{\epsilon}^2} &\stackrel{\eqref{eq:Q2_eps2}}{=} p_{\epsilon^2,i} \bm{e}_i \bm{e}_i^T\label{eq:Qtur_deps}\\
 \frac{\partial^2 f_{t,i}}{\partial \bm{\epsilon} \partial \bm{ h}_N} &= \left(\frac{\partial^2 f_{t,i}}{ \partial \bm{ h}_N \partial \bm{\epsilon}}\right)^T \stackrel{\eqref{eq:Q2_epsdh} \land \eqref{eq:Qtur_dedhN_vec}}{=}  -p_{\epsilon \Delta h,i} \bm{e}_i \bm{e}_i^T\bm{A}^T \bar{\bm{C}}_h^T 
 \label{eq:Qtur_dhNdeps}
\end{align}
\end{subequations}
%
$\forall i \in \mathfrak{P}$, i.e. along $i=1,2,\ldots,n_{\uell}$, can be presented compactly where nonlinearties are stored in vectors $\bm{p}_{\epsilon^2}, \bm{p}_{\epsilon \Delta h}$ and $\bm{p}_{\Delta h^2}$ which are functions on $\bm{\epsilon}$ and $\Delta \bm{h}$. However, the $i$-th component of these vector functions only depends on the $i$-th component of $\bm{\epsilon}$ and $\Delta \bm{h}$.

In analogy to the first turbulent flow derivatives in section \ref{sec:FirstDerivatives}, it can be recognized that the information along the different measurement-sets can be entirely stored in vector functions $\bm{p}_{\epsilon}, \bm{p}_{\Delta h}$ concerning \eqref{eq:dxQ_der} as well as $\bm{p}_{\epsilon^2}, \bm{p}_{\epsilon \Delta h}, \bm{p}_{\epsilon^2}$ concerning \eqref{eq:Qtur_d2}. This will be important for the next steps.

\section{Tensor Equation} \label{sec:TensorEquation}

\begin{lemma} \label{lemma:Tensor_A}
Writing $\bm{\mathcal{H}}_{k-1}(f) = \bm{\mathcal{H}}(f)(\bm{x}_{k-1})$ for any scalar function $f$, the Hessian term in the Tensor Equation \eqref{eq:TensorEquation} concerning the equation set \eqref{eq:calibrationTur} yields
\begin{equation}
\label{eq:TensorEquation2}
\left[
\begin{matrix}
\Delta \bm{x}_k^T \bm{\mathcal{H}}_{k-1}(f_{1}) \Delta \bm{x}_k\\
\Delta \bm{x}_k^T \bm{\mathcal{H}}_{k-1}(f_{2}) \Delta \bm{x}_k\\
\vdots\\
\Delta \bm{x}_k^T \bm{\mathcal{H}}_{k-1}(f_{n_{{\rm{m}}} n_{{\rm{j}}}}) \Delta \bm{x}_k\\
\end{matrix}
\right] = 
 \left[
\begin{matrix}
\bm{A} && &\\
 &\bm{A} &\\
 & & \ddots&\\
& & & \bm{A}
\end{matrix}
\right]
\left[
\begin{matrix}
\Delta \bm{x}_k^T \bm{\mathcal{H}}_{k-1}(Q^{(1)}_{1}) \Delta \bm{x}_k\\
\Delta \bm{x}_k^T \bm{\mathcal{H}}_{k-1}(Q^{(1)}_{2}) \Delta \bm{x}_k\\
\vdots\\
\Delta \bm{x}_k^T \bm{\mathcal{H}}_{k-1}(Q^{(1)}_{n_{\uell}}) \Delta \bm{x}_k\\
\Delta \bm{x}_k^T \bm{\mathcal{H}}_{k-1}(Q^{(2)}_{1}) \Delta \bm{x}_k\\
\vdots\\
\Delta \bm{x}_k^T \bm{\mathcal{H}}_{k-1}(Q^{(n_{{\rm{m}}})}_{n_{\uell}}) \Delta \bm{x}_k\\
\end{matrix}
\right]
\end{equation}
when considering pipe flow $Q^{(i)}_{j} = Q(\epsilon_j,\Delta h_j^{(i)}) \,\, \forall j\in\mathfrak{P} \land i \in \mathfrak{M}$ which is a function on the roughness and the head loss generally.
\end{lemma}
%
\begin{myproof*}
\normalfont
It is sufficient to show that 
\begin{equation}
\label{eq:proof_A_tensor}
\left[
\begin{matrix}
\Delta \bm{x}_k^T \bm{\mathcal{H}}_{k-1}(f_{1+(i-1)n_{{\rm{j}}}}) \Delta \bm{x}_k\\
\Delta \bm{x}_k^T \bm{\mathcal{H}}_{k-1}(f_{2+(i-1)n_{{\rm{j}}}}) \Delta \bm{x}_k\\
\vdots\\
\Delta \bm{x}_k^T \bm{\mathcal{H}}_{k-1}(f_{n_{{\rm{j}}}+(i-1)n_{{\rm{j}}}}) \Delta \bm{x}_k\\
\end{matrix}
\right]  \stackrel{!}{=} 
\bm{A}
\left[
\begin{matrix}
\Delta \bm{x}_k^T \bm{\mathcal{H}}_{k-1}(Q^{(i)}_{1}) \Delta \bm{x}_k\\
\Delta \bm{x}_k^T \bm{\mathcal{H}}_{k-1}(Q^{(i)}_{2}) \Delta \bm{x}_k\\
\vdots\\
\Delta \bm{x}_k^T \bm{\mathcal{H}}_{k-1}(Q^{(i)}_{n_{\uell}}) \Delta \bm{x}_k\\
\end{matrix}
\right] 
\end{equation}
for measurement-set $i=1$. The validity of \eqref{eq:proof_A_tensor} for remaining $i=2,3,\ldots,n_{{\rm{m}}}$ measurement-sets follows subsequently. Let $a_{ij} = [\bm{A}]_{ij}$ and
$
\bm{A} = [ 
\begin{matrix}
\bm{b}_1 &  \bm{b}_2 &\ldots & \bm{b}_{n_{{\rm{j}}}} 
\end{matrix}
]^T
$
according to \eqref{eq:A_split}. Hence $\bm{b}_j = [\begin{matrix} a_{j,1} &a_{j,2}&\ldots&a_{j,n_{\uell}}\end{matrix}]^T \, \forall j \in \mathfrak{I}$. Omitting the $1$-index in $\bm{A}\bm{x}_Q^{(1)} - \bar{\bm{q}}^{(1)} = [\begin{matrix}f_1 &\ldots &f_{n_{{\rm{j}}}} \end{matrix}]^T \stackrel{!}{=} \bm{0}$ concerning the first measurement-sets in favor of readability, one receives
\begin{gather}
 \bm{\mathcal{H}}_{k-1}(f_j) = \bm{\mathcal{H}}_{k-1}(\bm{b}^T_j \bm{x}_{Q})  =  a_{j,1} \bm{\mathcal{H}}_{k-1}(Q_{1}) + a_{j,2} \bm{\mathcal{H}}_{k-1}(Q_{2}) + \ldots
 +a_{j,n_{\uell}} \bm{\mathcal{H}}_{k-1}(Q_{n_{\uell}}) \nonumber \\
  \forall j \in \mathfrak{I}
 \label{eq:Tensor_Hessian_split}
\end{gather}
exploiting linearity of the \textit{Hessian} operator. Extended by the search direction from the left and right, \eqref{eq:Tensor_Hessian_split} yields
\begin{equation}
\Delta \bm{x}_k^T \bm{\mathcal{H}}_{k-1}(f_j) \Delta \bm{x}_k= \bm{b}^T_j
\left[
\begin{matrix}
\Delta \bm{x}_k^T \bm{\mathcal{H}}_{k-1}(Q_{1}) \Delta \bm{x}_k\\
\Delta \bm{x}_k^T \bm{\mathcal{H}}_{k-1}(Q_{2}) \Delta \bm{x}_k\\
\vdots\\
\Delta \bm{x}_k^T \bm{\mathcal{H}}_{k-1}(Q_{n_{\uell}}) \Delta \bm{x}_k\\
\end{matrix}
\right] \qquad \forall j \in \mathfrak{I} .
 \label{eq:Tensor_Hessian_splitFinal}
\end{equation}
By extending \eqref{eq:Tensor_Hessian_splitFinal} for each of node $j\in\mathfrak{I}$, one obtains \eqref{eq:proof_A_tensor}. Additionally, extended with the remaining measurements-sets $2,3,\ldots,n_{{\rm{m}}}$, it is apparent that \eqref{eq:TensorEquation2} is equivalent to the nonlinear term in \eqref{eq:TensorEquation}. This completes the proof.  \hfill$\qed$
\end{myproof*}
%
It is important to note that Lemma \ref{lemma:Tensor_A} does hold not only for the turbulent flow, but also for laminar and transitional flow.

\paragraph{Search Direction}
Knowing that $[\Delta \bm{x}_k]_i = \Delta x_{k,i}$ for $i=1,2,\ldots,n_{\uell}+n_{{\rm{m}}}(n_{{\rm{j}}}-n_{{\rm{p}}})$ a separation of the search direction\footnote{Please do not confuse the search direction $\bm{d}$ with the pipes` diameter $[\bm{\mathfrak{d}}]_i=d_i \,\, \forall i \in \mathfrak{P}$.} 
\begin{subequations}
\label{eq:searchDirection_not}
\begin{align}
\bm{d} &= \left[\begin{matrix} \bm{d}^T_\epsilon & \bm{d}^{(1)^T}_{h_N} & \bm{d}^{(2)^T}_{h_N}&\ldots&\bm{d}_{h_N}^{(n_{{\rm{m}}})^T}\end{matrix}\right]^T\\
\bm{d}_{\epsilon} & = \left[\begin{matrix} \Delta x_{k,1} & \Delta x_{k,2} & \ldots & \Delta x_{k,n_{\uell}}\end{matrix}\right]^T\\
\bm{d}_{h_N}^{(i)} &=  \left[\begin{matrix} \Delta x_{k,n_{\uell}+(i-1) (n_{{\rm{j}}}-n_{{\rm{p}}})+1} & \Delta x_{k,n_{\uell}+(i-1) (n_{{\rm{j}}}-n_{{\rm{p}}})+2} & \ldots & \Delta x_{k,n_{\uell}+i (n_{{\rm{j}}}-n_{{\rm{p}}})}\end{matrix}\right]^T \,\, \forall i\in \mathfrak{M}
\end{align}
between roughness and not-measured pressure heads is conducted. Analogously, function $\bm{f}(\bm{x}) = [\begin{matrix} f_1(\bm{x}) &f_2(\bm{x}) &\ldots &f_{n_{{\rm{m}}} n_{{\rm{j}}}}(\bm{x}) \end{matrix}]^T$ \eqref{eq:fx_xQTur_1} which describes the residual of set \eqref{eq:calibrationTur} is separated into
\begin{equation}
\bm{f}^{(i)}_{k-1} \defeq 
\left[
\begin{matrix}
f_{1+(i-1)n_{{\rm{j}}}}(\bm{x}_{k-1}) & f_{2+(i-1)n_{{\rm{j}}}}(\bm{x}_{k-1}) &\ldots &f_{n_{{\rm{j}}}+(i-1)n_{{\rm{j}}}}(\bm{x}_{k-1})
\end{matrix}
\right]^T \qquad \forall i \in \mathfrak{M}
\end{equation}
\end{subequations}
concerning the \textit{Tensor Method} iteration $k$ and measurement-set $i$.

%
\begin{theorem} \label{theorem:TensorEquation}
Let all assumptions in table \ref{tab:calibrationAssumptions} in addition to Assumption \ref{ass:FullTurbulent} hold.
Further apply notation \eqref{eq:searchDirection_not} as well as $\bm{p}_{\mathcal{X}}^{(i)} \defeq \bm{p}_{\mathcal{X}}(\bm{\epsilon}, \Delta \bm{h}^{(i)})$ for all the partial derivatives with respect to $\mathcal{X} \in \{\epsilon, \epsilon^2, \Delta h, \Delta h^2, \epsilon \Delta h \}$. 
Then, Tensor Equation \eqref{eq:TensorEquation} for set \eqref{eq:calibrationTur} is equivalent to 
\begin{align}
\bm{A} \bigg(& \overbrace{\frac{1}{2} \bm{p}^{(i)}_{\epsilon^2} \odot \bm{d}_{\epsilon}^{\odot^{2}} - \bm{d}_{\epsilon} \odot \bm{p}^{(i)}_{\epsilon \Delta h} \odot (\bm{A}^T \bar{\bm{C}}^T_h \bm{d}^{(i)}_{h_N}) + \frac{1}{2} \bm{p}^{(i)}_{\Delta h^2} \odot (\bm{A}^T \bar{\bm{C}}_h^T \bm{d}_{h_N}^{(i)})^{{\odot^{2}}}}^{\encircle{\tiny{\rm{III}}}}
  \nonumber\\
& + \underbrace{\bm{p}_{\epsilon}^{(i)} \odot \bm{d}_{\epsilon} - \bm{p}^{(i)}_{\Delta h} \odot (\bm{A}^T \bar{\bm{C}}_h^T \bm{d}^{(i)}_{h_N})}_{\encircle{\tiny{\rm{II}}}}
 + \underbrace{\DIAG(\bm{c}_l) \bm{A}^T \bm{L}^{-1} \bm{f}^{(i)}_{k-1}}_{\encircle{\tiny{\rm{I}}}}
 \bigg)  \eqdef \bm{A} \bar{\bm{m}}_k^{(i)} \stackrel{!}{=} \bm{0}
\label{eq:TensorEquation_Hadamard}
\end{align}
for all of the $i \in \mathfrak{M}$ measurement-sets, also considering abbreviation $\bm{L}=\bm{A}\DIAG(\bm{c}_l)\bm{A}^T$ including parameter vector $[\bm{c}_l]_j=g A_j/l_j$ $\forall j \in \mathfrak{P}$.
\end{theorem}
%
%
%
\begin{myproof*}
\normalfont
Under consideration of first turbulent flow derivatives \eqref{eq:dxQ_der}, Jacobian \eqref{eq:Jacobian} yields
\begin{equation}
\left[
\begin{matrix}
\DIAG(\bm{p}_{\epsilon}^{(1)}) & -\DIAG(\bm{p}_{\Delta h}^{(1)}) \bm{A}^T \bar{\bm{C}}_h^T &\bm{0} &\ldots &\bm{0} \\
\DIAG(\bm{p}_{\epsilon}^{(2)})  & \bm{0}& -\DIAG(\bm{p}_{\Delta h}^{(2)}) \bm{A}^T \bar{\bm{C}}_h^T   &\ldots &\bm{0}\\
\vdots & \vdots & &\ddots\\
\DIAG(\bm{p}_{\epsilon}^{(n_{{\rm{m}}})})  & \bm{0}&  \bm{0}&  \ldots & -\DIAG(\bm{p}_{\Delta h}^{(n_{{\rm{m}}})}) \bm{A}^T \bar{\bm{C}}_h^T 
\end{matrix}
\right] 
\end{equation}
leading to the following \textit{Jacobian} term in \textit{Tensor Equation} \eqref{eq:TensorEquation} when applying \eqref{eq:searchDirection_not}
\begin{equation}
\bm{A} \left(\bm{p}_{\epsilon}^{(i)} \odot \bm{d}_{\epsilon} - \bm{p}_{\Delta h}^{(i)} \odot (\bm{A}^T \bar{\bm{C}}_h^T \bm{d}_{h_N}^{(i)}) \right) \mathrel{\hat{=}} \bm{A} \, \encircle{\tiny{II}}. 
\label{eq:JacobianTerm}
\end{equation}
Compare this result with incidence matrix $\bm{A}$ times the term \encircle{\tiny{II}} in the second line of \eqref{eq:TensorEquation_Hadamard}.  Considering Lemma \ref{lemma:Tensor_A} concerning the \textit{Hessian} term with respect to the $j\in \mathfrak{P}$ turbulent pipe flow \eqref{eq:ft} (and the $k$-th iteration) in \textit{Tensor Equation} \eqref{eq:TensorEquation}, one obtains in the first $i=1$ measurement-set
\begin{subequations}
\begin{gather}
\Delta \bm{x}^T_k \bm{\mathcal{H}}_{k-1}(f^{(1)}_{t,j}) \Delta \bm{x}_k = 
\left[
\begin{matrix}
\bm{d}^T_{\epsilon}&
\bm{d}_{h_N}^{{(1)}^T}&
\ldots&
\bm{d}_{h_N}^{{(n_{{\rm{m}}})}^T}
\end{matrix}
\right]
\left[
\begin{matrix}
\frac{\partial^2 f^{(1)}_{t,j}}{\partial^2 \bm{\epsilon}^2} & \frac{\partial^2  f^{(1)}_{t,j}}{\partial \bm{\epsilon} \partial \bm{h}_N^{(1)}} & \bm{0} &\ldots\\
\frac{\partial^2  f^{(1)}_{t,j}}{\partial \bm{h}_N^{(1)} \partial \bm{\epsilon}} & \frac{\partial^2  f^{(1)}_{t,j}}{\partial \bm{h}_N^{{(1)}^2}} &\bm{0} & \ldots\\
\bm{0} & \bm{0} &\bm{0} & \ldots \\
\vdots &\vdots & \vdots &\ddots
\end{matrix}
\right]
\left[
\begin{matrix}
\bm{d}_{\epsilon}\\
\bm{d}_{h_N}^{(1)}\\
\vdots\\
\bm{d}_{h_N}^{(n_{{\rm{m}}})}
\end{matrix}
\right]\\
\stackrel{\eqref{eq:Qtur_d2}}{=}  p^{(1)}_{\epsilon^2,j} \bm{d}_{\epsilon}^T \bm{e}_j \bm{e}_j^T \bm{d}_{\epsilon} - 2 p^{(1)}_{\epsilon \Delta h,j} \bm{d}_{\epsilon}^T \bm{e}_j \bm{e}_j^T \bm{A}^T \bar{\bm{C}}_h^T \bm{d}_{h_N}^{(1)} + p_{\Delta h^2,j}^{(1)} \bm{d}_{h_N}^{(1)^T} \bar{\bm{C}}_h \bm{a}_j \bm{a}_j^T \bar{\bm{C}}_h^T\bm{d}_{h_N}^{(1)}
\end{gather}
\end{subequations}
and in the second measurement-set
\begin{subequations}
\begin{gather}
\Delta \bm{x}^T_k \bm{\mathcal{H}}_{k-1}(f^{(2)}_{t,j}) \Delta \bm{x}_k = 
\left[
\begin{matrix}
\bm{d}^T_{\epsilon}&
\bm{d}_{h_N}^{{(1)}^T}&
\ldots&
\bm{d}_{h_N}^{{(n_{{\rm{m}}})}^T}
\end{matrix}
\right]
\left[
\begin{matrix}
\frac{\partial^2 f^{(2)}_{t,j}}{\partial^2 \bm{\epsilon}^2} & \bm{0} & \frac{\partial^2  f^{(2)}_{t,j}}{\partial \bm{\epsilon} \partial \bm{h}_N^{(2)}} & \bm{0} &\ldots
\\
\bm{0} & \bm{0} &\bm{0}& \bm{0} & \ldots\\
\frac{\partial^2  f^{(2)}_{t,j}}{\partial \bm{h}_N^{(2)} \partial \bm{\epsilon}} & \bm{0} & \frac{\partial^2  f^{(2)}_{t,j}}{\partial \bm{h}_N^{{(2)}^2}} &\bm{0} & \ldots\\
\bm{0} & \bm{0} &\bm{0} &\bm{0}& \ldots \\
\vdots &\vdots & \vdots & \vdots &\ddots
\end{matrix}
\right]
\left[
\begin{matrix}
\bm{d}_{\epsilon}\\
\bm{d}_{h_N}^{(1)}\\
\vdots\\
\bm{d}_{h_N}^{(n_{{\rm{m}}})}
\end{matrix}
\right]\\
\stackrel{\eqref{eq:Qtur_d2}}{=}  p^{(2)}_{\epsilon^2,j} \bm{d}_{\epsilon}^T \bm{e}_j \bm{e}_j^T \bm{d}_{\epsilon} - 2 p^{(2)}_{\epsilon \Delta h,j} \bm{d}_{\epsilon}^T \bm{e}_j \bm{e}_j^T \bm{A}^T \bar{\bm{C}}_h^T \bm{d}_{h_N}^{(2)} + p_{\Delta h^2,j}^{(2)} \bm{d}_{h_N}^{(2)^T} \bar{\bm{C}}_h \bm{a}_j \bm{a}_j^T \bar{\bm{C}}_h^T\bm{d}_{h_N}^{(2)}
\end{gather}
\end{subequations}
which, ultimately, results in
\begin{gather}
\Delta \bm{x}^T_k \bm{\mathcal{H}}_{k-1}(f^{(i)}_{t,j}) \Delta \bm{x}_k =   p^{(i)}_{\epsilon^2,j} \bm{d}_{\epsilon}^T \bm{e}_j \bm{e}_j^T \bm{d}_{\epsilon} - 2 p^{(i)}_{\epsilon \Delta h,j} \bm{d}_{\epsilon}^T \bm{e}_j \bm{e}_j^T \bm{A}^T \bar{\bm{C}}_h^T \bm{d}_{h_N}^{(i)} \nonumber\\ 
+ p_{\Delta h^2,j}^{(i)} \bm{d}_{h_N}^{(i)^T} \bar{\bm{C}}_h \bm{a}_j \bm{a}_j^T \bar{\bm{C}}_h^T\bm{d}_{h_N}^{(i)} 
\qquad \forall i \in \mathfrak{M}\land j \in \mathfrak{P} .
\label{eq:hessian_term_hard}
\end{gather}
This is feasible as a result of Assumption 5 leading to
\begin{equation}
\frac{\partial^2 f_{t,j}^{(i)}}{\partial \bm{\epsilon} \partial \bm{h}_N^{(l)}} = \bm{0}, \quad \frac{\partial^2 f_{t,j}^{(i)}}{ \partial \bm{h}_N^{(l)^2}} = \bm{0} \quad \forall i \in \mathfrak{M} \ne l \in \mathfrak{M} \land j\in \mathfrak{P}.
\end{equation}
The application of the \textit{Hadamard} operator allows a compact representation of \eqref{eq:hessian_term_hard}
\begin{gather}
\bm{p}^{(i)}_{\epsilon^2} \odot \bm{d}^{\odot^2}_{\epsilon}- 2 \bm{d}_{\epsilon} \odot \bm{p}^{(i)}_{\epsilon \Delta h} \odot (\bm{A}^T \bar{\bm{C}}_h^T \bm{d}_{h_N}^{(i)}) + \bm{p}_{\Delta h^2}^{(i)} \odot (\bm{A}^T \bar{\bm{C}}^T_h\bm{d}_{h_N}^{(i)})^{\odot^2} \mathrel{\hat{=}}  2 \times \encircle{\tiny{III}}
\label{eq:hessian_term_hard_2}
\end{gather}
for all $i\in \{1,\ldots,n_{{\rm{m}}}\}=\mathfrak{M}$. In order for \eqref{eq:hessian_term_hard} to be equivalent to \eqref{eq:hessian_term_hard_2} along all $j\in\mathfrak{P}$ one has to prove that
\begin{gather}
\left[
\begin{matrix}
p_{\Delta h^2,1}^{(i)} \bm{d}_{h_N}^{(i)^T} \bar{\bm{C}}_h \bm{a}_1 \bm{a}_1^T \bar{\bm{C}}_h^T\bm{d}_{h_N}^{(i)} \\
p_{\Delta h^2,2}^{(i)} \bm{d}_{h_N}^{(i)^T} \bar{\bm{C}}_h \bm{a}_2 \bm{a}_2^T \bar{\bm{C}}_h^T\bm{d}_{h_N}^{(i)} \\
\vdots\\
p_{\Delta h^2,n_{\uell}}^{(i)} \bm{d}_{h_N}^{(i)^T} \bar{\bm{C}}_h \bm{a}_{n_{\uell}} \bm{a}_{n_{\uell}}^T \bar{\bm{C}}_h^T\bm{d}_{h_N}^{(i)} \\
\end{matrix}
\right]
=  \bm{p}_{\Delta h^2}^{(i)} \odot (\bm{A}^T \bar{\bm{C}}^T_h\bm{d}_{h_N}^{(i)})^{\odot^2}.
\end{gather}
Denoting $\bar{\bm{C}}^T_h\bm{d}_{h_N}^{(i)} = [\begin{matrix} c^{(i)}_1 & c^{(i)}_2& \ldots &c^{(i)}_{n_{{\rm{j}}}}\end{matrix}]^T$ and $a_{ij} = [\bm{A}]_{ij}$, this becomes apparent when writing
\begin{subequations}
\begin{gather}
\bm{d}_{h_N}^{(i)^T} \bar{\bm{C}}_h  \bm{a}_j = c^{(i)}_1 a_{j1} + c^{(i)}_2  a_{j1} + \ldots +  c^{(i)}_2  a_{j n_{{\rm{j}}}} \qquad \forall j\in \mathfrak{P} \land i\in \mathfrak{M} \\
\Rightarrow \bm{d}_{h_N}^{(i)^T} \bar{\bm{C}}_h \bm{a}_j \bm{a}_j^T \bar{\bm{C}}_h^T\bm{d}_{h_N}^{(i)} = \left( c^{(i)}_1 a_{j1} + c^{(i)}_2  a_{j1} + \ldots +  c^{(i)}_2  a_{j n_{{\rm{j}}}}\right)^2 = \left[(\bm{A}^T \bar{\bm{C}}_h^T \bm{d}_{h_N}^{(i)})^{\odot^2}\right]_j .
\end{gather}
\end{subequations}
The equivalence of the \textit{Jacobian} \eqref{eq:JacobianTerm} (see $\bm{A} \, \encircle{\tiny{II}}$) and the \textit{Hessian} term \eqref{eq:hessian_term_hard_2} (see \encircle{\tiny{III}}), under consideration of Lemma \ref{lemma:Tensor_A}, to the appropriate terms of the \textit{Tensor Equation} \eqref{eq:TensorEquation} has been proven. The equivalence of missing zero-order term $\bm{f}^{(i)}_{k-1}$ with $\bm{A} \, \encircle{\tiny{I}}$, i.e.
\begin{equation}
\bm{f}^{(i)}_{k-1} = \bm{A} \left( \DIAG(\bm{c}_l)  \bm{A}^T \bm{L}^{-1} \bm{f}^{(i)}_{k-1}\right) 
\label{eq:Tensor_0orderTerm}
\end{equation}
becomes apparent when considering abbreviation $\bm{L}=\bm{A}\DIAG(\bm{c}_l)\bm{A}^T$, which is positive definite as incidence matrix $\bm{A}$ has full rank \citep[Eq. (9)]{cite:ModelingHydraulicNetworks}. As a side note, one could certainly replace $\DIAG(\bm{c}_l) \bm{A}^T \bm{L}^{-1}$ in \eqref{eq:Tensor_0orderTerm} with, for instance, $\bm{A}^T(\bm{A}\bm{A}^T)^{-1}$. This usage of $\bm{L}$ has some historic meaning nonetheless, see \citep{cite:ModelingHydraulicNetworks}.

The combination of \eqref{eq:JacobianTerm}, \eqref{eq:hessian_term_hard_2}, \eqref{eq:Tensor_0orderTerm} under consideration of Lemma \ref{lemma:Tensor_A}, that is $\bm{A} \bar{\bm{m}}^{(i)}_k \mathrel{\hat{=}}  \bm{A} (\encircle{\tiny{I}} + \encircle{\tiny{II}} +\encircle{\tiny{III}}) = \bm{0}$, yields the \textit{Tensor Equation} \eqref{eq:TensorEquation} in measurement-set $i$.  This completes the proof. $\hfill \qed$
\end{myproof*}

In sum, Theorem \ref{theorem:TensorEquation} delivers the \textit{Tensor Equation} in a compact form which completely resolves the problem to store and evaluate $n_{{\rm{m}}} n_{{\rm{j}}}$ \textit{Hessians} $\bm{\mathcal{H}}_{k-1}(f_1), \bm{\mathcal{H}}_{k-1}(f_2), \ldots$ with $(n_{\uell}+n_{{\rm{m}}}(n_{{\rm{j}}}-n_{{\rm{p}}}))^2$ entries each. Effectively, only $5n_l n_{{\rm{m}}}$   second-turbulent-flow-derivatives need to be evaluated in comparison to $n_{{\rm{j}}} n_{{\rm{m}}} (n_{\uell}+n_{{\rm{m}}}(n_{{\rm{j}}}-n_{{\rm{p}}}))^2$. Actually, the linear \textit{Newton} equation for the search direction $\Delta \bm{x}_k = -\bm{J}_{k-1}^{-1}\bm{f}(\bm{x}_{k-1})$ considering the \textit{Jacobian}, already requires $n_{{\rm{m}}} n_{{\rm{j}}} \times (n_{\uell}+n_{{\rm{m}}}(n_{{\rm{j}}}-n_{{\rm{p}}}))$ first-turbulent-flow-derivative evaluations when not directly applying \eqref{eq:dxQ_der}. Overall, the application of the \textit{Hardamard} operator substantially improves computational effort in this context. Nonetheless, the question how to solve it, remains.

As the \textit{Tensor Equation} in all its derived forms, is nothing else but a set of $n_{{\rm{m}}} n_{{\rm{j}}}$ polynomials of degree 2 in $n_{\uell}+n_{{\rm{m}}}(n_{{\rm{j}}}-n_{{\rm{p}}})$ unknowns, the examples below will help to establish a connection between its solutions and the \textit{Hadamard} product.

\subsection{Motivating Examples} \label{sec:motivatingExample} 

Consider the set of two polynomial equations with degree 2 in two unknowns $x,y$
\begin{gather}
\begin{gathered}
2 x^2 + 2y^2 + 5x y-x +y -1 = 0\\
-3x^2-2y ^2 + 5xy + 10x -8y -8 = 0
\end{gathered}
\quad \Leftrightarrow \quad
\begin{gathered}
\overset{\encircle{\tiny{1a}}}{(x+ 2y -1)}  \overset{\encircle{\tiny{1b}}}{(2x + y +1)}= 0\\
\underset{\encircle{\tiny{2a}}}{(-x+y+2)} \underset{\encircle{\tiny{2b}}}{(3x -2y -4)}= 0
\end{gathered}
\label{eq:Tensor_motEx}
\end{gather}
which can be separated into linear terms each. The solutions of \eqref{eq:Tensor_motEx} can then be obtained by solving the linear equations
\begin{subequations}
\begin{gather}
\encircle{\tiny{1a}} \land \encircle{\tiny{2a}} \mathrel{\hat{=}} \left[
\begin{matrix}
1 &2\\
-1 &1
\end{matrix}
\right] \left[
\begin{matrix}
x\\
y
\end{matrix}
\right] = \left[
\begin{matrix}
1\\
-2
\end{matrix}
\right], \qquad
\encircle{\tiny{1b}} \land \encircle{\tiny{2b}} \mathrel{\hat{=}} \left[
\begin{matrix}
2 &1\\
3 &-2
\end{matrix}
\right] \left[
\begin{matrix}
x\\
y
\end{matrix}
\right] = \left[
\begin{matrix}
-1\\
4
\end{matrix}
\right], \\
\encircle{\tiny{1a}} \land \encircle{\tiny{2b}} \mathrel{\hat{=}} \left[
\begin{matrix}
1 &2\\
3 &-2
\end{matrix}
\right] \left[
\begin{matrix}
x\\
y
\end{matrix}
\right] = \left[
\begin{matrix}
1\\
4
\end{matrix}
\right], \qquad
\encircle{\tiny{1b}} \land \encircle{\tiny{2a}} \mathrel{\hat{=}} \left[
\begin{matrix}
2 &1\\
-1 &1
\end{matrix}
\right] \left[
\begin{matrix}
x\\
y
\end{matrix}
\right] = \left[
\begin{matrix}
-1\\
-2
\end{matrix}
\right]
\end{gather}
\end{subequations}
as combination of the linear terms in \eqref{eq:Tensor_motEx}, effectively, leading to 4 independent solutions. This means \eqref{eq:Tensor_motEx} can be rewritten in terms of (taking ``$(\encircle{\tiny{1a}} \land \encircle{\tiny{2a}}) \odot (\encircle{\tiny{1b}} \land \encircle{\tiny{2b}})$'')
\begingroup
\allowdisplaybreaks
\begin{subequations}
\begin{align}
\label{eq:TensorEx_Hadamard1}
\left( \left[
\begin{matrix}
1 &2\\
-1 &1
\end{matrix}
\right] \left[
\begin{matrix}
x\\
y
\end{matrix}
\right] - \left[
\begin{matrix}
1\\
-2
\end{matrix}
\right]\right)   \odot
\left( \left[
\begin{matrix}
2 &1\\
3 &-2
\end{matrix}
\right] \left[
\begin{matrix}
x\\
y
\end{matrix}
\right] - \left[
\begin{matrix}
-1\\
4
\end{matrix}
\right] \right) &= \bm{0}\\
\intertext{or (taking ``$(\encircle{\tiny{1a}} \land \encircle{\tiny{2b}}) \odot (\encircle{\tiny{1b}} \land \encircle{\tiny{2a}})$'')}
\label{eq:TensorEx_Hadamard2}
\left( \left[
\begin{matrix}
1 &2\\
3 &-2
\end{matrix}
\right] \left[
\begin{matrix}
x\\
y
\end{matrix}
\right] - \left[
\begin{matrix}
1\\
4
\end{matrix}
\right] \right)  
\odot
\left( \left[
\begin{matrix}
2 &1\\
-1 &1
\end{matrix}
\right] \left[
\begin{matrix}
x\\
y
\end{matrix}
\right] - \left[
\begin{matrix}
-1\\
-2
\end{matrix}
\right] \right) &= \bm{0}
\end{align}
\end{subequations}
\endgroup
when applying the \textit{Hadamard} product. According to \textit{Bézout}'s theorem \cite[p. 10]{cite:Bezout}, a set of $n$ independent polynomials $\{\mathfrak{f}_1,\mathfrak{f}_2,\ldots, \mathfrak{f}_n\}$ in $n$ unknowns has at most 
\begin{equation}
\Pi_{i=1}^{n} \text{deg}(\mathfrak{f}_i)
\end{equation}
(i.e. the product of the degrees of each polynomial) solutions as it is the case in the present example. There possibly exist infinitely many solutions if the equations are linear dependent \cite[p. 10]{cite:Bezout}.
However, by selecting representation \eqref{eq:TensorEx_Hadamard1} or \eqref{eq:TensorEx_Hadamard2} one loses 2 solutions when applying the \textit{Hadamard} product.

The factorization into linear terms as in \eqref{eq:Tensor_motEx} is not always possible for a polynomial with more than one variable, for instance
\begin{gather}
\mathfrak{f}(x,y) = xy+1
\end{gather}
is not factorizable into linear terms. For the general case, consider a general polynomial of degree 2 in two unknowns, also known as conic section
\begin{equation}
\label{eq:conic}
\mathfrak{f}(x,y) = a x^2 + 2 h xy + b y^2 + 2 f x + 2 g y + c = 0 
\end{equation}
with coefficients $a,h,b,f,g,c$. The following properties of \eqref{eq:conic} can be found in literature. Property \ref{prop:conic} and \ref{prop:conic_real} were found in \cite[p.63]{conic_1} as well as \cite[p.40]{conic_2}, though no proofs are given therein. Other mentions of Property \ref{prop:conic} can be found in, for instance, \citep{conic_3}.

%
%
\begin{property} \label{prop:conic}
Let coefficients $a,b,h,g,f,c \in \mathbb{R}$. Then, the conic section $\mathfrak{f}$ \eqref{eq:conic} is factorizable into the linear-pair
\begin{equation}
\label{eq:conic_factors}
\mathfrak{f}(x,y) = (A x + By + C) (D x + E y + F) = 0
\end{equation}
with coefficients $A,B,C,D,E,F \in \mathbb{C}$ if and only if
\begin{align}
\label{eq:conic_determinante}
\Delta \defeq \left|
\begin{matrix}
a &h &f\\
h &b &g\\
f &g &c
\end{matrix}
\right| = 0
\qquad \land \qquad 
\hat{\Delta} \defeq \left| \begin{matrix} a &h\\ h&b \end{matrix}\right| = ab - h^2 \le 0
\end{align}
the determinant $\Delta = 0$ and its sub determinant $\hat{\Delta} \le 0$. In addition to that, these two linear factors as in \eqref{eq:conic_factors} represent
\begin{itemize}
\item two intersecting lines if and only if $\hat{\Delta} < 0$
\item two parallel lines \ep{or a single line if $a=b=h=0$} if and only if $\hat{\Delta} = 0$
\end{itemize}
However, in case $\Delta = 0$ and $\hat{\Delta} > 0$,  only a single point concerning $x,y$ satisfies the conic section \eqref{eq:conic} \ep{interpreted graphically, it is an ellipse with zero radius}.
\end{property}
%
%

In the general case where $\Delta = 0$, conic section \eqref{eq:conic} is called degenerate as the coefficient matrix to build determinant $\Delta$ is singular.

%
%
\begin{property} \label{prop:conic_real}
\sloppy The factorization of conic section \eqref{eq:conic} into \eqref{eq:conic_factors} yields real-valued $A,B,C,D,E,F$ if and only if 
\begin{subequations}
\begin{align}
\Delta = 0 \qquad &\land \qquad \hat{\Delta} < 0\\
\text{or} \quad \Delta = 0 \qquad &\land \qquad \hat{\Delta} = 0 \qquad \land \qquad f^2+ g^2 \ge c(a+b). \label{eq:cond_2_conicReal}
\end{align}
\end{subequations}
\end{property}
%
%
\begin{lemma} \label{lemma:conic_reals}
Let $\Delta = 0$ and $\hat{\Delta} = 0$, concerning the degenerate conic section $\mathfrak{f}(x,y)$ \eqref{eq:conic}. Then,
\begin{equation}
f^2 + g^2 \ge c (a+b) \qquad \Leftrightarrow \qquad f^2 \ge ac \land g^2 \ge bc 
\end{equation}
in reference to \eqref{eq:cond_2_conicReal}, i.e. the second case of  Property \ref{prop:conic_real}.
\end{lemma}
%
\begin{myproof*}
\normalfont
Knowing that $\hat{\Delta} = 0  \Leftrightarrow h^2 = a b$, the determinant $\Delta$ yields
\begin{subequations}
\begin{align}
\Delta &= abc + 2fgh - ag^2 - bf^2 -ch^2 = 0\\
         &= 2fgh - ag^2 - bf^2 = 0. \label{eq:conic_quadratics}
\end{align}
Subsequently, 
\begin{align}
g &= \frac{1}{-2a} \left( 2fh \pm \sqrt{4 (fh)^2 - 4 abf^2} \right) \stackrel{\hat{\Delta} = 0}{=} -\frac{f h}{a}\\
g^2 &= \left( \frac{fh}{a}\right)^2 = f^2 \frac{b}{a} \label{eq:conic_proof_g}\\
 f^2 &= g^2 \frac{a}{b}.  \label{eq:conic_proof_f}
\end{align}
This means that 
\begin{align}
f^2 + g^2 \stackrel{ \eqref{eq:conic_proof_g}}{=} f^2 \left( \frac{a+b}{a}\right) \ge c (a+b)  \qquad &\Rightarrow \qquad f^2 \ge ac\\
f^2 + g^2 \stackrel{ \eqref{eq:conic_proof_f}}{=} g^2 \left( \frac{a+b}{b}\right) \ge c (a+b)  \qquad &\Rightarrow \qquad g^2 \ge bc.
\end{align}
\end{subequations}
From the other side of the argumentation, it is known that $f^2 \ge ac \land g^2 \ge bc$ and thus $f^2 + g^2 \ge ac + bc$. This completes the proof. $\hfill\qed$
\end{myproof*}

\paragraph{Plausibility Check of Property \ref{prop:conic}}
Conducting a comparison of coefficients of \eqref{eq:conic} and \eqref{eq:conic_factors}, the six conditions 
\begin{subequations}
\label{eq:conic_coefficient_comparision}
\begin{gather}
AD = a, \qquad BE = b, \qquad CF = c  \label{eq:conic_comparision_1}\\
AE + DB = 2h, \qquad BF+EC=2g, \qquad AF+DC = 2f \label{eq:conic_comparision_2}
\end{gather}
\end{subequations}
must be met if $\mathfrak{f}(x,y)$ ought to be factorizable into linear terms. This is the case (necessary condition) if the set \eqref{eq:conic_coefficient_comparision} of six equations in six unknowns is consistent. Consistency can be verified when multiplying conditions \eqref{eq:conic_comparision_2} among each other such that
\begin{align}
 8hgf =& \, (AE+DB) (BF+EC) (AF+DC) \nonumber\\
 =& \, 2ADBECF + AD(B^2 F^2 + E^2 C^2) + BE (A^2 F^2 + D^2 C^2)  \label{eq:conic_blabla}\\
    &+ CF(A^2 E^2 + B^2 D^2) .\nonumber
\end{align}
By reformulating the square of conditions \eqref{eq:conic_comparision_2} respectively, one obtains
\begin{subequations}
\label{eq:conic_plausability_squares}
\begin{align}
(AE)^2  + (DB)^2= 4 h^2 - 2AEDB \stackrel{\eqref{eq:conic_comparision_1}}{=}& 4 h^2 - 2 a b\\
(BF)^2 + (EC)^2 = 4g^2 - 2 BF EC \stackrel{\eqref{eq:conic_comparision_1}}{=}& 4 g^2 - 2 bc\\
(AF)^2 + (DC)^2 = 4f^2 - 2AFDC \stackrel{\eqref{eq:conic_comparision_1}}{=}& 4 f^2 - 2ac
\end{align}
\end{subequations}
which is inserted into the right hand side of \eqref{eq:conic_blabla}. To conclude, consistency of set \eqref{eq:conic_coefficient_comparision} is preserved if 
\begin{gather}
\label{eq:conic_proof_final}
8hgf \stackrel{\eqref{eq:conic_comparision_1} \land \eqref{eq:conic_plausability_squares}}{=} 2abc  + a (4g^2 - 2 bc) + b (4f^2 - 2ac) + c (4 h^2- 2ab)
\end{gather}
holds. Bringing $8ghf$ of \eqref{eq:conic_proof_final} on the right hand side while dividing by $-4$, condition \eqref{eq:conic_proof_final} yields
\begin{equation}
0 \stackrel{!}{=} 2ghf + abc - a g^2 - b f^2 - c h^2 = \Delta
\end{equation}
the determinant $\Delta$ \eqref{eq:conic_determinante}.

As a remark on polynomials of degree $n$, it is possible to show that if a homogeneous polynomial, that is $\mathfrak{f}(\mu \bm{x}) = \mu \mathfrak{f}(\bm{x})$ for any $\bm{x} \in \mathbb{C}^n, \mu \in \mathbb{C}$ has a factorization, then its factors must be homogeneous too. Consider, for instance, the polynomial in variables $x,y,z$
\begin{equation}
x^3 + y^3 + z^3 = (x+y+z)(x+wy+w2z)(x+w2y+wz) \qquad \text{for} \quad w = e^{2i \pi/3}
\end{equation}
(credit to \citep{mathstack:PolynomialFactor}) where $i$ denotes the imaginary unit satisfying $i^2 = -1$.

%
\subsection{Degenerate Tensor Equation} \label{sec:TensorDegenerate}
%
\begin{remark} \label{remark:alpha_kernel}
Looking for a search direction $\bm{d}_{\epsilon}$ and $\bm{d}_{h_N}^{(i)}$ for measurement-set $i \in \mathfrak{M}$ which solves Tensor Equation \eqref{eq:TensorEquation_Hadamard}, i.e. $\bm{A} \bar{\bm{m}}^{(i)}_k(\bm{d}_{\epsilon},\bm{d}_{h_N}^{(i)}) = \bm{0} \,\,\forall i \in \mathfrak{M}$, the confinement to look for a search direction which solves $\bar{\bm{m}}_k^{(i)}(\bm{d}_{\epsilon},\bm{d}_{h_N}^{(i)}) = \bm{0} \,\, \forall i \in \mathfrak{M}$ only is eligible yet far too conservative. This is the consequence of incidence matrix $\bm{A} \in \mathbb{Z}_{\{-1,0,1\}}^{n_{{\rm{j}}} \times n_{\uell}}$ being fat, i.e. $n_{\uell} > n_{{\rm{j}}}$ \ep{see \cite{PipeRoughness_arxiv}}.
\end{remark}
%

To circumvent the problem that the incidence matrix $\bm{A}$ can not be directly considered when factorizing $\bar{\bm{m}}_k^{(i)}=\bm{0}$ (as function of $\bm{d}_{\epsilon}$ and $\bm{d}_{h_N}^{(i)}$)  into linear terms, the problem is projected into the kernel. Knowing that $\bm{A}\bm{S}^T \equiv \bm{0}$ when considering cycle matrix $\bm{S} \in \mathbb{Z}^{(n_{\rm{j}}-n_{\uell}) \times n_{\uell}}_{\{-1,0,1\}}$, one can write
\begin{equation}
\label{eq:alpha_intro}
\bm{A} \bar{\bm{m}}^{(i)}_k(\bm{d}_{\epsilon},\bm{d}_{h_N}^{(i)}) = \bm{0}  \qquad \Leftrightarrow \qquad \bm{m}_k \defeq \bar{\bm{m}}_{k}^{(i)}(\bm{d}_{\epsilon},\bm{d}_{h_N}^{(i)})  - \bm{S}^T \bm{\alpha}^{(i)} = \bm{0} \quad \forall i \in \mathfrak{M}
\end{equation}
for some $\bm{\alpha}^{(i)} \in \mathbb{C}^{n_{\uell} - n_{{\rm{j}}}}$ along all measurement-set $i \in \mathfrak{M}$, where $\bm{m}_k^{(i)}(\bm{d}_{\epsilon},\bm{d}_{h_N}^{(i)}, \bm{\alpha}^{(i)}) \,\, \forall i\in \mathfrak{M}$ is now considered to be a function on $\bm{\alpha}^{(i)}$ too.

%
%
\begin{theorem} \label{theorem:TensorSeparation}
Denoting $p^{(i)}_{\mathcal{X}_j} = \big[\bm{p}_{\mathcal{X}}^{(i)}(\bm{\epsilon}_k, \Delta \bm{h}^{(i)}_k)\big]_j$ for all pipes $j \in \mathfrak{P}$ and measurement-sets $i\in \mathfrak{M}$ concerning all partial derivatives of turbulent flow \eqref{eq:ft} with respect to $\mathcal{X} \in \{\epsilon, \epsilon^2, \Delta h, \Delta h^2, \epsilon \Delta h \}$ in the solving-iteration $k$. 
Then, Tensor Equation \eqref{eq:TensorEquation_Hadamard} in the measurement-set $i \in \mathfrak{M}$ can be factorized into linear terms
\begin{align}  \label{eq:TensorSeparation}
\bm{m}_k^{(i)} =(\bm{B}^{(i)} \bm{d}_{\epsilon} + \bm{C}^{(i)} \bm{d}_{h_N}^{(i)} + \bm{v}^{(i)}) \odot (\bm{E}^{(i)}\bm{d}_{\epsilon} + \bm{F}^{(i)} \bm{d}_{h_N}^{(i)} + \bm{w}^{(i)}) \quad i \in \mathfrak{M}
\end{align}
connected via the Hadamard product if and only if $\Delta_j^{(i)}(\bm{\alpha}^{(i)}) = 0$ and $\hat{\Delta}_j^{(i)} \le 0 \quad \forall j \in \mathfrak{P}$ for a specific $\bm{\alpha}^{(i)} \in \mathbb{C}^{n_{\uell}-n_{{\rm{j}}}}$, same as in \eqref{eq:alpha_intro}, concerning the determinant
\begin{equation}
\label{eq:Tensor_determinants}
\Delta_j^{(i)}(\bm{\alpha}^{(i)}) \defeq \frac{1}{2}
\left|
\begin{matrix}
 p_{\epsilon^2_j}^{(i)} & - p_{\epsilon \Delta h_j}^{(i)} & p_{\epsilon_j}^{(i)}\\
 - p_{\epsilon \Delta h_j}^{(i)} & p_{\Delta h^2_j}^{(i)} & -p_{\Delta h_j}^{(i)} \\
p_{\epsilon_j}^{(i)} & - p_{\Delta h_j^{(i)}} & 2 \bar{f}^{(i)}_{0,j}(\bm{\alpha}^{(i)})
\end{matrix}
\right| \quad \text{and} \quad \hat{\Delta}_j^{(i)} \defeq 
\left|
\begin{matrix}
p_{\epsilon^2_j}^{(i)} & -p_{\epsilon \Delta h_j}^{(i)} \\
 -p_{\epsilon \Delta h_j}^{(i)} & p_{\Delta h^2_j}^{(i)} \\
\end{matrix}
\right|.
\end{equation}
Also applying
\begin{equation}
\label{eq:f0bar}
\bar{f}_{0,j}^{(i)}(\bm{\alpha}^{(i)}) = \big[\bar{\bm{f}}^{(i)}_0 (\bm{\alpha}^{(i)})\big]_j \defeq \Big[ \DIAG(\bm{c}_l) \bm{A} \bm{L}^{-1} \bm{f}_{k-1}^{(i)} - \bm{S}^T \bm{\alpha}^{(i)}\Big]_j \qquad \forall j\in \mathfrak{P} \land i \in \mathfrak{M},
\end{equation}
matrices in \eqref{eq:TensorSeparation} yield the diagonal forms $\bm{B}^{(i)}=\DIAG(\bm{b}^{(i)}), \bm{E}^{(i)}=\DIAG(\bm{e}^{(i)})$, whereas $\bm{C}^{(i)} = \DIAG(\bm{c}^{(i)})\bm{A}^T \bar{\bm{C}}_h^T$ and $\bm{F}^{(i)} = \DIAG(\tilde{\bm{f}}^{(i)}) \bm{A}^T \bar{\bm{C}}_h^T$ with $\bm{b}^{(i)},\bm{e}^{(i)}, \bm{c}^{(i)}, \tilde{\bm{f}}^{(i)}, \bm{v}^{(i)} ,\bm{w}^{(i)} \in \mathbb{C}^{n_{\uell}}$ respectively which will be referred to as Tensor Separators. Thereby, $\bm{v}^{(i)}, \bm{w}^{(i)} \in \mathbb{C}^{n_{\uell}}$ are functions on $\bm{\alpha}^{(i)} \in \mathbb{C}^{n_{\uell}-n_{{\rm{j}}}}$ which lie inside the kernel of the incidence matrix $\bm{A}$.
\end{theorem}
%
%
\begin{myproof*}
\normalfont
For the sake of readability, the index $(i)$ along the measurement-sets $i=1,2,\ldots,n_{{\rm{m}}}$ is not displayed. 
Comparing terms of \eqref{eq:TensorEquation_Hadamard} with the ones of \eqref{eq:TensorSeparation} one obtains:
\begingroup\leqnos
\begin{gather}
\bm{B}\bm{d}_{\epsilon} \odot \bm{E}\bm{d}_{\epsilon} \stackrel{!}{=} \frac{1}{2} \bm{p}_{\epsilon^2} \odot \bm{d}_{\epsilon}^{\odot^2}  \tag{\textit{\textbf{i}}} \label{eq:T_i}\\
\bm{C}\bm{d}_{h_N} \odot \bm{F} \bm{d}_{h_N} \stackrel{!}{=} \frac{1}{2} \bm{p}_{\Delta h^2} \odot (\bm{A}^T \bar{\bm{C}}_h^T \bm{d}_{h_N})^{\odot^{2}}  \tag{\textit{\textbf{ii}}} \label{eq:T_ii}\\
\bm{B} \bm{d}_{\epsilon} \odot \bm{F}\bm{d}_{h_N} + \bm{C} \bm{d}_{h_N} \odot \bm{E} \bm{d}_{\epsilon} \stackrel{!}{=} - \bm{d}_{\epsilon} \odot \bm{p}_{\epsilon \Delta h}\odot (\bm{A}^T \bar{\bm{C}}_h^T \bm{d}_{h_N})\tag{\textit{\textbf{iii}}} \label{eq:T_iii}\\
\bm{B} \bm{d}_{\epsilon} \odot  \bm{w} + \bm{v} \odot \bm{E}\bm{d}_{\epsilon} \stackrel{!}{=} \bm{p}_{\epsilon} \odot \bm{d}_{\epsilon}  \tag{\textit{\textbf{iv}}} \label{eq:T_iv}\\
\bm{C} \bm{d}_{h_N} \odot \bm{w} + \bm{v} \odot \bm{F}\bm{d}_{h_N} \stackrel{!}{=} - \bm{p}_{\Delta h}   \odot (\bm{A}^T \bar{\bm{C}}_h^T \bm{d}_{h_N})\tag{\textit{\textbf{v}}} \label{eq:T_v}\\
  \bm{v} \odot \bm{w} \stackrel{!}{=} \DIAG(\bm{c}_l) \bm{A}^T \bm{L}^{-1} \bm{f}_{k-1} - \bm{S}^T \bm{\alpha} \eqdef \bar{\bm{f}}_0(\bm{\alpha})  \tag{\textit{\textbf{vi}}} \label{eq:T_vi}
\end{gather} 
\endgroup

The simplification starts with the premise that the above conditions should hold for every $\bm{d}_{\epsilon}$ and $\bm{d}_{h_N}$, not only for ones to find. As a result, $\bm{B} = \DIAG(\bm{b})$ and $\bm{E} = \DIAG(\bm{e})$ in \eqref{eq:T_i} are diagonal and lead to $\bm{b} \odot \bm{e} = \frac{1}{2} \bm{p}_{\epsilon^2}$. Subsequently, condition \eqref{eq:T_iv} yields
\begin{equation}
(\DIAG(\bm{w})\bm{B} + \DIAG(\bm{v})\bm{E}) \bm{d}_{\epsilon} = \bm{p}_{\epsilon} \odot \bm{d}_{\epsilon} 
\qquad \Leftrightarrow \qquad
 \bm{w} \odot \bm{b} + \bm{v} \odot \bm{e} = \bm{p}_{\epsilon}.
\end{equation}
In analogy, \eqref{eq:T_iii} yields 
\begin{subequations}
\begin{gather}
\bm{d}_{\epsilon} \odot (\DIAG(\bm{b}) \bm{F} \bm{d}_{h_N} + \DIAG(\bm{e}) \bm{C}\bm{d}_{h_N}) = -\bm{d}_{\epsilon} \odot (\DIAG(\bm{p}_{\epsilon \Delta h}) \bm{A}^T \bar{\bm{C}}_h^T \bm{d}_{h_N})\\
\Rightarrow \DIAG(\bm{b}) \bm{F}+ \DIAG(\bm{e}) \bm{C}= -\DIAG(\bm{p}_{\epsilon \Delta h}) \bm{A}^T \bar{\bm{C}}_h^T. \label{eq:tensor_theo_sep}
\end{gather}
\end{subequations}
Considering \eqref{eq:tensor_theo_sep} and condition \eqref{eq:T_v}, which can be rewritten in terms of $\DIAG(\bm{w})\bm{C} + \DIAG(\bm{v})\bm{F} = -\DIAG(\bm{p}_{\Delta h})\bm{A}^T \bar{\bm{C}}_h^T$, one can conclude that (also concerning \eqref{eq:T_ii})
\begin{equation}
\label{eq:con_FC}
\bm{F} = \DIAG(\tilde{\bm{f}}) \bm{A}^T \bar{\bm{C}}_h^T, \qquad \bm{C} = \DIAG(\bm{c}) \bm{A}^T \bar{\bm{C}}_h^T
\end{equation}
is the only feasible choice for matrices $\bm{F},\bm{C}$. Conditions \eqref{eq:T_i},\eqref{eq:T_iii}, \eqref{eq:T_iv},\eqref{eq:T_v} and \eqref{eq:T_vi} are already  independent from the search direction.  With \eqref{eq:con_FC}, \eqref{eq:T_ii} yields $\tilde{\bm{f}} \odot \bm{c} = \frac{1}{2} \bm{p}_{\Delta h^2}$, also leading to $\bm{b}\odot \tilde{\bm{f}} + \bm{e}\odot \bm{c} = -\bm{p}_{\epsilon \Delta h}$ concerning \eqref{eq:T_iii} or \eqref{eq:tensor_theo_sep}. In sum, one receives
\begingroup\leqnos
\begin{align}
 \bm{b} \odot \bm{e} &\stackrel{!}{=} \frac{1}{2} \bm{p}_{\epsilon^2}\tag{\textit{\textbf{i}}} \label{eq:H_i}\\
\bm{c} \odot \tilde{\bm{f}} &\stackrel{!}{=} \frac{1}{2} \bm{p}_{\Delta h^2}  \tag{\textit{\textbf{ii}}} \label{eq:H_ii}\\
\bm{b} \odot \tilde{\bm{f}} + \bm{c} \odot \bm{e} &\stackrel{!}{=} - \bm{p}_{\epsilon \Delta h}\tag{\textit{\textbf{iii}}} \label{eq:H_iii}\\
\bm{b} \odot \bm{w} + \bm{v} \odot \bm{e} &\stackrel{!}{=} \bm{p}_{\epsilon}\tag{\textit{\textbf{iv}}} \label{eq:H_iv}\\
\bm{c} \odot \bm{w}  + \bm{v} \odot \tilde{\bm{f}} &\stackrel{!}{=} - \bm{p}_{\Delta h}  \tag{\textit{\textbf{v}}} \label{eq:H_v}\\
  \bm{v} \odot \bm{w} &\stackrel{!}{=}  \bar{\bm{f}}_0(\bm{\alpha})  \tag{\textit{\textbf{vi}}} \label{eq:H_vi}
\end{align} 
\endgroup
this utterly appealing, decoupled form. Considering set \eqref{eq:H_i}-\eqref{eq:H_vi} componentwise, meaning for all its $n_{\uell}$ components, it is the equivalent condition-set obtained when factorizing conic \eqref{eq:conic} into \eqref{eq:conic_factors}. This can be seen when comparing set \eqref{eq:H_i}-\eqref{eq:H_vi} with condition-set \eqref{eq:conic_coefficient_comparision}. Therefore the same requirements for its linear factorization apply. Applying Property \ref{prop:conic} completes the proof. $\hfill \qed$

\end{myproof*}
%
%
%

In Theorem \ref{theorem:TensorSeparation} all $\bm{\alpha}^{(i)}$ along $i \in \mathfrak{M}$ are still unknown. However, in case $\bm{\alpha}^{(i)} = \bm{0} \,\, \forall i \in \mathfrak{M}$ the determinant $\Delta_j^{(i)}$ can be calculated.  The case $\bm{\alpha}^{(i)}=\bm{0}$ corresponds with $\bar{\bm{m}}^{(i)}_k=\bm{0}$.


Following Theorem \ref{theorem:TensorSeparation}, by assuming $\Delta_j^{(i)} = 0$ and $\hat{\Delta}_j^{(i)} \le 0 \,\, \forall i \in \mathfrak{M} \land j \in \mathfrak{P}$, one takes a closer look on the special case where  $n_{{\rm{m}}}=2\ge n_{{\rm{m,min}}}$, 
then \eqref{eq:TensorSeparation} yields
\begin{subequations}
\label{eq:sol_nM_2}
\begin{gather}
\label{eq:sol_nM_2_A}
\left[
\begin{matrix}
\bm{B}^{(1)} & \bm{C}^{(1)} & \\
\bm{B}^{(2)} &  & \bm{C}^{(2)}
\end{matrix}
\right] \bm{d} = -\left[
\begin{matrix}
\bm{v}^{(1)}\\
\bm{v}^{(2)}
\end{matrix}
\right], \qquad
\left[
\begin{matrix}
\bm{E}^{(1)} & \bm{F}^{(1)} & \\
\bm{E}^{(2)} &  & \bm{F}^{(2)}
\end{matrix}
\right] \bm{d} = -\left[
\begin{matrix}
\bm{w}^{(1)}\\
\bm{w}^{(2)}
\end{matrix}
\right] \\
\left[
\begin{matrix}
\bm{B}^{(1)} & \bm{C}^{(1)} & \\
\bm{E}^{(2)} &  & \bm{F}^{(2)}
\end{matrix}
\right] \bm{d} = -\left[
\begin{matrix}
\bm{v}^{(1)}\\
\bm{w}^{(2)}
\end{matrix}
\right], \qquad
\left[
\begin{matrix}
\bm{E}^{(1)} & \bm{F}^{(1)} & \\
\bm{B}^{(2)} &  & \bm{C}^{(2)}
\end{matrix}
\right] \bm{d} = -\left[
\begin{matrix}
\bm{w}^{(1)}\\
\bm{v}^{(2)}
\end{matrix}
\right] 
\label{eq:sol_nM_2_B}
\end{gather}
\end{subequations}
meaning four search directions $\bm{d}$  \eqref{eq:searchDirection_not} may be obtained. This provides that \sloppy $\bm{B}^{(i)},\bm{C}^{(i)},\bm{v}^{(i)},$ $\bm{E}^{(i)},\bm{F}^{(i)}, \bm{w}^{(i)}$, which will be referred to as \textit{Tensor Separators}, have been found as a solution of set \eqref{eq:H_i}-\eqref{eq:H_vi}.
 %
\begin{corollary} \label{corollary:numberSolutions}
Let all assumptions in table \ref{tab:calibrationAssumptions} in addition to Assumption \ref{ass:FullTurbulent} hold. 
Following the necessary and sufficient conditions of  Property \ref{prop:conic}, one can deduce the following statements about the number of search directions solving Tensor-Equation \eqref{eq:TensorEquation} thus \eqref{eq:TensorEquation_Hadamard} for all $i\in \mathfrak{M}$ exactly or at least in approximation:
\begin{enumerate}
\item if  $\Delta_j^{(i)}(\bm{\alpha}^{(i)})  \ne 0$ for any $ i \in \mathfrak{M}$ or $j \in \mathfrak{P}$ there exist at most $2^{n_{{\rm{j}}} n_{{\rm{m}}}}$ exact solutions \ep{Bezout's Theorem}.
\item if  $\Delta_j^{(i)}(\bm{\alpha}^{(i)})  = 0 \,\, \forall i \in \mathfrak{M} \land j \in \mathfrak{P}$ there exist at most $2^{n_{{\rm{m}}}}$ exact solutions and if \ep{additionally}
\begin{enumerate}[label=\ep{\alph*}]
	\item 	$\hat{\Delta}_j^{(i)} < 0 \,\, \forall i \in \mathfrak{M} \land j \in \mathfrak{P}$, \eqref{eq:TensorEquation_Hadamard} yields precisely $2^{n_{{\rm{m}}}}$ different linear matrix-equations such as \eqref{eq:sol_nM_2}. As they are over-determined by nature $n_{{\rm{m}}}\ge n_{{\rm{m,min}}}$, precisely $2^{n_{{\rm{m}}}}$  different $\bm{d}$ result when applying, e.g., the pseudo-inverse.
	\item \label{corollary:case_unique_tensor}  $\hat{\Delta}_j^{(i)} = 0 \,\, \forall i \in \mathfrak{M} \land j \in \mathfrak{P}$ and $\bm{v}^{(i)}=\bm{w}^{(i)} \,\, \forall i \in \mathfrak{M}$, Tensor-Equation \eqref{eq:TensorEquation_Hadamard} yields a single linear matrix-equation \ep{e.g. one of \eqref{eq:sol_nM_2}} over all $i\in\mathfrak{M}$. Given $n_{{\rm{m}}} \ge n_{{\rm{m,min}}}$, there either exits a unique or no exact solution for $\bm{d}$ due to linear independency \ep{Assumption 5}.
	
	\item $\hat{\Delta}_j^{(i)} >0  \,\, \forall j \in \mathfrak{P}$, one obtains $\bm{d}_{\epsilon}$ and $\bm{d}_{h_N}^{(i)}$ as parts of the search direction $\bm{d}$ already when only considering a single measurement-set by solving one of the $i \in \mathfrak{M}$ in \eqref{eq:TensorEquation_Hadamard}. However, the roughness part $\bm{d}_{\epsilon}$ may be incompatible with other measurement-sets in $\mathfrak{M}$  concerning the satisfaction of \eqref{eq:TensorEquation_Hadamard}.
	
\end{enumerate}
\end{enumerate}

 \end{corollary}
%
Corollary \ref{corollary:numberSolutions} underlines the difficulty to find a supposedly unique roughness $\bm{\epsilon}$ with growing number of measurements as the number of feasible search directions in each iteration step for solving set \eqref{eq:calibrationTur} grows exponentially with the number of measurements $n_{{\rm{m}}}$.

 %
\begin{theorem} \label{theorem:TensorSolution}
Let all assumptions in table \ref{tab:calibrationAssumptions} in addition to Assumption \ref{ass:FullTurbulent} hold. Further suppose that $\Delta_j^{(i)}(\bm{\alpha}^{(i)})=0$ and $\hat{\Delta}_j^{(i)} \le 0 \,\, \forall j \in \mathfrak{P}$ in measurement-set $i\in \mathfrak{M}$. Then, Tensor Separators as a solution of set \eqref{eq:H_i} to \eqref{eq:H_vi} appear in pairs and can be found among
\begingroup
\allowdisplaybreaks
\begin{subequations}
\label{eq:TensorSeparator_Solution}
\begin{align}
ec &= \frac{1}{2}\left( -p_{\epsilon \Delta h} \pm_1 \sqrt{p_{\epsilon \Delta h}^2 - p_{\epsilon^2} p_{\Delta h^2}}\right) \label{eq:ec}  \\
bw &=  \frac{1}{2} \left( p_{\epsilon} \pm_2 \sqrt{p_{\epsilon}^2 - 2 \bar{f}_0 p_{\epsilon^2}} \right) \label{eq:bw}  \\
\tilde{f}v &=  \frac{1}{2} \left( -p_{\Delta h} \pm_3 \sqrt{p_{\Delta h}^2 - 2 \bar{f}_0 p_{\Delta h^2}} \right) \label{eq:ftildev} \\
b\tilde{f} &=  \frac{1}{2}\left( -p_{\epsilon \Delta h} \mp_1 \sqrt{p_{\epsilon \Delta h}^2 - p_{\epsilon^2} p_{\Delta h^2}}\right) = 
\frac{p_{\epsilon^2} p_{\Delta h^2}}{2 \left(-p_{\epsilon \Delta h} \pm_1 \sqrt{p_{\epsilon \Delta h}^2 - p_{\epsilon^2} p_{\Delta h^2}}\right)} \label{eq:bftilde}\\
ev &=  \frac{1}{2} \left( p_{\epsilon} \mp_2 \sqrt{p_{\epsilon}^2 - 2 \bar{f}_0 p_{\epsilon^2}} \right) =
 \frac{p_{\epsilon^2} \bar{f}_{0}}{p_{\epsilon} \pm_2 \sqrt{p_{\epsilon}^2 - 2 \bar{f}_0 p_{\epsilon^2}}} \label{eq:ev}\\
cw &=  \frac{1}{2} \left( -p_{\Delta h} \mp_3 \sqrt{p_{\Delta h}^2 - 2 \bar{f}_0 p_{\Delta h^2}} \right) =
\frac{p_{\Delta h^2} \bar{f}_0}{-p_{\Delta h} \pm_3 \sqrt{p_{\Delta h}^2 - 2 \bar{f}_0 p_{\Delta h^2}}}
\label{eq:cw}
\end{align}
\end{subequations}
\endgroup
\noindent presented in scalar form. The current measurement-set index $i\in\mathfrak{M}$ as well as the index along the pipes $j\in\mathfrak{P}$ is not displayed for the sake of readability, e.g. $ec \mathrel{\hat{=}} [\bm{e}^{(i)} \odot \bm{c}^{(i)}]_j $ \eqref{eq:ec}. Solution pair \eqref{eq:TensorSeparator_Solution} has 3 different locations of the $\pm$ sign presented with indices which indicate that their are $2^3=8$ per $j\in\mathfrak{P}$ possibilities per $j\in\mathfrak{P}$.
 \end{theorem}
%
\begin{remark} \label{remark:TensorSolution_consistency}
In reference to the solution pairs \eqref{eq:TensorSeparator_Solution} in Theorem \ref{theorem:TensorSolution}, only those $\pm_l$ combinations concerning $l=1,2,3$ are feasible which satisfy
\begin{gather}
\left(-p_{\epsilon \Delta h} \pm_1 \sqrt{p_{\epsilon \Delta h}^2 - p_{\epsilon^2} p_{\Delta h^2}}\right) \left(p_{\epsilon} \pm_2 \sqrt{p_{\epsilon}^2 - 2 \bar{f}_0 p_{\epsilon^2}}\right)
\left(-p_{\Delta h} \pm_3 \sqrt{p_{\Delta h}^2 - 2 \bar{f}_0 p_{\Delta h^2}}\right) \nonumber\\
= 2 p_{\epsilon^2} p_{\Delta h^2} \bar{f}_0.
\label{eq:TensorSolution_consistency}
\end{gather}
Thereby \eqref{eq:TensorSolution_consistency} can be considered as multiplication of \eqref{eq:H_i}, \eqref{eq:H_ii} and \eqref{eq:H_vi} which subsequently has to equal the multiplication of \eqref{eq:ec}, \eqref{eq:bw} and \eqref{eq:ftildev}. According to Theorem \ref{theorem:TensorSeparation}, there exists at least one feasible $\pm_l$ combination per $j\in \mathfrak{P}$ with respect to $l=1,2,3$ such that \eqref{eq:TensorSolution_consistency} holds, provided that $\Delta_j^{(i)}(\bm{\alpha}^{(i)}) = 0$ and $\hat{\Delta}^{(i)}_j \le 0$ concerning determinants \eqref{eq:Tensor_determinants}.
\end{remark}
%
%
\begin{myproof*}
\normalfont
\sloppy Analogously to \eqref{eq:TensorSeparator_Solution}, partial turbulent-flow derivatives $p_{\mathcal{X}}$ (see \eqref{eq:dxQ_der} and \eqref{eq:Qtur_d2}) with respect to $\mathcal{X} \in \{\epsilon, \epsilon^2, \Delta h, \Delta h^2, \epsilon \Delta h \}$ and $b,c,e,\tilde{f},v,w$ as well as $\bar{f}_0$ (for the zero order term) are considered to be scalar temporarily. Using \eqref{eq:H_i},\eqref{eq:H_ii},\eqref{eq:H_vi} to express $e,c,v$ by means of $b, \tilde{f}, w$ (or vice versa) such that
\begingroup\leqnos
\begin{align}
e = \frac{1}{2} \frac{p_{\epsilon^2}}{b}, \quad b = \frac{1}{2} \frac{p_{\epsilon^2}}{e}\tag{\textit{\textbf{i}*}} \label{eq:H_i*}\\
c = \frac{1}{2} \frac{p_{\Delta h^2}}{\tilde{f}}, \quad \tilde{f} = \frac{1}{2} \frac{p_{\Delta h^2}}{c}\tag{\textit{\textbf{ii}*}} \label{eq:H_ii*}\\
v  = \frac{\bar{f}_0}{w}, \quad  w = \frac{\bar{f}_0}{v}\tag{\textit{\textbf{vi}*}} \label{eq:H_vi*}
\end{align}
\endgroup
 one receives three independent quadratic equations in the context of \eqref{eq:H_iii}, \eqref{eq:H_iv} and \eqref{eq:H_v}
\begingroup\leqnos
\begin{align}
(ec)^2+ p_{\epsilon \Delta h} e c+ \frac{1}{4} p_{\epsilon^2} p_{\Delta h^2} = 0\tag{\textit{\textbf{iii}*}} \label{eq:H_iii*} \\
(bw)^2 - p_{\epsilon} bw + \frac{1}{2} \bar{f}_0 p_{\epsilon^2} \tag{\textit{\textbf{iv}*}} = 0 \label{eq:H_iv*}\\
(\tilde{f}v)^2 + p_{\Delta h} \tilde{f}v + \frac{1}{2} \bar{f}_0 p_{\Delta h^2}\tag{\textit{\textbf{v}*}} = 0 \label{eq:H_v*} .
\end{align}
\endgroup
The set \eqref{eq:H_i} to \eqref{eq:H_vi} has no solution for individual $b,e,\tilde{f},e,v,w$, but one can obtain the pairs
\begingroup\leqnos
\begin{align}
ec &= \frac{-p_{\epsilon \Delta h} \pm_1 \sqrt{p_{\epsilon \Delta h}^2 - p_{\epsilon^2} p_{\Delta h^2}}}{2} \tag{\textit{\textbf{iii}*}} \\
bw &=  \frac{p_{\epsilon} \pm_2 \sqrt{p_{\epsilon}^2 - 2 \bar{f}_0 p_{\epsilon^2}}}{2} \tag{\textit{\textbf{iv}*}} \\
\tilde{f}v &=  \frac{-p_{\Delta h} \pm_3 \sqrt{p_{\Delta h}^2 - 2 \bar{f}_0 p_{\Delta h^2}}}{2} \tag{\textit{\textbf{v}*}} .
\end{align}
\endgroup
Although the diagonal elements of the \textit{Tensor Separators} appear in pairs, it will be sufficient for the separation of the \textit{Tensor} equation \eqref{eq:TensorEquation_Hadamard}. The index $l=1,2,3$ of the ``$\pm_l$'' symbols thereby indicate the location of different signs, resulting in $2^{3} = 8$ potential possibilities for the \textit{Tensor Separators}. 

The investigation of the final solution of \eqref{eq:sol_nM_2} for the case $n_{{\rm{m}}}=2$ will be helpful to clarify what is actually needed for the separation \eqref{eq:TensorSeparation} of the \textit{Tensor} equation. Applying the diagonal forms of \textit{Tensor Separators} and \eqref{eq:con_FC} on the left hand side \eqref{eq:sol_nM_2_A} while multiplying with $\DIAG([\begin{matrix} \bm{e}^{(1)^T} & \bm{e}^{(2)^T} \end{matrix}])$
\begin{subequations}
\label{eq:TensorLin_sol}
\begin{gather}
\left[
\begin{matrix}
\DIAG( \bm{e}^{(1)} \odot \bm{b}^{(1)}) & \DIAG(\bm{e}^{(1)} \odot \bm{c}^{(1)})\bm{A}^T \bar{\bm{C}}_h^T&\\
 \DIAG( \bm{e}^{(2)} \odot \bm{b}^{(2)}) & &  \DIAG(\bm{e}^{(2)} \odot {\bm{c}}^{(2)}) \bm{A}^T \bar{\bm{C}}_h^T\\
\end{matrix}
\right] \bm{d} = -\left[
\begin{matrix}
 \bm{e}^{(1)} \odot \bm{v}^{(1)}\\
 \bm{e}^{(2)} \odot\bm{v}^{(2)}
\end{matrix}
\right]
\label{eq:TenSol_1}
\end{gather}
the right hand side of \eqref{eq:sol_nM_2_A} while multiplying with $\DIAG([\begin{matrix} \bm{b}^{(1)^T} & \bm{b}^{(2)^T} \end{matrix}])$
\begin{gather}
\left[
\begin{matrix}
\DIAG( \bm{b}^{(1)} \odot \bm{e}^{(1)}) & \DIAG( \bm{b}^{(1)} \odot \tilde{\bm{f}}^{(1)})\bm{A}^T \bar{\bm{C}}_h^T&\\
 \DIAG( \bm{b}^{(2)} \odot\bm{e}^{(2)}) & &  \DIAG(\bm{b}^{(2)} \odot \tilde{\bm{f}}^{(2)}) \bm{A}^T \bar{\bm{C}}_h^T\\
\end{matrix}
\right] \bm{d} = -\left[
\begin{matrix}
 \bm{b}^{(1)} \odot \bm{w}^{(1)}\\
 \bm{b}^{(2)} \odot\bm{w}^{(2)}
\end{matrix}
\right]
\label{eq:TenSol_2}
\end{gather}
the left hand side of \eqref{eq:sol_nM_2_B} while multiplying with $\DIAG([\begin{matrix} \bm{e}^{(1)^T} & \bm{b}^{(2)^T} \end{matrix}])$
\begin{gather}
\left[
\begin{matrix}
\DIAG( \bm{e}^{(1)} \odot \bm{b}^{(1)}) & \DIAG(\bm{e}^{(1)} \odot \bm{c}^{(1)})\bm{A}^T \bar{\bm{C}}_h^T&\\
 \DIAG( \bm{b}^{(2)} \odot\bm{e}^{(2)}) & &  \DIAG(\bm{b}^{(2)} \odot \tilde{\bm{f}}^{(2)}) \bm{A}^T \bar{\bm{C}}_h^T\\
\end{matrix}
\right] \bm{d} = -\left[
\begin{matrix}
 \bm{e}^{(1)} \odot \bm{v}^{(1)}\\
 \bm{b}^{(2)} \odot\bm{w}^{(2)}
\end{matrix}
\right]
\label{eq:TenSol_3}
\end{gather}
and the right hand side of \eqref{eq:sol_nM_2_B} while multiplying with $\DIAG([\begin{matrix} \bm{b}^{(1)^T} & \bm{e}^{(2)^T} \end{matrix}])$
\begin{gather}
\left[
\begin{matrix}
 \DIAG( \bm{b}^{(1)} \odot\bm{e}^{(1)}) &  \DIAG(\bm{b}^{(1)} \odot \tilde{\bm{f}}^{(1)}) \bm{A}^T \bar{\bm{C}}_h^T\\
 \DIAG( \bm{e}^{(2)} \odot \bm{b}^{(2)}) & &\DIAG(\bm{e}^{(2)} \odot \bm{c}^{(2)})\bm{A}^T \bar{\bm{C}}_h^T\\
\end{matrix}
\right] \bm{d} = -\left[
\begin{matrix}
 \bm{e}^{(1)} \odot \bm{v}^{(1)} \\
 \bm{b}^{(2)} \odot\bm{w}^{(2)}
\end{matrix}
\right]
\label{eq:TenSol_4}
\end{gather}
\end{subequations}
one has to determine the ``scalar'' terms $eb, ec,ev,b\tilde{f},bw$ concerning \eqref{eq:H_i} to \eqref{eq:H_vi}. However, as $eb$ (see \eqref{eq:H_i}), $ec,bw$ are already known, only $ev, b\tilde{f}$ and $cw$ need to be found to show \eqref{eq:TensorSeparator_Solution}.

Looking for $b\tilde{f}$, one makes use of the first solution of the quadratic equation from before $e = \frac{1}{2c} \Big(-p_{\epsilon \Delta h} \pm_1 \sqrt{p_{\epsilon \Delta h}^2 - p_{\epsilon^2} p_{\Delta h^2}}\Big)$ where $c$ is expressed by means of \eqref{eq:H_ii*} $c = \frac{1}{2} \frac{p_{\Delta h^2}}{\tilde{f}}$ leading to 
\begin{equation}
e = \frac{\tilde{f}}{p_{\Delta h^2}} \left(-p_{\epsilon \Delta h} \pm_1 \sqrt{p_{\epsilon \Delta h}^2 - p_{\epsilon^2} p_{\Delta h^2}}\right)
\end{equation}
which is then inserted into \eqref{eq:H_i*} $b = \frac{1}{2 e} p_{\epsilon^2}$ yielding in
\begin{equation}
b\tilde{f} =\frac{p_{\epsilon^2} p_{\Delta h^2}}{2 \left(-p_{\epsilon \Delta h} \pm_1 \sqrt{p_{\epsilon \Delta h}^2 - p_{\epsilon^2} p_{\Delta h^2}}\right)} .
\end{equation}
The same result for $b \tilde{f}$ is obtained when using $b \tilde{f} = - p_{\epsilon \Delta h} - ec$ according to \eqref{eq:H_iii} and applying $ec$ as a result of \eqref{eq:H_iii*}, see \eqref{eq:bftilde}.

Looking for $ev$, one uses \eqref{eq:H_vi*} in terms of $v = \frac{\bar{f}_0}{w}$ to be inserted in $bw$ such that $b = \frac{v}{2 \bar{f}_0} \left(p_{\epsilon} \pm_2 \sqrt{p_{\epsilon}^2 - 2 \bar{f}_0 p_{\epsilon^2}}\right)$ which  is then inserted into \eqref{eq:H_i*} yielding
\begin{equation}
ev = \frac{p_{\epsilon^2} \bar{f}_{0}}{p_{\epsilon} \pm_2 \sqrt{p_{\epsilon}^2 - 2 \bar{f}_0 p_{\epsilon^2}}}.
\end{equation}
The same result for $ev$ is obtained when using $ev = - p_{\epsilon} - bw$ according to \eqref{eq:H_iv} and applying $bw$ as a result of \eqref{eq:H_iv*}, see \eqref{eq:ev}.

The solution-pairs of equation set \eqref{eq:H_i} to \eqref{eq:H_vi} are now summarized in the scalar form \eqref{eq:TensorSeparator_Solution}
knowing that $\bm{b},\bm{e},\tilde{\bm{f}},\bm{e},\bm{v},\bm{w}$, meaning their element-wise combinations, have length $n_{\uell}$ actually and have to be considered along the different measurement-sets $1,2,\ldots,n_{{\rm{m}}}$. The determination of pair $cw$ was thereby accomplished analogously to $b\tilde{f}$ and $ev$. 

When inserting pairs  \eqref{eq:TensorSeparator_Solution} back into set \eqref{eq:H_i} to \eqref{eq:H_vi}, one will recognize that  solution \eqref{eq:TensorSeparator_Solution} is only consistent if \eqref{eq:TensorSolution_consistency} is satisfied. In reference to Remark \ref{remark:TensorSolution_consistency}, one needs to find a feasible $\pm_l$ combination concerning $l=1,2,3$ where \eqref{eq:TensorSolution_consistency} holds. $\hfill \qed$
\end{myproof*}
%
%
%
\section{The Factorization of a Conic Section as Exemplification}

In this section the scalar coefficients $A,B,\ldots,F$ denote those which separate conic section \eqref{eq:conic} into linear terms \eqref{eq:conic_factors}.
Following Theorem \ref{theorem:TensorSolution}, the factorization \eqref{eq:conic_factors} of the conic section \eqref{eq:conic} can be accomplished by the pairs

\begin{subequations}
\begin{minipage}{.45\linewidth}
\begin{align}
DB  &= h \pm_1 \sqrt{h^2 - ab}  \label{eq:DB}\\
AF &= f \pm_2 \sqrt{f^2 - ac}  \label{eq:AF}\\
EC &= g \pm_3 \sqrt{g^2 - bc}  \label{eq:EC}\\ \nonumber
\end{align}
\end{minipage}
\begin{minipage}{.45\linewidth}
\begin{align}
AE  &= h \mp_1 \sqrt{h^2 - ab} \label{eq:AE}\\
DC &= f \mp_2 \sqrt{f^2 - ac} \label{eq:DC}\\
BF &= g \mp_3 \sqrt{g^2 - bc} \label{eq:BF}.
\end{align}
\vspace{0.1cm} 
\end{minipage}
\end{subequations}

\noindent The multiplication of  \eqref{eq:conic_factors}  with $D$ and $B$ yields
\begin{equation}
\label{eq:conic_factors_nu}
\mathfrak{f}(x,y) = \nu (DA x + DBy + DC) (BD x + BE y + BF) = 0
\end{equation}
where factor $\nu = \frac{1}{DB}$. Pairs $DA = a$ and $BE = b$ are given by \eqref{eq:conic_comparision_1}. Subsequently,
\begin{align}
\label{eq:conic_solved_factors}
\mathfrak{f}(x,y) =& \, \nu \left( a x + \Big( h \pm_1 \sqrt{h^2 - ab}\Big) y + f \mp_2 \sqrt{f^2 - ac}\right) \nonumber\\
                                & \, \times \left( \Big( h \pm_1 \sqrt{h^2 - ab}\Big)x + by + g \mp_3 \sqrt{g^2 - bc} \right).                                     
\end{align}
Note that this is only possible if and only if $\Delta = 0$ and $\hat{\Delta} \le 0$ \eqref{eq:conic_determinante}. From representation \eqref{eq:conic_solved_factors} it is also clear that in case $\hat{\Delta} = 0$ the coefficients of the linear factorization are real  if $f^2\ge ac \land g^2\ge bc$ which is equivalent to the requirement that $f^2+g^2 \ge c (a+b)$ (Lemma \ref{lemma:conic_reals}).

\paragraph{Example} Consider the conic section
\begin{align}
\label{eq:conic_example_factor}
\mathfrak{f}(x,y) &= 2 x^2 + 2y^2 + 5xy -x + y-1=0,
\end{align}
then one has to determine for which $\pm$ combinations
\begin{equation}
\label{eq:conic_ex_consistent}
\left( h \pm_1 \sqrt{h^2 - ab}\right)
\left( f \pm_2 \sqrt{f^2 - ac}\right)
\left(g \pm_3 \sqrt{g^2 - bc} \right) = a bc 
\end{equation}
holds. Verifying consistency of set \eqref{eq:conic_coefficient_comparision}, \eqref{eq:conic_ex_consistent} is the multiplication of \eqref{eq:DB}, \eqref{eq:AF} and \eqref{eq:EC}, i.e. $(DB) (AF) (EC)$, which can also be written in terms of $(AD) (BE) (CF)= abc$ according to \eqref{eq:conic_comparision_1} (in comparison to \eqref{eq:TensorSolution_consistency} for consistency of set \eqref{eq:H_i} to \eqref{eq:H_vi}).
In analogy, the multiplication of \eqref{eq:AE}, \eqref{eq:DC} and \eqref{eq:BF}, i.e. $(AE) (DC) (BF)$, equals $(AD)(BE)(CF)=abc$ reversing all $l=1,2,3$ of $\pm_l$ in \eqref{eq:conic_ex_consistent} .
%
%

Referring to table \ref{table:conic_example}, one can see that only ``$++-$'' and ``$--+$'' are feasible combinations for the conic section \eqref{eq:conic_example_factor}. Note that there is a seeming conflict between Corollary \ref{corollary:numberSolutions} and Remark \ref{remark:TensorSolution_consistency}. In Corollary \ref{corollary:numberSolutions} it is said that there is \textit{exactly} one linear factorization if $\Delta  = 0$ and $\hat{\Delta} \le 0$ whereas in Remark \ref{remark:TensorSolution_consistency} it is said that there exist possibly more than one $\pm_l$ combinations concerning $l=1,2,3$ to be feasible in terms of \eqref{eq:conic_ex_consistent} or \eqref{eq:TensorSolution_consistency}. It will be shown by means of the present example that these combinations result in the identical factorization with different $\nu$ \eqref{eq:conic_factors_nu} (which is not always $\frac{1}{DB}$) to be determined only.

\vspace{0.3cm}
\begin{table}[H]                                                                       
\centering                                                                                
\begin{tabular}{c|c|c|c||c|c|c}   
$2 \left( h \pm_1 \sqrt{h^2 - ab}\right)$ & $2\left( f \pm_2 \sqrt{f^2 - ac}\right)$ & $2 \left(g \pm_3 \sqrt{g^2 - bc} \right) $ &$8abc$ &$\pm_1$ &$\pm_2$ & $\pm_3$     \\
\hline
$5 \pm_1 3$ & $-1\pm_2 3$ & $1 \pm_3 3$ &$-32$ & && \\
\hline \hline          
8 &2 & 4  &  64                   &$+$&$+$&$+$ \\ \hline
8 &2 & -2 &\cellcolor{olive!20}$-32$                &$+$&$+$&$-$ \\ \hline
8 &-4 & 4 &    -128                &$+$&$-$ &$+$ \\ \hline
8 &-4 & -2&       64             &$+$&$-$&$-$ \\ \hline
2 &2 & 4  &         16           &$-$&$+$&$+$ \\ \hline
2 &2 & -2&           -8          &$-$&$+$&$-$ \\ \hline
2 &-4 & 4&\cellcolor{olive!20}$-32$  &$-$&$-$&$+$ \\ \hline
2 &-4 & -2&           16         &$-$&$-$&$-$ \\ \hline
\end{tabular}                                                                             
\caption{Combinations of $\pm$ which factorize conic section \eqref{eq:conic_example_factor} to preserve consistency of set \eqref{eq:conic_coefficient_comparision} .}  
\label{table:conic_example}
\end{table}
This becomes apparent when applying the feasible combination ``$++-$''
\begin{subequations}
\begin{align}
\mathfrak{f}(x,y) &= \nu \left( 2x + \frac{1}{2}\Big( 5 + 3\Big)y + \frac{1}{2} \Big( -1 - 3\Big) \right) 
			    \left( \frac{1}{2}\Big( 5 + 3\Big)x + 2y + \frac{1}{2}\Big( 1 + 3\Big)\right) \nonumber\\
	&=  \frac{1}{2} (2x + 4y -2) (4x + 2y + 2)	    
\end{align}
as well as the combination ``$--+$'' obtained in table \ref{table:conic_example} on \eqref{eq:conic_solved_factors}
\begin{align}
\mathfrak{f}(x,y) &= \nu \left( 2x + \frac{1}{2}\Big( 5 - 3\Big)y + \frac{1}{2} \Big( -1 + 3\Big) \right) 
			    \left( \frac{1}{2}\Big( 5 - 3\Big)x + 2y + \frac{1}{2}\Big( 1 - 3\Big)\right) \nonumber\\
	&=  (2x + y + 1) (x + 2y - 1) .	    
\end{align}
\end{subequations}
%

\section{Final Observations} \label{sec:FinalObservations}

The reader shall be reminded at this point that pairs $\bm{b} \odot \bm{w},\tilde{\bm{f}} \odot \bm{v},\bm{e} \odot \bm{v}, \bm{c} \odot \bm{w}$, given by \eqref{eq:TensorSeparator_Solution}, comprise $\bar{\bm{f}}_0$ which itself is a function on $\bm{\alpha}$ in the corresponding measurement-set. Those $\bm{\alpha}^{(i)}$ for $i \in \mathfrak{M} =  \{1,2,\ldots,n_{{\rm{m}}}\}$ lie in the nullspace of incidence matrix $\bm{A}$, i.e. $\bm{A} \bar{\bm{m}}_k^{(i)}=\bm{0} \Leftrightarrow \bar{\bm{m}}_k^{(i)}-\bm{S}^T \bm{\alpha}^{(i)} = \bm{0}$. Applying the proposed solution \eqref{eq:TensorSeparator_Solution} (also using $\bm{b}\odot \bm{e}$ \eqref{eq:H_i}) for all the $n_{\uell}$ components of the \textit{Tensor Separators} on the deceivingly linear forms \eqref{eq:TensorLin_sol}, one will actually recognize that in all 4 representations \eqref{eq:TensorLin_sol} only the right hand sides (those independent from the search direction $\bm{d}$) depend on $\bm{\alpha}^{(i)}$.

To continue the line of thought, which started with Theorem \ref{theorem:TensorSeparation}, the proposed forms \eqref{eq:TensorLin_sol}, solving the degenerate \textit{Tensor Equation} \eqref{eq:TensorEquation_Hadamard}, can actually be transformed. Taking a closer look at \eqref{eq:TenSol_1} comprising pairs $\bm{b}^{(i)} \odot \bm{e}^{(i)} = \frac{1}{2} \bm{p}^{(i)}_{\epsilon^2}$ \eqref{eq:H_i}, $\bm{e}^{(i)}\odot\bm{c}^{(i)}$  \eqref{eq:ec} and $\bm{e}^{(i)} \odot \bm{v}^{(i)}$ \eqref{eq:ev} in measurement-set $i$, the $\bm{\alpha}^{(i)}$ only appear in the numerator of the right hand side (see \eqref{eq:ev}).
\begin{subequations}
\label{eq:MI_eq_gugu}
\begin{gather}
\left[
\begin{matrix}
\DIAG(\bm{e}^{(1)} \odot \bm{b}^{(1)})& \DIAG(\bm{e}^{(1)}\odot \bm{c}^{(1)}) & & \\
& & \DIAG(\bm{e}^{(2)}\odot \bm{b}^{(2)})& \DIAG(\bm{e}^{(2)}\odot \bm{c}^{(2)})
\end{matrix}
\right]
\left[
\begin{matrix}
\bm{I} &&\\
& \bm{A}^T \bar{\bm{C}}_h^T &\\
\bm{I} & & \\
& & \bm{A}^T \bar{\bm{C}}_h^T\\
\end{matrix}
\right] \nonumber\\
\times \bm{d} \stackrel{\eqref{eq:ev}}{=} - \frac{1}{2}\left[
\begin{matrix}
\bm{p}_{\epsilon}^{(1)} \mp^{(1)}_2 \left( \bm{p}_{\epsilon}^{(1)^2} - 2 \bar{\bm{f}}^{(1)}_0 \odot \bm{p}_{\epsilon^2}^{(1)} \right)^{\odot^{1/2}}\\
\bm{p}_{\epsilon}^{(2)} \mp^{(2)}_2 \left( \bm{p}_{\epsilon}^{(2)^2} - 2 \bar{\bm{f}}^{(2)}_0 \odot \bm{p}_{\epsilon^2}^{(2)} \right)^{\odot^{1/2}}
\end{matrix}
\right].
\label{eq:MI_eq}
\end{gather}
In reference to Remark \ref{remark:TensorSolution_consistency} and table \ref{table:conic_example}, signs ``$\pm$'' may also vary along the measurement-sets and  potentially along all $\mathfrak{P}$, i.e. along all components.
%
%
%
\begin{remark} \label{remark:pm_signs}
Please mind that ``$\pm^{(i)}_{2}$'' signs utilized in \eqref{eq:MI_eq_gugu} can also vary along $j\in\mathfrak{P}$, i.e. along all of the partial derivative's $[\bm{p}_{\mathcal{X}}]_j$ components $\forall j\in\mathfrak{P}$ with respect to $\mathcal{X} \in \{\epsilon, \epsilon^2, \Delta h, \Delta h^2, \epsilon \Delta h \}$. However, only those combinations are feasible which satisfy \eqref{eq:TensorSolution_consistency} in the corresponding measurement-set $i\in\mathfrak{M}$ and in $j\in \mathfrak{P}$.
\end{remark}
%
%
%
Ultimately, this will turn out irrelevant for the next considerations when taking the element-wise square.
One can write for the right hand side of \eqref{eq:TenSol_2} (applying \eqref{eq:bw})
\begin{equation}
- \frac{1}{2}\left[
\begin{matrix}
\bm{p}_{\epsilon}^{(1)} \pm^{(1)}_2 \left( \bm{p}_{\epsilon}^{(1)^2} - 2 \bar{\bm{f}}^{(1)}_0 \odot \bm{p}_{\epsilon^2}^{(1)} \right)^{\odot^{1/2}}\\
\bm{p}_{\epsilon}^{(2)} \pm^{(2)}_2 \left( \bm{p}_{\epsilon}^{(2)^2} - 2 \bar{\bm{f}}^{(2)}_0 \odot \bm{p}_{\epsilon^2}^{(2)} \right)^{\odot^{1/2}}
\end{matrix}
\right].
\end{equation}
This is can be done completely analogously for the mixed terms  \eqref{eq:TenSol_3} and  \eqref{eq:TenSol_4} by simply changing the $\pm_2$ signs. Thereby, the $\bm{\alpha}^{(i)}$ solely appear in the numerator of the right hand side.

\end{subequations}

The following notation is applied:
\begin{subequations}
\label{eq:M}
\begin{gather}
\bm{M}_{1} \stackrel{\eqref{eq:H_i} \land \eqref{eq:ec}}{\defeq} 
2 \left[
\begin{matrix}
\diag{\bm{p}_{\epsilon^2}^{(1)^{\odot^{-1/2}}}}&  \\
& \diag{\bm{p}_{\epsilon^2}^{(2)^{\odot^{-1/2}}}}
\end{matrix}
\right] \times \label{eq:MI} \\
\left[
\begin{matrix}
\DIAG(\bm{e}^{(1)} \odot \bm{b}^{(1)})& \DIAG(\bm{e}^{(1)}\odot \bm{c}^{(1)}) & & \\
& & \DIAG(\bm{e}^{(2)}\odot \bm{b}^{(2)})& \DIAG(\bm{e}^{(2)}\odot \bm{c}^{(2)})
\end{matrix}
\right] 
\left[
\begin{matrix}
\bm{I} &&\\
& \bm{A}^T \bar{\bm{C}}_h^T &\\
\bm{I} & & \\
& & \bm{A}^T \bar{\bm{C}}_h^T\\
\end{matrix}
\right]
\nonumber
\end{gather}
\begin{gather}
\bm{M}_{2}  \stackrel{\eqref{eq:H_i} \land \eqref{eq:bftilde}}{\defeq} 
2 \left[
\begin{matrix}
\diag{\bm{p}_{\epsilon^2}^{(1)^{\odot^{-1/2}}}}&  \\
& \diag{\bm{p}_{\epsilon^2}^{(2)^{\odot^{-1/2}}}}
\end{matrix}
\right] \times \label{eq:MII} \\
\left[
\begin{matrix}
\DIAG(\bm{b}^{(1)} \odot \bm{e}^{(1)})& \DIAG(\bm{b}^{(1)}\odot \tilde{\bm{f}}^{(1)}) & & \\
& & \DIAG(\bm{b}^{(2)}\odot \bm{e}^{(2)})& \DIAG(\bm{b}^{(2)}\odot \tilde{\bm{f}}^{(2)})
\end{matrix}
\right] 
\left[
\begin{matrix}
\bm{I} &&\\
& \bm{A}^T \bar{\bm{C}}_h^T &\\
\bm{I} & & \\
& & \bm{A}^T \bar{\bm{C}}_h^T\\
\end{matrix}
\right]
\nonumber
\end{gather}
\begin{gather}
\bm{M}_{3}  \stackrel{\eqref{eq:H_i} \land \eqref{eq:ec} \land \eqref{eq:bftilde}}{\defeq} 
2 \left[
\begin{matrix}
\diag{\bm{p}_{\epsilon^2}^{(1)^{\odot^{-1/2}}}}&  \\
& \diag{\bm{p}_{\epsilon^2}^{(2)^{\odot^{-1/2}}}}
\end{matrix}
\right] \times \label{eq:MIII} \\
\left[
\begin{matrix}
\DIAG(\bm{e}^{(1)} \odot \bm{b}^{(1)})& \DIAG(\bm{e}^{(1)}\odot \bm{c}^{(1)}) & & \\
& & \DIAG(\bm{e}^{(2)}\odot \bm{b}^{(2)})& \DIAG(\bm{b}^{(2)}\odot \tilde{\bm{f}}^{(2)})
\end{matrix}
\right] 
\left[
\begin{matrix}
\bm{I} &&\\
& \bm{A}^T \bar{\bm{C}}_h^T &\\
\bm{I} & & \\
& & \bm{A}^T \bar{\bm{C}}_h^T\\
\end{matrix}
\right]
\nonumber
\end{gather}
\begin{gather}
\bm{M}_{4}  \stackrel{\eqref{eq:H_i} \land \eqref{eq:ec} \land \eqref{eq:bftilde}}{\defeq}
2 \left[
\begin{matrix}
\diag{\bm{p}_{\epsilon^2}^{(1)^{\odot^{-1/2}}}}&  \\
& \diag{\bm{p}_{\epsilon^2}^{(2)^{\odot^{-1/2}}}}
\end{matrix}
\right] \times  \label{eq:MIV} \\
\left[
\begin{matrix}
\DIAG(\bm{b}^{(1)} \odot \bm{e}^{(1)})& \DIAG(\bm{b}^{(1)}\odot \tilde{\bm{f}}^{(1)}) & & \\
& & \DIAG(\bm{e}^{(2)}\odot \bm{b}^{(2)})& \DIAG(\bm{e}^{(2)}\odot {\bm{c}}^{(2)})
\end{matrix}
\right] 
\left[
\begin{matrix}
\bm{I} &&\\
& \bm{A}^T \bar{\bm{C}}_h^T &\\
\bm{I} & & \\
& & \bm{A}^T \bar{\bm{C}}_h^T\\
\end{matrix}
\right]
\nonumber
\end{gather}
\end{subequations}

\begin{equation}
 \bm{s} \defeq \left[
\begin{matrix}
\bm{p}_{\epsilon}^{(1)} \odot \bm{p}_{\epsilon^2}^{(1)^{\odot^{-1/2}}}\\
\bm{p}_{\epsilon}^{(2)} \odot \bm{p}_{\epsilon^2}^{(2)^{\odot^{-1/2}}}
\end{matrix}
\right]
\label{eq:s}
\end{equation}
%
%
\begin{remark} \label{remark:notation_TensorTrans}
The notation applied on \eqref{eq:M} and \eqref{eq:s} is only valid for $n_{{\rm{m}}}=2$ measurement-sets resulting in $2^{n_{{\rm{m}}}}=4$ different variants of matrix $\bm{M}_{\mathfrak{Y}}$ for $\mathfrak{Y} \in \{1,2,\ldots,2^{n_{{\rm{m}}}}\}$. The difference is only given by varying $\pm_1$ along all $j\in\mathfrak{P}$ in pairs $[\bm{e} \odot \bm{c}]_j$ and $[\bm{b}\odot \tilde{\bm{f}}]_j$. However, notation $\bm{M}_{\mathfrak{Y}}$ and $\bm{s}$ is also applied for all other $n_{{\rm{m}}}$ considering additional measurement-sets, leading to $2^{n_{{\rm{m}}}}$ variants concerning $\mathfrak{Y}$.
\end{remark}
%
%
\begin{proposition} \label{prop:TensorTransKernel}
Let $\Delta_j^{(i)}(\bm{\alpha}^{(i)}) = 0$ and $\hat{\Delta}_j^{(i)} \le 0 \,\, \forall j\in\mathfrak{P} \land i \in \mathfrak{M}$ concerning \eqref{eq:Tensor_determinants}. Knowing that $\bm{f}_{k-1} = \big[\begin{matrix} \bm{f}_{k-1}^{(1)^T} &  \bm{f}_{k-1}^{(2)^T} &\ldots& \bm{f}_{k-1}^{(n_{{\rm{m}}})^T}  \end{matrix}\big]^T$ while applying notation \eqref{eq:M} and \eqref{eq:s} in reference to Remark \ref{remark:notation_TensorTrans}, Tensor Equation \eqref{eq:TensorEquation_Hadamard} is equivalent to 
\begin{align}
\frac{1}{2} \left[
\begin{matrix}
\bm{A} & &\\
& \ddots &\\
 & &\bm{A}
\end{matrix}
\right]
\left( \left(\bm{M}_{\mathfrak{Y}} \bm{d}\right)^{\odot^2} + 2 \bm{M}_{\mathfrak{Y}}\bm{d}  \odot \bm{s} \right) +\bm{f}_{k-1}= \bm{0}
\label{eq:M_eq_kernel}
\end{align}
for all $\mathfrak{Y} \in \{1,2,\ldots,2^{n_{{\rm{m}}}}\}$ in Tensor-Method iteration $k$.
\end{proposition}
%
%
\begin{myproof*}
\normalfont
\eqref{eq:TensorLin_sol} extended to arbitrary $n_{{\rm{m}}}$ can be written as 
\begin{equation}
\label{eq:TenTrans_1}
\bm{M}_{\mathfrak{Y}} \bm{d} =-\left[
\begin{matrix}
\bm{p}_{\epsilon^2}^{(1)^{\odot^{-1/2}}} \odot \bm{p}_{\epsilon}^{(1)} \pm^{(1)}_2 \left(  \bm{p}_{\epsilon^2}^{(1)^{\odot^{-1}}} \odot \bm{p}_{\epsilon}^{(1)^2} - 2 \bar{\bm{f}}^{(1)}_0 \right)^{\odot^{1/2}}\\
\vdots\\
\bm{p}_{\epsilon^2}^{(n_{{\rm{m}}})^{\odot^{-1/2}}} \odot \bm{p}_{\epsilon}^{(n_{{\rm{m}}})} \pm^{(n_{{\rm{m}}})}_2 \left( \bm{p}_{\epsilon^2}^{(n_{{\rm{m}}})^{\odot^{-1}}} \odot  \bm{p}_{\epsilon}^{(n_{{\rm{m}}})^2} - 2 \bar{\bm{f}}^{(n_{{\rm{m}}})}_0 \right)^{\odot^{1/2}}
\end{matrix}
\right].
\end{equation}
whereas the $\pm_2^{(i)}$ sign is actually only valid for the pair $\bm{b}^{(i)} \odot \bm{w}^{(i)}$. Effectively, this will turn out irrelevant and thus also eligible for pair $\bm{e}^{(i)} \odot \bm{v}^{(i)}$ after bringing the $\bm{p}_{\epsilon^2}^{(i)^{\odot^{-1/2}}} \odot \bm{p}_{\epsilon}^{(i)}$ term of \eqref{eq:TenTrans_1} on the left hand side and then taking the element-wise square (consider Remark \ref{remark:pm_signs})
\begin{gather}
\vast(
\bm{M}_{\mathfrak{Y}} \bm{d} +  \underbrace{\left[
\begin{matrix}
\bm{p}_{\epsilon}^{(1)} \odot \bm{p}_{\epsilon^2}^{(1)^{\odot^{-1/2}}}\\
\vdots\\
\bm{p}_{\epsilon}^{(n_{{\rm{m}}})} \odot \bm{p}_{\epsilon^2}^{(n_{{\rm{m}}})^{\odot^{-1/2}}}
\end{matrix}
\right]}_{\bm{s}} \vast)^{\odot^2} - \bm{s}^{\odot^2}= -2\left[
\begin{matrix}
\bar{\bm{f}}_0^{(1)}\\
\vdots\\
\bar{\bm{f}}_0^{(n_{{\rm{m}}})}
\end{matrix}
\right] \nonumber\\
= -2\left[
\begin{matrix}
\diag{\bm{c}_l} \bm{A}^T \bm{L}^{-1} \bm{f}^{(1)}_{k-1} - \bm{S}^T \bm{\alpha}^{(1)}\\
\vdots\\
\diag{\bm{c}_l} \bm{A}^T \bm{L}^{-1} \bm{f}^{(n_{{\rm{m}}})}_{k-1} - \bm{S}^T \bm{\alpha}^{(n_{{\rm{m}}})}
\end{matrix}
\right]
\label{eq:sII}
\end{gather}
yielding
\begin{equation}
\left(\bm{M}_{\mathfrak{Y}} \bm{d} +\bm{s} \right)^{\odot^2} - \bm{s}^{\odot^2}
+ 2 \underbrace{\left[
\begin{matrix}
\diag{\bm{c}_l} \bm{A}^T \bm{L}^{-1} \bm{f}^{(1)}_{k-1}\\
\vdots\\
\diag{\bm{c}_l} \bm{A}^T \bm{L}^{-1} \bm{f}^{(n_{{\rm{m}}})}_{k-1}
\end{matrix}
\right]}_{\bm{r}_{f}}=
2 \left[
\begin{matrix}
\bm{S}^T & &\\
& \ddots&\\
&&\bm{S}^T
\end{matrix}
\right]
\left[
\begin{matrix}
 \bm{\alpha}^{(1)}\\
 \vdots\\
\bm{\alpha}^{(n_{{\rm{m}}})}
\end{matrix}
\right]
\end{equation}
which ultimately results in
\begin{equation}
\frac{1}{2}\left(\bm{M}_{\mathfrak{Y}} \bm{d}\right)^{\odot^2} +  \bm{M}_{\mathfrak{Y}}\bm{d}  \odot \bm{s}+\bm{r}_{f}=
 \left[
\begin{matrix}
\bm{S}^T & &\\
& \ddots&\\
&&\bm{S}^T
\end{matrix}
\right]
\left[
\begin{matrix}
 \bm{\alpha}^{(1)}\\
 \vdots\\
\bm{\alpha}^{(n_{{\rm{m}}})}
\end{matrix}
\right]
\label{eq:MII_Eq}
\end{equation}
where all $\bm{\alpha}^{(i)}$ and the search direction $\bm{d}$ are unknown. The projection in the kernel of incidence matrix $\bm{A}$ conducted in Theorem \ref{theorem:TensorSeparation} is now reversed knowing that $\bm{A}\bm{S}^T\equiv \bm{0}$ and that $\bm{A} (\diag{\bm{c}_l}\bm{A}^T \bm{L}\bm{f}_{k-1}^{(i)}) = \bm{f}_{k-1}^{(i)}$ as a consequence of  $\bm{L}=\bm{A} \DIAG(\bm{c}_l)\bm{A}^T\succ 0$ (see \citep{cite:ModelingHydraulicNetworks}). Thus \eqref{eq:MII_Eq} yields \eqref{eq:M_eq_kernel}. $\hfill\qed$
\end{myproof*}
%
%

Note that $\bm{M}_{\mathfrak{Y}}$ depends on all the partial derivatives $\bm{p}^{(i)}_{\mathcal{X}}$ with respect to $\mathcal{X} \in \{\epsilon, \epsilon^2, \Delta h, \Delta h^2, \epsilon \Delta h \}$ ($\bm{b} \odot \bm{e}$ in \eqref{eq:H_i} and pair $b\tilde{f}$ in \eqref{eq:bftilde} in the scalar case) whereas $\bm{s}$ depends on $\bm{p}^{(i)}_{\epsilon}$ and $\bm{p}^{(i)}_{\epsilon^2} $ for $i=1,2,\ldots,n_{{\rm{m}}}$. 
%
\begin{proposition} \label{prop:zero_searchDirection}
Let all assumptions in table \ref{tab:calibrationAssumptions} hold. 
Further suppose that $\bm{x}_{k-1} = \bm{x}^*$ one already converged to the real root $\bm{x}^*$ in the previous iteration step $k-1$. Then, the zero-search direction $\bm{d} = \bm{0}$ is the only solution for \eqref{eq:M_eq_kernel} for all $\mathfrak{Y} \in \{1,2,\ldots,2^{n_{{\rm{m}}}}\}$.
\end{proposition}
%
\begin{myproof*}
\normalfont
As $\bm{x}_{k-1} = \bm{x}^*$, $\bm{f}_{k-1} = \bm{f}(\bm{x}_{k-1}) =\bm{f}(\bm{x}^*) = \bm{0}$ \eqref{eq:fx_xQTur_1} per definition. Subsequently, 
\begin{equation}
 \left[
\begin{matrix}
\bm{A} &&\\
& \ddots &\\
  &&\bm{A}\\
\end{matrix}
\right]
  \bigg(\bm{M}_{\mathfrak{Y}} \bm{d} \odot \left( \bm{M}_{\mathfrak{Y}} \bm{d}+ 2 \bm{s} \right) \bigg) = \bm{0}
  \label{eq:d_anal}
\end{equation}
meaning that the search direction $\bm{d} = \bm{0}$ is a feasible solution of \eqref{eq:d_anal}. As a result, $\bm{\alpha}^{(i)} = \bm{0}$ $\forall i$ in this special case in order for $\bm{d} = \bm{0}$ to be feasible in $\bm{M}_{\mathfrak{Y}} \bm{d} \odot \left( \bm{M}_{\mathfrak{Y}} \bm{d}+ 2 \bm{s} \right)= 
 [\begin{matrix} \bm{\alpha}^{(1)^T}\bm{S} & \ldots & \bm{\alpha}^{(n_{{\rm{m}}})^T}\bm{S} \end{matrix}]^T$ which is formally equivalent to \eqref{eq:d_anal}. This means that the remaining solutions ought to satisfy $\bm{M}_{\mathfrak{Y}} \bm{d}=- 2 \bm{s}$ if existent. Knowing that $\bm{M}_{\mathfrak{Y}} \in \mathbb{C}^{n_{{\rm{m}}} n_{\uell} \times (n_{\uell} + n_{{\rm{m}}} (n_{\rm{j}} - n_{\rm{p}}))}$ has more rows than columns which are expected to be linear independent according to Assumption 5 as part of table \ref{tab:calibrationAssumptions}, there is no $\bm{d}$ which solves $\bm{M}_{\mathfrak{Y}} \bm{d}=- 2 \bm{s}$ exactly. As a result, $\bm{d}=\bm{0}$ is the unique solution of  \eqref{eq:d_anal}. $\hfill \qed$
 \end{myproof*}
%
Note that the partial derivatives  $\bm{p}^{(i)}_{\mathcal{X}}=\bm{p}_{\mathcal{X}}(\bm{\epsilon},\Delta \bm{h}^{(i)})$ with respect to $\mathcal{X} \in \{\epsilon, \epsilon^2, \Delta h, \Delta h^2, \epsilon \Delta h \}$ are functions on $\Delta \bm{h}^{(i)}$ and $\bm{\epsilon}$, whereas $\Delta \bm{h}^{(i)} =  \bm{C}_s\bm{h}_s^{(i)} - \bm{A}^T ( \bm{C}_h^T \bm{y}_h^{(i)} + \bar{\bm{C}}_h^T \bm{h}_N^{(i)}+\bm{z})$ \eqref{eq:calibrationTur_2} is considered as a function on $\bm{h}_N^{(i)}$ in measurement-set $i \in \mathfrak{M}$.
%
\begin{corollary} \label{corollary:Tensor_main}
Suppose that $\big[\bm{p}_{\mathcal{X}}^{*(i)}\big]_j=\big[\bm{p}_{\mathcal{X}}(\bm{\epsilon}^*,\bm{h}_N^{*(i)})\big]_j = p_{\mathcal{X},j}^{*(i)} \,\, \forall j\in \mathfrak{P} \land i \in \mathfrak{M}$ with respect to  $\mathcal{X} \in \{\epsilon, \epsilon^2, \Delta h, \Delta h^2, \epsilon \Delta h \}$ in the real root $\bm{x}^* = \big[\begin{matrix} \bm{\epsilon}^{*^T} &\bm{h}_N^{*(1)^T} &\ldots &\bm{h}_N^{*(n_{{\rm{m}}})^T}\end{matrix}\big]^T$. Following Proposition \ref{prop:zero_searchDirection} and Corollary \ref{corollary:numberSolutions} \ep{case 2\ref{corollary:case_unique_tensor}}, one can construct the following \textit{argumentum e contrario}. As $\bm{d}=\bm{0}$ is the unique solution if $\bm{x}_{k-1}=\bm{x}^*$, the only feasible factorization \eqref{eq:TensorSeparation} of Tensor Equation \eqref{eq:TensorEquation_Hadamard} ought to satisfy $\Delta_j^{(i)}(\bm{0}) = 0 \land \hat{\Delta}_j^{(i)} = 0$ \eqref{eq:Tensor_determinants} and $\bm{v}^{(i)}(\bm{\alpha}^{(i)})=\bm{w}^{(i)}(\bm{\alpha}^{(i)})$ \ep{compare with \eqref{eq:sol_nM_2}} with $\bm{\alpha}^{(i)} = \bm{0} \,\, \forall j\in \mathfrak{P} \land i \in \mathfrak{M}$ in the real root $\bm{x}^*$. Hence $\bar{\bm{f}}_0 = \bm{0}$ \eqref{eq:f0bar} in $\bm{x}^*$. Subsequently, the sufficient conditions
\begin{equation}
\label{eq:FinalCorollary}
\begin{gathered}
\Delta_j^{(i)} \stackrel{\eqref{eq:Tensor_determinants}}{=}  2 p_{\Delta h,j}^{*(i)} p_{\epsilon \Delta h,j}^{*(i)} p_{\epsilon,j}^{*(i)} - p_{\epsilon^2,j}^{*(i)} \left(p_{\Delta h,j}^{*(i)}\right)^2 - p_{\Delta h^2,j}^{*(i)} \left( p_{\epsilon,j}^{*(i)} \right)^2 = 0 \\
\hat{\Delta}_j^{(i)} \stackrel{\eqref{eq:Tensor_determinants}}{=} \left(p_{\epsilon \Delta h,j}^{*(i)}\right)^2 - p_{\epsilon^2,j}^{*(i)} p_{\Delta h^2,j}^{*(i)} = 0
\end{gathered}
\end{equation}
$\forall j\in \mathfrak{P} \land i \in \mathfrak{M}$ for $\bm{x}^*$ solving \eqref{eq:calibrationTur} must hold.
\end{corollary}
%

\subsection{Transformation inside of the Kernel}
The block diagonal of cycle matrices is denoted by $\bm{S}_b = \DIAG([\begin{matrix} \bm{S}^T &\ldots& \bm{S}^T \end{matrix}])$ and $\bm{\alpha} = [\begin{matrix} \bm{\alpha}^{(1)^T} &\ldots & \bm{\alpha}^{(n_{{\rm{m}}})^T} \end{matrix}]^T$, then \eqref{eq:MII_Eq} can be rewritten in terms of 
\begin{equation}
\frac{1}{2}\left(
\left[
\begin{matrix}
\bm{M}_{\mathfrak{Y}} & \bm{0}
\end{matrix}
\right]
\left[
\begin{matrix}
 \bm{d}\\
 \bm{\alpha}
 \end{matrix}
 \right]
\right)^{\odot^2} + 
\left[
\begin{matrix}
\DIAG( \bm{s}) \bm{M}_{\mathfrak{Y}} & -\bm{S}_b
\end{matrix}
\right]
\left[
\begin{matrix}
 \bm{d}\\
 \bm{\alpha}
 \end{matrix}
 \right]
+\bm{r}_{f}= \bm{0}
\label{eq:Tensor_II_final}
\end{equation}
$\forall \mathfrak{Y} \in \{1,2,\ldots,2^{n_{{\rm{m}}}}\}$ for a arbitrary number of measurement-sets when also accounting for the appropriate extensions in $\bm{M}_{\mathfrak{Y}}$ \eqref{eq:MII} and $\bm{s}$ \eqref{eq:sII}. In this representation it is visually clear that not only the search direction $\bm{d}$ but also $\bm{\alpha}$ is unknown. 
%
\begin{proposition} \label{prop:Tensor_kernelTrans}
With the assumption that there exists a perfect inversion  of
$
\tilde{\bm{M}} = 
\big[
\begin{matrix}
\DIAG( \bm{s}) \bm{M}_{\mathfrak{Y}} & -\bm{S}_b
\end{matrix}
\big]
$
such that $\tilde{\bm{M}}^{\#}\tilde{\bm{M}} = \bm{I}$ with a generalized inverse $\tilde{\bm{M}}^{\#}$, one can transform the problem \eqref{eq:Tensor_II_final} to unknowns $\bm{\beta}  =  \bm{M}_{\mathfrak{Y}} \bm{d}$  solving
\begin{align}
\bm{W} \bm{\beta}^{\odot^2} + 2\bm{\beta}+2\bm{W} \bm{r}_{f}= \bm{0}
\label{eq:TensorFinal}
\end{align}
where
$\bm{W} = \left[
\begin{matrix}
\bm{M}_{\mathfrak{Y}} & \bm{0}
\end{matrix}
\right] \tilde{\bm{M}}^{\#}$. However, if one is able to obtain a $\bm{\beta}$ satisfying \eqref{eq:TensorFinal}, one simultaneously solves \eqref{eq:Tensor_II_final} and thus \eqref{eq:M_eq_kernel} with
\begin{equation}
\left[
\begin{matrix}
\bm{d}\\
\bm{\alpha}
\end{matrix}
\right] = 
-\tilde{\bm{M}}^{\#} \left(\frac{1}{2}\bm{\beta}^{\odot^2} + \bm{r}_f\right).
\label{eq:TensorTrans_sol}
\end{equation}
\end{proposition}
%
\begin{myproof*}
\normalfont
 Denoting $\tilde{\bm{M}} [\begin{matrix} \bm{d}^T & \bm{\alpha}^T \end{matrix}]^T = \bm{y}$, \eqref{eq:Tensor_II_final} yields
\begin{equation}
\bigg(
\underbrace{\left[
\begin{matrix}
\bm{M}_{\mathfrak{Y}} & \bm{0}
\end{matrix}
\right] \tilde{\bm{M}}^{\#}}_{\bm{W}}
\bm{y}
\bigg)^{\odot^2} + 2
\bm{y}
+2\bm{r}_{f}= \bm{0}.
\label{eq:Tensor_II_final_2}
\end{equation}
With the aim to transform the problem to unknowns $\bm{\beta} =\bm{W} \bm{y} = \left[ \begin{matrix} \bm{M}_{\mathfrak{Y}} & \bm{0} 
\end{matrix}\right] [\begin{matrix}\bm{d}^T &\bm{\alpha}^T \end{matrix}]^T =  \bm{M}_{\mathfrak{Y}} \bm{d}$, \eqref{eq:Tensor_II_final_2} is multiplied with $\bm{W}$ from the left resulting in $\bm{W} \bm{\beta}^{\odot^2} + 2\bm{\beta}+2\bm{W}\bm{r}_{f}= \bm{0}$. Matrix $\bm{W}$ does certainly not have full rank in this context. Provided that a $\bm{\beta}$ satisfying \eqref{eq:TensorFinal} has been found, one reformulates \eqref{eq:Tensor_II_final_2}, which is equivalent to $\bm{\beta}^{\odot^2} + 2 \bm{y} + 2\bm{r}_f=\bm{0}$. Knowing that $\bm{y} =\tilde{\bm{M}} [\begin{matrix} \bm{d}^T & \bm{\alpha}^T \end{matrix}]^T$
one receives \eqref{eq:TensorTrans_sol}. Keep in mind that this is only feasible if a transformation $\bm{y} = \tilde{\bm{M}} [\begin{matrix} \bm{d}^T & \bm{\alpha}^T \end{matrix}]^T$ $\Leftrightarrow$ $[\begin{matrix} \bm{d}^T & \bm{\alpha}^T \end{matrix}]^T= \tilde{\bm{M}}^{\#} \bm{y}$ satisfying $\tilde{\bm{M}}^{\#} \tilde{\bm{M}} = \bm{I}$ is found.
 $\hfill\qed$
\end{myproof*}

\section{Applied Methodology and Algorithms} \label{sec:illustrativeComparison}

This section serves the purpose to present the applied algorithms on the basis of \textit{Newton-Raphson} and the \textit{Tensor-Method} in a comprehensive manner. These algorithms are then applied on a selected simulation example in the next section.

Since there are no explicit solutions of the \textit{Tensor Equation} available, although such possibly exist, one has to solve it iteratively. 
Applying iterative methods, one will not, in the majority of cases, receive search directions $\bm{d}_{\epsilon}, \bm{d}_{h_N}^{(i)}$ perfectly satisfying $\bm{m}_k^{(i)}(\bm{d}_{\epsilon},\bm{d}_{h_N}^{(i)}) = \bm{0}$ $\forall i$ \eqref{eq:TensorEquation_Hadamard}, but an approximation.
However, in order to keep the comparison as fair as possible, the authors apply the MATLAB built-in \textit{fsolve}(.) function, which itself makes use of the \textit{Levenberg–Marquardt} algorithm with standard settings to solve \eqref{eq:TensorEquation_Hadamard}. The authors thereby also supply the \textit{Jacobian} of \eqref{eq:TensorEquation_Hadamard} to \textit{fsolve}(.) which can be classified in terms of blocks
\begin{gather}
\frac{\partial \bm{m}^{(i)}_{k}}{\partial \bm{d}} = \bm{A}
\kbordermatrix{
        & & & & &i=j& \cr
        &\frac{\partial  \bar{\bm{m}}_k^{(i)}}{\partial \bm{d}_{\epsilon}} &\bm{0} &\ldots &\bm{0}& \frac{\partial \bar{\bm{m}}_k^{(i)}}{\partial \bm{d}_{h_N}^{(j)}}&\bm{0} &\ldots
	} 
	\qquad \forall i,j \in \mathfrak{M}
\label{eq:TensorJac_m}
\end{gather}
along its rows. Denoting $\bar{\bm{J}}_{T,\epsilon}^{(i)}(\bm{d}_{\epsilon}, \bm{d}^{(i)}_{h_N}) \defeq \frac{\partial \bar{\bm{m}}_k^{(i)}}{\partial \bm{d}_{\epsilon}}$ and $\bar{\bm{J}}^{(i)}_{T,h_N}(\bm{d}_{\epsilon}, \bm{d}^{(i)}_{h_N})\defeq \frac{\partial \bar{\bm{m}}_k^{(i)}}{\partial \bm{d}_{h_N}^{(i)}}$ (although neglected, arguments $\bm{d}_{\epsilon}$ and $\bm{d}^{(i)}_{h_N}$ would need another iteration index as in the outer iteration concerning $k$) one receives
\begin{subequations}
\label{eq:partialJacobians}
\begin{gather}
\bar{\bm{J}}_{T,\epsilon}^{(i)} = 
\diag{\bm{p}^{(i)}_{\epsilon^2} \odot \bm{d}_{\epsilon} + \bm{p}^{(i)}_{\epsilon} - \bm{p}^{(i)}_{\epsilon \Delta h} \odot \bm{A}^T \bar{\bm{C}}_h^T\bm{d}^{(i)}_{h_N}}\\
\bar{\bm{J}}^{(i)}_{T,h_N}  = \frac{1}{2} \DIAG( \bm{p}^{(i)}_{h_N^2}) \frac{\partial}{ \partial \bm{d}^{(i)}_{h_N}} \left( (\bm{A}^T \bar{\bm{C}}_h^T \bm{d}^{(i)}_{h_{\mathstrut \mathsmaller N}})^{\odot^2}\right) -\diag{\bm{p}^{(i)}_{\Delta h} + \bm{p}^{(i)}_{\epsilon \Delta h} \odot \bm{d}_{\epsilon}} \bm{A}^T \bar{\bm{C}}_h^T 
\label{eq:TensorJacobian}
\end{gather}
\end{subequations}
$\forall i \in \mathfrak{M}$ comprising partial derivatives $\bm{p}_{\mathcal{X}}^{(i)} = \bm{p}_{\mathcal{X}}(\bm{\epsilon}_k, \Delta \bm{h}^{(i)}_k)$ with respect to $\mathcal{X} \in \{\epsilon, \epsilon^2, \Delta h, \Delta h^2, \epsilon \Delta h \}$ in the outer iteration step $k$. Denoting $\bm{A}^T \bar{\bm{C}}_h^T = \big[ \begin{matrix} \bm{\mathfrak{c}}_1 & \bm{\mathfrak{c}}_2 &\ldots &\bm{\mathfrak{c}}_{n_{\uell}} \end{matrix}\big]^T$ , the remaining derivative reads as
 \begin{equation}
\frac{1}{2} \frac{\partial}{ \partial \bm{d}^{(i)}_{h_N}} \left( (\bm{A}^T \bar{\bm{C}}_h^T \bm{d}^{(i)}_{h_{\mathstrut \mathsmaller N}})^{\odot^2}\right) = 
\left[
\begin{matrix}
\left( \bm{\mathfrak{c}}^T_{1} \bm{d}_{h_N}^{(i)} \right) \bm{\mathfrak{c}}^T_{1}\\
\vdots\\
\left( \bm{\mathfrak{c}}^T_{n_{\uell}} \bm{d}_{h_N}^{(i)} \right) \bm{\mathfrak{c}}^T_{n_{\uell}}
\end{matrix}
\right].
 \end{equation}
Extending \eqref{eq:TensorJac_m} to arbitrary $n_{{\rm{m}}}$, one receives 
 \begin{equation}
 \bm{J}_T = \left[
 \begin{matrix}
 \frac{\partial \bm{m}^{(1)}_{k}}{\partial \bm{d}}\\
 \vdots\\
 \frac{\partial \bm{m}^{(n_{{\rm{m}}})}_{k}}{\partial \bm{d}}
 \end{matrix}
 \right] \stackrel{\eqref{eq:partialJacobians}}{=} 
  \left[
\begin{matrix}
\bm{A} &&\\
& \ddots &\\
  &&\bm{A}\\
\end{matrix}
\right]
 \left[
 \begin{matrix}
 \bar{\bm{J}}_{T,\epsilon}^{(1)} & \bar{\bm{J}}^{(1)}_{T,h_N} & & \\
 \bar{\bm{J}}_{T,\epsilon}^{(2)} & &\bar{\bm{J}}^{(2)}_{T,h_N} & \\
 \vdots & & &\ddots &\\
 \bar{\bm{J}}_{T,\epsilon}^{(n_{{\rm{m}}})} & & & &\bar{\bm{J}}^{(n_{{\rm{m}}})}_{T,h_N}
 \end{matrix}
 \label{eq:TensorJacobian_final}
 \right]
 \end{equation}
which is utilized for the present comparison. The initial value for the search direction $\bm{d}_0$ supplied to \textit{fsolve}(.) is selected as 10\% of the \textit{Newton} direction $\bm{d}_0 = \bm{d} =  -0.1 \times (\bm{J}_k^T\bm{J}_k)^{-1} \bm{J}_k^T \bm{f}_{k}$ in following algorithm.

\subsection{Customized Algorithms}
The following definitions shall help to properly distinguish between the applied algorithms which can be distinguished between two different types. Type (I) handles the classical iterative solving of a set of nonlinear equations with step length variation as proposed in \citep[Algorithm 1]{PipeRoughness_arxiv}. Since this algorithms of type (I) still require to start in the vicinity of the real root $\bm{x}^*$ for convergence, a type (II) algorithm then launches the type (I) algorithm (e.g. \citep[Algorithm 1]{PipeRoughness_arxiv}) several times with different initial values. The intermediate best result (in terms of the residual $v(\bm{x}) = \Vert f(\bm{x}) \Vert_{\mathcal{L}_1}$) along the different initial values is denoted by $\bm{x}^+$.

Please mind that Algorithm \ref{alg:Tensor} (of type (I)) as well as Algorithm \ref{alg:IC_Tensor} (of type (II)) proposed in this paper rely on the modification of algorithms proposed in \citep{PipeRoughness_arxiv}, that are \citep[Algorithm 1]{PipeRoughness_arxiv} and \citep[Algorithm 2]{PipeRoughness_arxiv}. The authors thereby directly refer to specific algorithm-lines in \citep[Algorithm 1]{PipeRoughness_arxiv} and \citep[Algorithm 2]{PipeRoughness_arxiv} which have to be adapted.
%
%
\setcounter{algorithm}{2}
\begin{algorithm}[H] 
\caption{Modification of \citep[Algorithm 1]{PipeRoughness_arxiv} applying \textit{Tensor Equation} \eqref{eq:TensorEquation_Hadamard}}  \label{alg:Tensor}
\begin{algorithmic}
\Procedure{Tensor}{$\text{FUN}, \bm{x}_0$}
   \State{It is identical to  \textsc{Newton}(\text{FUN}, $\bm{x}_0)$ except two changes (a,b).} \Comment{  \citep[Algorithm 1]{PipeRoughness_arxiv}}
   \Statex{ }  \vspace{-0.3cm}\hspace{.6cm}\hrulefill  
    \LineComment{(a,b) \textsc{Replace \citep[Algorithm 1, algorithm-lines 5,10]{PipeRoughness_arxiv} with:}}
   \State{$\Delta \bm{x}_k \gets \bm{d}$} \Comment{\eqref{eq:searchDirection_not} as a (approximated) solution of \eqref{eq:TensorEquation_Hadamard}}
    \LineComment{solving \eqref{eq:TensorEquation_Hadamard} iteratively by means of, e.g., ``$\bm{d}=$\textit{fsolve}(\eqref{eq:TensorEquation_Hadamard}, \eqref{eq:TensorJacobian_final}, $\bm{d}_0$)'', take}
   \State{$\bm{d}_0 \gets  - 0.1 \times (\bm{J}_k^T\bm{J}_k)^{-1} \bm{J}_k^T \bm{f}_{k}$} \Comment{as initial value}
   \Statex{ }  \vspace{-0.3cm}\hspace{.6cm}\hrulefill  
\EndProcedure
\end{algorithmic}
\end{algorithm}
%
%


\paragraph[Modification of sans]{Modification of \citep[Algorithm 2]{PipeRoughness_arxiv} of Type \ep{\emph{II}}}
A considerably small modification of \citep[Algorithm 2]{PipeRoughness_arxiv} turned out to be particularly well received in combination with Algorithm \ref{alg:Tensor} when solving problem \eqref{eq:calibrationTur}. This modification only concerns the variation of roughness values in each iteration of \citep[Algorithm 2]{PipeRoughness_arxiv}, where only those roughnesses are varied which exceed the 5\% mark of the corresponding pipes' diameter.
When also varying (although less aggressively) those roughness values in $\bm{x}^+$, used for the initial value $\bm{x}_0$ in the next iteration, which do \textit{not} exceed the $5\%$ mark of the corresponding pipe's diameter $d_i$\footnote{Please do not confuse the pipe's diameter $d_i$ of pipe $i\in\mathfrak{P}$ with the search direction $\bm{d}$.}, improvements with regard to convergence have been noticed. In this context, a normal distribution with zero mean and a standard deviation of $0.05\%=0.0005$ of the pipe's diameter is applied.

%
\begin{algorithm}[H] 
\caption{Modification of \citep[Algorithm 2]{PipeRoughness_arxiv} for launching \citep[Algorithm 1]{PipeRoughness_arxiv} or \ref{alg:Tensor}}  \label{alg:IC_Tensor}
\begin{algorithmic}
\Procedure{TensorCalibration}{$\text{FUN}, \bm{x}_0$, \text{TensorMethod}}
 \LineComment{Input variable ``TensorMethod'' is true if the \textit{Tensor Method} is applied.}
   \State{It is identical to  \textsc{NetCalibration}(\text{FUN},$\bm{x}_0)$,  i.e. \citep[Algorithm 2]{PipeRoughness_arxiv}, with one }  
   \State{addition (a) and one change (b) in \citep[Algorithm 2]{PipeRoughness_arxiv}. }
   \Statex{ }  \vspace{-0.3cm}\hspace{.6cm}\hrulefill  
    \LineComment{(a) \textsc{Add between \citep[Algorithm 2, algorithm-line 9 and 10]{PipeRoughness_arxiv}:}}
   \State{find $[\begin{matrix}\bar{i}_1 &\bar{i}_2 &\ldots &\bar{i}_{\bar{n}_{\epsilon}} \end{matrix}]$ such that $[\bm{x}^+]_{i\in \mathfrak{P}} = \epsilon_i \le 0.05 d_i$  $\Rightarrow n_{\epsilon} + \bar{n}_{\epsilon} = n_{\uell}$} 
   \State{$[\bm{x}_0]_i \gets [\bm{x}_0]_i +\text{normal}(0,0.0005 d_i) \quad \text{for} \quad i=\bar{i}_1,\bar{i}_2,\ldots,\bar{i}_{\bar{n}_{\epsilon}}$} 
   \Statex{ }  \vspace{-0.3cm}\hspace{.6cm}\hrulefill      
   \LineComment{(b) \textsc{Replace function call in \citep[Algorithm 2, algorithm-line 2]{PipeRoughness_arxiv} with:}}
           \If{TensorMethod is \textbf{true}}
		\State{$[\bm{x}^{+},\bm{f}^{+}] \gets \textsc{Tensor}(\text{FUN},\bm{x}_0)$}	 \Comment{call Algorithm \ref{alg:Tensor}}
		\Else
		\State{$[\bm{x}^{+},\bm{f}^{+}] \gets \textsc{Newton}(\text{FUN},\bm{x}_0)$}	 \Comment{call  \citep[Algorithm 1]{PipeRoughness_arxiv}}
        \EndIf  
     \Statex{ }  \vspace{-0.3cm}\hspace{.6cm}\hrulefill          
\EndProcedure
\end{algorithmic}
\end{algorithm}
%
%

Also, one has to specify whether to use the \textit{Tensor Method} (subject to this paper) or the original \textit{Newton-Raphson} method (subject to \citep{PipeRoughness_arxiv}) method to determine the search direction by adding an additional input variable denoted with ``TensorMethod'' which is either \textbf{true} or \textbf{false}.

\section{Simulation Example} \label{sec:sim_example}

The same network with the same measurement-sets as in the simulation example in \citep{PipeRoughness_arxiv} are applied to compare a \textit{Newton-Raphson} solving approach with the \textit{Tensor} method, i.e. an algorithm which solves \textit{Tensor Equation} \eqref{eq:TensorEquation_Hadamard} to obtain a search direction.
This network, illustrated in figure \ref{fig:3CycleNet}, features $n_{{\rm{j}}}  = 5$ nodes, $n_{\uell}=8$ pipes, hence $n_{\uell}-n_{{\rm{j}}} = 3$  independent cycles, $n_{\rm{p}}=3$ pressure sensors and the requirement of at least $n_{\rm{m,min}} = \ceil{n_{\uell}/n_{\rm{p}}} = 3$ independent measurement-sets.
This network has $n_{{\rm{q}}}=3$ consumers at nodes $k=2,3,4$ and $n_{{\rm{s}}}=1$ constant pressure source. 
\tikzset{
  on each segment/.style={
    decorate,
    decoration={
      show path construction,
      moveto code={},
      lineto code={
        \path [#1]
        (\tikzinputsegmentfirst) -- (\tikzinputsegmentlast);
      },
      curveto code={
        \path [#1] (\tikzinputsegmentfirst)
        .. controls
        (\tikzinputsegmentsupporta) and (\tikzinputsegmentsupportb)
        ..
        (\tikzinputsegmentlast);
      },
      closepath code={
        \path [#1]
        (\tikzinputsegmentfirst) -- (\tikzinputsegmentlast);
      },
    },
  },
  mid arrow/.style={postaction={decorate,decoration={
        markings,
        mark=at position .5 with {\arrow[#1]{stealth}}
      }}},
}

\tikzstyle{bigblock} = [draw, rectangle, minimum height=5.5cm, minimum width=10cm,rounded corners,>=latex']
   
\begingroup
\setlength{\intextsep}{2pt}%
\setlength{\columnsep}{0pt}%
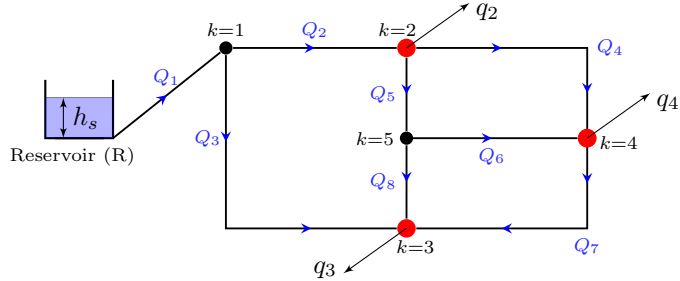
\begin{figure}[H]
\begin{center}
\begin{tikzpicture}[box/.style={draw,rounded corners,text width=3cm,align=center},scale=1.2]
\node[] (N1) {};
\node[] at ([yshift=-2cm]N1) (N2) {};
\node[] at ([xshift=2cm,yshift=-2cm]N1) (N3){};
\node[] at ([xshift=2cm]N1) (N4){};
\node[] at ([xshift=2cm]N4) (N5){};
\node[] at ([xshift=2cm,yshift=-2cm]N4) (N6){};
\node[] at ([yshift=-1cm]N4) (N7){};
\node[] at ([yshift=-1cm]N5) (N8){};
\node[] at ([xshift=-1.5cm,yshift=-1cm]N1) (R) {};
\node[] at ([xshift=-0.5cm]R) (R_LD) {};
\node[] at ([xshift=-0.5cm,yshift=0.75cm]R) (R_LU) {};
\node[] at ([xshift=0.25cm,yshift=0cm]R) (R_RD) {};
\node[] at ([xshift=0.25cm,yshift=0.75cm]R) (R_RU) {};
\node[] at ([yshift=-0.3cm]R_LU) (R_dummy) {};
\draw[thick] (R_LU) |- (R_RD);
\draw[thick] (R_RU)|- (R_LD);
\draw[fill=blue,opacity=0.3] (R_LD) rectangle ([yshift=-0.3cm]R_RU);
\draw[draw,<->,>=latex'] ($(R_LD) + (0.2,0)$) -- node[right] {{$h_s$}} ($(R_LD)+(0.2,0.45)$);
\node[] at ([yshift=0.2cm]N1) {\scriptsize{$k$=1}};
\node[] at ([yshift=-0.2cm,xshift=0.1cm]N3) {\scriptsize{$k$=3}};
\node[] at ([yshift=0.2cm,xshift=-0.1cm]N4) {\scriptsize{$k$=2}};
\node[] at ([xshift=0.35cm,yshift=-0.05cm]N8) {\scriptsize{$k$=4}};
\node[] at ([xshift=-0.35cm]N7) {\scriptsize{$k$=5}};
\node[] at ([xshift=-0.2cm,yshift=-0.2cm]R) {\scriptsize{Reservoir (R)}};

\node[circle,fill=black,inner sep=0pt,minimum size=5pt] (C1) at (N1) {};
\node[circle,fill=red,inner sep=0pt,minimum size=7pt] (C3) at (N3) {};
\node[circle,fill=red,inner sep=0pt,minimum size=7pt] (C4) at (N4) {};
\node[circle,fill=black,inner sep=0pt,minimum size=5pt] (C7) at (N7) {};
\node[circle,fill=red,inner sep=0pt,minimum size=7pt] (C8) at (N8) {};

\path [draw=black,line width= 0.025cm,postaction={on each segment={mid arrow=blue}}] (C1) -- node[above] {{\color{blue!85}\scriptsize{$Q_2$}}}(C4); 
\path [draw=black,line width= 0.025cm,postaction={on each segment={mid arrow=blue}}] (C1) |- node[] {}(C3); 
\node[] at ([xshift=-0.2cm,yshift=1cm]N2) {{\color{blue!85}\scriptsize{$Q_3$}}};

\path [draw=black,line width= 0.025cm,postaction={on each segment={mid arrow=blue}}] (C7) -- node[left] {{\color{blue!85}\scriptsize{$Q_8$}}}(C3); 
\path [draw=black,line width= 0.025cm,postaction={on each segment={mid arrow=blue}}] (C4) -- node[left] {{\color{blue!85}\scriptsize{$Q_5$}}}(C7);
\path [draw=black,line width= 0.025cm,postaction={on each segment={mid arrow=blue}}] (C4) -|   node[right] {{\color{blue!85}\scriptsize{$Q_4$}}}($(C8)$); 
\path [draw=black,line width= 0.025cm,postaction={on each segment={mid arrow=blue}}] ($(C8)$) |-  node[below] {{\color{blue!85}\scriptsize{$Q_7$}}}(C3); 
\path [draw=black,line width= 0.025cm,postaction={on each segment={mid arrow=blue}}] ($(C7)$) --  node[below] {{\color{blue!85}\scriptsize{$Q_6$}}}(C8); 
\path [draw=black,line width= 0.025cm,postaction={on each segment={mid arrow=blue}}] ($(R_RD)$) --  node[above] {{\color{blue!85}\scriptsize{$Q_1$}}}(C1); 

\node[circle,fill=red,inner sep=0pt,minimum size=7pt] (C8) at (N8) {};

\draw[draw,->,>=latex'] ($(N4)$) --  ($(N4)+(0.7,0.5)$);
\node[above] at ($(N4)+(0.9,0.2)$) {{$q_2$}};
\draw[draw,->,>=latex'] ($(N3)$) --  ($(N3)+(-0.7,-0.5)$);
\node[above] at ($(N3)+(-0.9,-0.65)$) {{$q_3$}};

\draw[draw,->,>=latex'] ($(N8)$) --  ($(N8)+(0.7,0.5)$);
\node[above] at ($(N8)+(0.9,0.2)$) {{$q_4$}};

\end{tikzpicture}
\caption{3-cycle network with pressure sensors at red colored nodes $k=2,3,4$.}
\label{fig:3CycleNet}
\end{center}
\end{figure}
%
\vspace{-0.3cm}

The incidence matrix
\begin{equation}
\bm{A} = \left[
\begin{matrix}
1 & -1 & -1 & 0&0&0&0&0\\
0& 1&0&-1&-1&0&0&0\\
0&0&1&0&0&0&1&1\\
0&0&0&1&0&1&-1&0\\
0&0&0&0&1&-1&0&-1
\end{matrix}
\right] \in \mathbb{Z}_{\{-1,0,1\}}^{n_{{\rm{j}}} \times n_{\uell}},
\end{equation}
as well as the following matrices are utilized
\begin{equation}
\bm{C}_h = \left[
\begin{matrix}
     0     &1     &0     &0    & 0\\
     0     &0    & 1     &0     &0\\
     0     &0     &0     &1    &0
\end{matrix}
\right], \qquad 
\bar{\bm{C}}_h = 
\left[
\begin{matrix}
     1     &0     &0     &0     &0\\
     0     &0    & 0     &0    & 1
\end{matrix}
\right], \qquad
\bm{C}_s = 
\left[
\begin{matrix}
1 &\bm{0}_{7}^T
\end{matrix}
\right]^T.
\end{equation}
The nodal elevation $\bm{z} = [\begin{matrix}0 &10 &5 &0 &0 \end{matrix}]^T$ (in m), the pipes' diameter $\bm{\mathfrak{d}} = 0.04 \times \bm{1}_{n_{\uell}}$ (in m) (i.e. $d_i = [\bm{\mathfrak{d}}]_i \,\, \forall i \in \mathfrak{P}$), their length $\bm{l} = [\begin{matrix} 10 &10&20&15&5&10&15&5\end{matrix}]^T$ (in m), roughnesses $\bm{\epsilon} = [\begin{matrix}2 &1.75 &1.5&1.25&1&0.75&0.5&0.25 \end{matrix}]^T \times 10^{-3}$ (in m) are chosen, whereas minor losses are set to zero. In order to produce an independent set of steady-state configurations (``measurements''), a dynamic model is utilized which has been derived in  \cite{cite:ModelingHydraulicNetworks,cite:DynamicModel_CCWI} while varying the consumption $\bar{\bm{q}}$. For more details about the creation of the steady-state measurement-sets, please have a look at \citep{PipeRoughness_arxiv} but also at \citep{cite:ModelingHydraulicNetworks}.

\subsection{Measurement-Sets}

The non-zero components of the nodal consumption are denoted by $\bm{q} = [\begin{matrix} q_2 &q_3 &q_4\end{matrix}]^T$ and can be obtained by $\bm{q} = \bm{C}_h \bar{\bm{q}}$ in this example.
As mentioned in \cite{PipeRoughness_arxiv}, the selected configuration led to sensed head values $\bm{y}_h+ \bm{C}_h \bm{z}$, which can barely be distinguished among each other \citep[Figure 3c]{PipeRoughness_arxiv}. Due to very little difference among the $\bm{y}_h + \bm{C}_h \bm{z}$, numerical inaccuracies are sufficient to cause serious difficulty to restore the roughness $\bm{\epsilon}$ with $\bm{y}_h$ and $\bm{q}$ when applying \eqref{eq:calibrationTur}, presumably violating Assumption 5. The quantities to set up \eqref{eq:calibrationTur} as well as its \textit{Jacobian} \eqref{eq:Jacobian} are summarized in the following table.

\vspace{0.3cm}
\begin{table}[H]                                                                       
\centering                                                                                
\begin{tabular}{c|c|c|c||c}   
set &1 &2 &3  &unit                   \\                             
\hline             \hline  
   &90.9743   &85.0087   &77.5380&\\
$\bm{y}_h$   &90.8720   &84.8200   &77.2370& m\\
   &90.8339   &84.7638   &77.1594 &\\
   \hline
    &0.9002    &1.1001    &1.3000&\\
 $\bm{q}$   &1.5002    &2.0001    &2.5000& l/s\\
    &1.0502    &1.3501    &1.6500& \\  
    \hline
$\bm{h}_s$    &100 &100 &100&m                                         
\end{tabular}                                                                             
\caption{Measurement-sets (see \citep[Table 2]{PipeRoughness_arxiv}).}  
\label{tab:3cycle_config1}                                                                
\end{table}
%

\paragraph{Initial Values}

The initial roughness value is chosen as 1\% of the pipes' diameter whereas details about the choice of the initial not-measured pressure heads can be found in \citep[Section 6.2]{PipeRoughness_arxiv}. Effectively, the initial vector
\begin{align}
\bm{x}_0 &= \left[
\begin{matrix}
 \bm{\epsilon}_0^T & h_{N_0,1}^{(1)}& h_{N_0,5}^{(1)} & h_{N_0,1}^{(2)}&  h_{N_0,5}^{(2)} &h_{N_0,1}^{(3)} & h_{N_0,5}^{(3)}
\end{matrix}
\right]^T \\
&=
\left[
\begin{matrix}
0.0004  \times \bm{1}^T_{n_{\uell}} 
  &93.9488
  &90.8934
   &89.9429
   &84.8642
   &84.9250
   &77.3115
\end{matrix}
\right]^T \nonumber
\label{eq:x0_3cycle_calibrate}
\end{align}
is utilized. The minimal and maximal value of all surrounding pressure heads  in the corresponding measurement-set is chosen for lower and upper boundary concerning $\underline{\bm{h}_N}$ and $\overline{\bm{h}_N}$ such that $[\underline{\bm{h}_N}]_{(i-1) n_{\rm{j}} + j} \le [\bm{h}_N^{(i)}]_j \le [\overline{\bm{h}_N}]_{(i-1) n_{\rm{j}} + j}$  for all $j \in \bar{\mathcal{P}} \, \land \, i \in \mathfrak{M}$, leading, for instance, to a maximal value of the pressure at node 1 of $h^{(i)}_{N,1} \le \overline{h_{N}}_{1}^{(i)} = h_s^{(i)} = 100 \, \forall i$. As it will turn out that the presented $[\bm{x}_{h_N}]_{(i-1) n_{\rm{j}} + j} = [\bm{h}_N^{(i)}]_j$ for all $j \in \bar{\mathcal{P}} \, \land \, i \in \mathfrak{M}$ results never leave their defined physically relevant range, these boundaries are not important for the present example.

\subsection{Results and Discussion}

\paragraph{Configuration concerning Table \ref{tab:3cycle_tensor}}
The authors use  a fixed number of 13 launches of Algorithm \ref{alg:IC_Tensor} in the outer loop which launches Algorithm \ref{alg:Tensor} (i.e. ``TensorMethod = \textbf{true}'') 50 times each (fixed number).
In each of these 50 calls of Algorithm \ref{alg:Tensor}, \eqref{eq:calibrationTur} is attempted to solve with the \textit{Tensor Method}, resulting in a total of $13 \times 50 = 650$ launches of \eqref{eq:calibrationTur} with different initial conditions each. The best (concerning the smallest $v(\bm{x}^+) = \Vert f(\bm{x}^+)\Vert_{\mathcal{L}_1}$) among those 50 launches, denoted by $\bm{x}^+$ can be found in table \ref{tab:3cycle_tensor}. 

\paragraph{Configuration concerning Table \ref{tab:3cycle_newton}}
The same initial conditions, loading parameters etc. are applied as before. The only difference concerns Algorithm \ref{alg:IC_Tensor} which now calls \citep[Algorithm 1]{PipeRoughness_arxiv} (i.e. input variable ``TensorMethod = \textbf{false}'') 50 times. For clarification, the 3-cycle network in figure \ref{fig:3CycleNet} is calibrated with a fixed number of 13 launches of Algorithm \ref{alg:IC_Tensor} in the outer loop which launches \citep[Algorithm 1]{PipeRoughness_arxiv} (i.e. ``TensorMethod = \textbf{false}'') 50 times each (fixed number).
Important to note is that for \citep[Algorithm 1]{PipeRoughness_arxiv} a scaling of $\bm{x}_k$ in reference to \citep[Remark 5]{PipeRoughness_arxiv} is conducted although not specifically indicated in \citep[Algorithm 1]{PipeRoughness_arxiv}. This is contrary to Algorithm \ref{alg:Tensor} which does not scale $\bm{x}_k$ due to the necessity to solve the nonlinear \textit{Tensor Equation}.

\paragraph{Row Notation}
The following applies to both tables \ref{tab:3cycle_tensor} and \ref{tab:3cycle_newton}. The row denoted with ``iter of $\bm{x}^+$'' indicates the iteration among the 50 (calling either Algorithm \ref{alg:Tensor} for table \ref{tab:3cycle_tensor}, or \citep[Algorithm 1]{PipeRoughness_arxiv} for table \ref{tab:3cycle_newton}) where the intermediate best result $\bm{x}^+$ concerning Algorithm \ref{alg:IC_Tensor} was obtained. 

 Concerning the last row, denoted with ``average \# iter to $\bm{x}^+$'', the authors took the mean along iterations of Algorithm \ref{alg:IC_Tensor} of the number of iterations needed in Algorithm \ref{alg:Tensor} and \citep[Algorithm 1, algorithm-line 14]{PipeRoughness_arxiv} to converge or abort until $\bm{x}^+$ is obtained. For instance, for the 3rd entry in table \ref{tab:3cycle_tensor} in row ``average \# iter to $\bm{x}^+$'' it took 5 iterations of Algorithm \ref{alg:IC_Tensor} to obtain $\bm{x}^+$, in these 5 runs Algorithm \ref{alg:Tensor} took 
\vspace{0.2cm}
\begin{table}[H]               
\centering                  
\begin{tabular}{c|c|c|c|c|c}
run & 1&2 &3&4&5 \\  
\hline\hline                
\makecell{\# Algorithm \ref{alg:Tensor} \\iterations}&4 & 23 & 14 & 3 & 1 \\      
\end{tabular}               
\caption{Number of iterations of Algorithm \ref{alg:Tensor} to converge or abort concerning the 3rd column of table \ref{tab:3cycle_tensor}. Iteration variable can be found in \citep[Algorithm 1, algorithm-line 14]{PipeRoughness_arxiv}.}                                                                                                          
\end{table}  
\noindent iterations respectively to converge or abort. The 3rd entry (meaning the 3rd column) in table \ref{tab:3cycle_tensor} in row ``average \# iter to $\bm{x}^+$'', for instance, is then the mean over the values 
\begin{equation}
9.00 = \frac{1}{5} \left( 4 +23 +14 + 3+1\right).
\end{equation}
The same methodology was applied for the row ``average \# iter to $\bm{x}^+$'' in table \ref{tab:3cycle_newton}.

\paragraph{Colored Table Entries and Range}
Blue colored entries in table \ref{tab:3cycle_tensor} and \ref{tab:3cycle_newton} indicate that the corresponding result left the physical relevant range. Thereby, only roughness values occasionally left the 5\% mark of the corresponding pipe's diameter. In this context all $\bm{x}_{h_N}$, meaning the subset of non-measured pressure heads in the solution, remained in their physically relevant range. The olive-colored results highlight the $\bm{x}^+$ of those launches which have the smallest $v(\bm{x}^+)$ in the entire table.

\clearpage
\begin{landscape}
\begin{table}                                                                                                                            
\centering                                                                                                                                      
\begin{tabular}{c||c|c|c|c|c|c|c|c|c|c|c|c|c||c}                                                                                                
launch & 1 & 2 & 3 & 4 & 5 & 6 & 7 & 8 & 9 & 10 & 11 & 12 & 13 & $\bm{x}^*$ \\                                                               
\hline \hline                                                                                                                                      
$\epsilon_{1}$ & 1.969 & 1.977 & 1.946 &  \cellcolor{blue!15}2.091 & 1.955 & 1.943 & 1.925 & 1.954 & 1.963 & 1.988 & 1.974 & 1.375 & 1.923 & 2.000 \\               
\hline                                                                                                                                          
$\epsilon_{2}$ & 1.839 & 1.816 & 1.906 & 1.498 & 1.879 & 1.914 & 1.968 & 1.881 & 1.856 & 1.784 & 1.824 &  \cellcolor{blue!15}3.878 & 1.976 & 1.750 \\               
\hline                                                                                                                                          
$\epsilon_{3}$ & 1.583 & 1.561 & 1.645 & 1.270 & 1.619 & 1.651 & 1.700 & 1.622 & 1.597 & 1.532 & 1.569 &  \cellcolor{blue!15}3.478 & 1.708 & 1.500 \\               
\hline                                                                                                                                          
$\epsilon_{4}$ & 1.180 & 1.177 & 1.187 & 1.128 & 1.186 & 1.190 & 1.196 & 1.184 & 1.182 & 1.171 & 1.178 & 1.333 & 1.196 & 1.250 \\               
\hline                                                                                                                                          
$\epsilon_{5}$ & 1.037 & 1.041 & 1.028 & 1.111 & 1.029 & 1.023 & 1.016 & 1.032 & 1.034 & 1.049 & 1.040 & 0.864 & 1.016 & 1.000 \\               
\hline                                                                                                                                          
$\epsilon_{6}$ & 0.817 & 0.818 & 0.806 & 0.806 & 0.826 & 0.829 & 0.822 & 0.804 & 0.822 & 0.808 & 0.818 & 0.722 & 0.811 & 0.750 \\               
\hline                                                                                                                                          
$\epsilon_{7}$ & 0.497 & 0.498 & 0.502 & 0.544 & 0.486 & 0.480 & 0.482 & 0.506 & 0.491 & 0.511 & 0.498 & 0.483 & 0.491 & 0.500 \\               
\hline                                                                                                                                          
$\epsilon_{8}$ & 0.235 & 0.235 & 0.234 & 0.241 & 0.235 & 0.234 & 0.233 & 0.234 & 0.235 & 0.236 & 0.235 & 0.217 & 0.233 & 0.250 \\               
\hline                                                                                                                                          
$h_{N,1}^{(1)}$ & 93.154 & 93.141 & 93.191 & 92.961 & 93.176 & 93.195 & 93.224 & 93.177 & 93.163 & 93.123 & 93.146 & 94.157 & 93.228 & 93.104 \\
\hline                                                                                                                                          
$h_{N,5}^{(1)}$ & 90.885 & 90.885 & 90.885 & 90.884 & 90.885 & 90.885 & 90.885 & 90.885 & 90.885 & 90.885 & 90.885 & 90.886 & 90.885 & 90.885 \\
\hline                                                                                                                                          
$h_{N,1}^{(2)}$ & 88.621 & 88.599 & 88.682 & 88.298 & 88.657 & 88.689 & 88.737 & 88.660 & 88.636 & 88.570 & 88.607 & 90.291 & 88.744 & 88.538 \\
\hline                                                                                                                                          
$h_{N,5}^{(2)}$ & 84.846 & 84.846 & 84.846 & 84.845 & 84.846 & 84.846 & 84.846 & 84.846 & 84.846 & 84.846 & 84.846 & 84.847 & 84.846 & 84.846 \\
\hline                                                                                                                                          
$h_{N,1}^{(3)}$ & 82.942 & 82.909 & 83.033 & 82.458 & 82.996 & 83.044 & 83.116 & 83.000 & 82.964 & 82.865 & 82.921 & 85.448 & 83.128 & 82.818 \\
\hline                                                                                                                                          
$h_{N,5}^{(3)}$ & 77.280 & 77.280 & 77.280 & 77.279 & 77.280 & 77.280 & 77.280 & 77.280 & 77.280 & 77.280 & 77.280 & 77.282 & 77.280 & 77.280 \\
\hline  \hline                                                                                                             
$v(\bm{x}^+) \times 10^{7}$ & \cellcolor{olive!20}0.484 & 0.637 & 0.655 & 1.067 & 0.547 & 0.569 & 0.690 & 0.669 & 0.553 & 0.553 & 0.532 & 0.770 & 0.710 & 1.082 \\    
\hline                                                                                                                                          
iter of $\bm{x}^+$ & 29 & 10 & 5 & 45 & 23 & 6 & 49 & 27 & 50 & 42 & 20 & 42 & 34 &  \\                                                         
\hline                                                                                                                                          
\makecell{average \# \\ iter to $\bm{x}^+$} & 3.79 & 8.30 & 9.00 & 7.16 & 7.91 & 10.67 & 8.55 & 6.37 & 3.90 & 4.31 & 8.25 & 9.55 & 9.97 &  \\                                                                                                                                                             
\end{tabular}                                                                                                                                                                                                                                   
\caption{\textsc{Tensor-Results.}\\ Results of Algorithm \ref{alg:IC_Tensor} along 13 launches calibrating the 3-cycle network (figure \ref{fig:3CycleNet}). Thereby Algorithm \ref{alg:IC_Tensor} calls Algorithm \ref{alg:Tensor} (``TensorMethod$=$\textbf{true}'') for a fixed number of 50 times by tightening $\epsilon_f$ and $\epsilon_x$, which belong to Algorithm \ref{alg:IC_Tensor} and occur in \citep[Algorithm 2, algorithm-line 6]{PipeRoughness_arxiv}, appropriately.  Roughnesses $\epsilon_i \,\, \forall i \in \mathfrak{P}$ are presented in mm, whereas pressure heads $h_{N,j}^{(i)} \,\, \forall i\in \mathfrak{M} \land j \in \bar{\mathcal{P}}$ are presented in m. Computational duration: 1310.185976s.}     
\label{tab:3cycle_tensor}                                                                                                                   
\end{table}                                                                                                                          
\clearpage
\end{landscape}
\begin{landscape}
\begin{table}                                                                                                                                           
\centering                                                                                                                                              
\begin{tabular}{c||c|c|c|c|c|c|c|c|c|c|c|c|c||c}                                                                                                        
launch & 1 & 2 & 3 & 4 & 5 & 6 & 7 & 8 & 9 & 10 & 11 & 12 & 13 & $\bm{x}^*$ \\                                                                       
\hline                 \hline                                                                                                                                 
$\epsilon_{1}$ & 1.273 & 1.266 & 1.254 &  \cellcolor{blue!15}2.007 &  \cellcolor{blue!15}2.004 &  \cellcolor{blue!15}2.004 & 1.270 &  \cellcolor{blue!15}2.004 &  \cellcolor{blue!15}2.004 & 1.273 &  \cellcolor{blue!15}2.004 & 1.268 &  \cellcolor{blue!15}2.019 & 2.000 \\                       
\hline                                                                                                                                                  
$\epsilon_{2}$ &  \cellcolor{blue!15}4.277 &  \cellcolor{blue!15}4.302 & \cellcolor{blue!15} 4.329 & 1.723 & 1.734 & 1.732 &  \cellcolor{blue!15}4.294 & 1.732 & 1.733 & 4.277 & 1.732 &  \cellcolor{blue!15}4.300 & 1.677 & 1.750 \\                       
\hline                                                                                                                                                  
$\epsilon_{3}$ &  \cellcolor{blue!15}3.873 & \cellcolor{blue!15} 3.906 & \cellcolor{blue!15} 3.973 & 1.487 & 1.495 & 1.496 & \cellcolor{blue!15} 3.876 & 1.495 & 1.496 & 3.872 & 1.496 & \cellcolor{blue!15} 3.883 & 1.472 & 1.500 \\                       
\hline                                                                                                                                                  
$\epsilon_{4}$ & 1.360 & 1.344 & 1.262 & 1.220 & 1.231 & 1.220 & 1.420 & 1.221 & 1.224 & 1.363 & 1.218 & 1.420 & 1.121 & 1.250 \\                       
\hline                                                                                                                                                  
$\epsilon_{5}$ & 0.827 & 0.827 & 0.838 & 1.001 & 0.997 & 1.002 & 0.803 & 1.000 & 0.999 & 0.826 & 1.002 & 0.800 & 1.043 & 1.000 \\                       
\hline                                                                                                                                                  
$\epsilon_{6}$ & 0.698 & 0.724 & 0.792 & 0.761 & 0.759 & 0.761 & 0.670 & 0.760 & 0.761 & 0.697 & 0.761 & 0.670 & 0.865 & 0.750 \\                       
\hline                                                                                                                                                  
$\epsilon_{7}$ & 0.473 & 0.484 & 0.507 & 0.497 & 0.496 & 0.497 & 0.457 & 0.497 & 0.496 & 0.473 & 0.497 & 0.458 & 0.523 & 0.500 \\                       
\hline                                                                                                                                                  
$\epsilon_{8}$ & 0.226 & 0.232 & 0.251 & 0.263 & 0.262 & 0.263 & 0.238 & 0.263 & 0.262 & 0.227 & 0.264 & 0.239 & 0.249 & 0.250 \\                       
\hline                                                                                                                                                  
$h_{N,1}^{(1)}$ & 94.342 & 94.355 & 94.376 & 93.093 & 93.098 & 93.098 & 94.347 & 93.097 & 93.098 & 94.342 & 93.098 & 94.350 & 93.075 & 93.104 \\        
\hline                                                                                                                                                  
$h_{N,5}^{(1)}$ & 90.886 & 90.886 & 90.886 & 90.885 & 90.885 & 90.885 & 90.886 & 90.885 & 90.885 & 90.886 & 90.885 & 90.886 & 90.885 & 90.885 \\        
\hline                                                                                                                                                  
$h_{N,1}^{(2)}$ & 90.598 & 90.621 & 90.655 & 88.519 & 88.528 & 88.527 & 90.607 & 88.526 & 88.527 & 90.598 & 88.527 & 90.612 & 88.488 & 88.538 \\        
\hline                                                                                                                                                  
$h_{N,5}^{(2)}$ & 84.848 & 84.848 & 84.848 & 84.846 & 84.846 & 84.846 & 84.848 & 84.846 & 84.846 & 84.848 & 84.846 & 84.848 & 84.846 & 84.846 \\        
\hline                                                                                                                                                  
$h_{N,1}^{(3)}$ & 85.910 & 85.943 & 85.995 & 82.789 & 82.802 & 82.801 & 85.923 & 82.800 & 82.802 & 85.909 & 82.801 & 85.931 & 82.743 & 82.818 \\        
\hline                                                                                                                                                  
$h_{N,5}^{(3)}$ & 77.283 & 77.283 & 77.284 & 77.281 & 77.281 & 77.281 & 77.283 & 77.281 & 77.281 & 77.283 & 77.281 & 77.283 & 77.281 & 77.280 \\        
\hline     \hline                                                                                                                                             
$v(\bm{x}^+) \times 10^{6}$ & 16.361 & \cellcolor{olive!20}9.826 & 10.342 & 17.003 & 12.034 & 15.932 & 15.785 & 16.376 & 14.440 & 15.690 & 17.275 & 16.439 & 10.889 & 0.108 \\
\hline                                                                                                                                                  
iter of $\bm{x}^+$ & 15 & 11 & 36 & 8 & 15 & 13 & 5 & 6 & 22 & 40 & 44 & 12 & 8 &  \\                                                                   
\hline                                                                                                                                                  
\makecell{average \# \\ iter to $\bm{x}^+$} & 11.00 & 27.09 & 123.56 & 264.25 & 13.33 & 253.38 & 216.80 & 243.50 & 63.91 & 140.40 & 241.89 & 12.25 & 16.25 &  \\        
\end{tabular}                                                                                                                                           
\caption{\textsc{Newton-Results.}\\ Results of Algorithm \ref{alg:IC_Tensor} along 13 launches calibrating the 3-cycle network (figure \ref{fig:3CycleNet}). Thereby Algorithm \ref{alg:IC_Tensor} calls \citep[Algorithm 1]{PipeRoughness_arxiv} (``TensorMethod$=$\textbf{false}'') for a fixed number of 50 times by tightening $\epsilon_f$ and $\epsilon_x$, which belong to Algorithm \ref{alg:IC_Tensor} and occur in \citep[Algorithm 2, algorithm-line 6]{PipeRoughness_arxiv}, appropriately.
 Roughnesses $\epsilon_i \,\, \forall i \in \mathfrak{P}$ are presented in mm, whereas pressure heads $h_{N,j}^{(i)} \,\, \forall i\in \mathfrak{M} \land j \in \bar{\mathcal{P}}$ are presented in m. 
 Computational duration: 157.195594s.}                                                                                                                                
\label{tab:3cycle_newton}                                                                                                                              
\end{table}  
\end{landscape}

\paragraph{Discussion}
When comparing the residuals $v(\bm{x}^+)$ of table \ref{tab:3cycle_tensor} and \ref{tab:3cycle_newton} it is abundantly clear that the \textit{Tensor-Results} are significantly better than the \textit{Newton-Results} in terms of $v(\bm{x}^+)$. This is particularly true when considering that every single result $\bm{x}^+$ in table \ref{tab:3cycle_tensor} has a smaller residual than the actual root $\bm{x}^*$, whereas all of the results in table \ref{tab:3cycle_newton} have a residual $v(\bm{x}^+)$ which is approximately two orders of magnitude larger than $v(\bm{x}^*)$.

Also, the number of iterations needed for Algorithm \ref{alg:Tensor} to converge or abort is on average only a fraction of the number of \textit{``Newton steps''}, meaning the numbers of iterations needed for \citep[Algorithm 1]{PipeRoughness_arxiv} to converge or abort. This can be seen when comparing the row ``average \# iter to $\bm{x}^+$'' of the both above tables. With that being said it should be emphasized that since the \textit{Tensor Equation} \eqref{eq:TensorEquation_Hadamard} is solved iteratively with the MATLAB built-in \textit{fsolve}(.) function, it takes several times longer to finish the computation of \textit{Tensor-Results} (table \ref{tab:3cycle_tensor}) compared to the \textit{Newton-Results} (table \ref{tab:3cycle_newton}).

While the \textit{Tensor-Results} are certainly superior to the \textit{Newton-Results} in terms of the residual $v(\bm{x}^+)$, one will recognize that when comparing the results of, for instance, launch $\{4,5,6,8,9\}$ in table \ref{tab:3cycle_newton} to \textit{Tensor-Results}, the deviation of \textit{Newton-Results} $\bm{x}^+$ to $\bm{x}^*$ is visibly smaller. In opinion of the authors, there are two possible explanations for this. First, the scaling interferes with the results leading to numerical inaccuracies \citep[Remark 5]{PipeRoughness_arxiv}. Second, the $n_{{\rm{m}}}=3$ measurement-sets are not independent from each other in order for the real root $\bm{x}^*$ to be properly distinguished in the solution space (in reference to Assumption 5). The truth presumably lies somewhere in between these two explanations.

One observation is particularly eye-catching, the \textit{Tensor-Results} tend to remain in their physical range (in terms of roughnesses)  whereas \textit{Newton-Results} tend to leave it. Interestingly, roughnesses in \textit{Newton-Results} or even in \textit{Tensor-Results} which exceed their physical limits are always those which are connected to pipe 1 ($Q_1$ in figure \ref{fig:3CycleNet}) which really has a chosen relative roughness of $\epsilon_1/d_1 = 5\%$ (meaning it was placed at the 5\% boundary intentionally). Therefore, the results consistently locate node $k=1$ (see figure \ref{fig:3CycleNet}) which is adjacent to the pipe with the highest roughness. Due to this observation, the authors expect the presented roughness scheme to be eligible for leak localization and detection, although simulations and real-world test are yet to be made.

\section{Conclusion and Outlook}

In conclusion to the attempt to find explicit solutions for the \textit{Tensor Equation}, the authors are convinced that the applied methods using the \textit{Hadamard} product for the separation of terms, in reference to Theorem \ref{theorem:TensorSeparation}, are well suited to develop an algorithm for solving problem \eqref{eq:calibrationTur} which converges faster and more reliably than the standard \textit{Newton-Raphson} type approach, and potentially provides multiple solutions for the search direction. For the attentive reader, algorithms solving quadratic equations, also similar to \eqref{eq:M_eq_kernel} or \eqref{eq:TensorFinal}, filled the PhD thesis \citep{quadraticEquations_phdThesis} underling the difficulty of the subject. 
%
Alternatively to the presented \textit{Tensor-Method}, it would be interesting to apply \textit{Halley}'s method \citep{HalleyMethod} on \eqref{eq:calibrationTur}-type problems as it also applies second order derivatives to obtain a search direction for solving nonlinear systems of equations iteratively.


Before the proposed algorithms can be applied to real-world networks, the original problem formulation \eqref{eq:calibrationTur} has to be extended to also allow pipe flows in the laminar and transitional \textit{Reynolds} area. Ultimately, the sufficiently smooth and explicit description of the flow in the transitional \textit{Reynolds} proposed in \cite{transitionalWaterFlow} has to be applied to complete the turbulent problem formulation.




%
%
\begin{appendices}
\setcounter{equation}{0}
\renewcommand{\theequation}{\thesection.\arabic{equation}}

\section{Derivatives Along Vectors} \label{app:derivatives}

\paragraph{First-Order Derivatives}
The derivative of the scalar function $h(\bm{x}) : \mathbb{K}^{n} \rightarrow \mathbb{K}$ operating on a field $\mathbb{K}$ with respect to vector $\bm{x} = [\begin{matrix} x_1 &\ldots &x_n \end{matrix}]^T$ is defined to result in the row-vector
\begin{equation*}
\frac{\partial h (\bm{x})}{\partial \bm{x}} := \Big[
\begin{matrix}
\frac{\partial h}{\partial x_1} & \frac{\partial h}{\partial x_2} &\ldots & \frac{\partial h}{\partial x_n} 
\end{matrix}
\Big] \eqdef \left( \nabla_{\bm{x}} h \right)^T
\end{equation*} 
whereas its gradient $\nabla_{\bm{x}} h$, which is exclusively denoted by the \textit{Nabla} operator $\nabla$, is defined to yield a column-vector. As a remark, the function argument of $h$, that is $\bm{x}$, is only provided in the operator $\nabla_{\bm{x}} \mathrel{\hat{=}} \nabla$ if clarification about the partial derivatives is needed.
When considering the vector-field $\bm{f}(\bm{x}) = [\begin{matrix} f_1(\bm{x}) &\ldots &f_m(\bm{x}) \end{matrix}]^T : \mathbb{K}^n \rightarrow \mathbb{K}^m$ the derivative along the vector $\bm{x}$ is defined to result in
\begin{equation*}
\frac{\partial \bm{f}(\bm{x})}{\partial \bm{x}} := \left[
\begin{matrix}
\frac{\partial f_1(\bm{x})}{\partial x_1} & \frac{\partial f_1(\bm{x})}{\partial x_2} & \ldots  & \frac{\partial f_1(\bm{x})}{\partial x_n} \\
\frac{\partial f_2(\bm{x})}{\partial x_1} & \frac{\partial f_2(\bm{x})}{\partial x_2} & \ldots  & \frac{\partial f_2(\bm{x})}{\partial x_n} \\
\vdots & \vdots &\ddots & \vdots \\
\frac{\partial f_m(\bm{x})}{\partial x_1} & \frac{\partial f_m(\bm{x})}{\partial x_2} & \ldots  & \frac{\partial f_m(\bm{x})}{\partial x_n} \\
\end{matrix}
\right],
\end{equation*}
the \textit{Jacobian}. 

\paragraph{Second-Order Derivatives}
The second-order derivative of the scalar function $h(\bm{x})$ with respect to $\bm{x}$ yields 
\begin{equation*}
\frac{\partial^2 h(\bm{x})}{\partial \bm{x}^2} = \mathcal{H}(h)(\bm{x}) \defeq  
\left[
\begin{matrix}
\frac{\partial^2 h}{\partial x_1^2} & \frac{\partial^2 h}{\partial x_2 \partial x_1 } & \ldots & \frac{\partial^2 h}{ \partial x_{n} \partial x_1}\\
\frac{\partial^2 h}{\partial x_1 \partial x_2} & \frac{\partial^2 h}{\partial x_2^2} & \ldots & \frac{\partial^2 h}{\partial x_n \partial x_{2}}\\
\vdots &\vdots & \ddots &\vdots\\
\frac{\partial^2 h}{\partial x_1\partial x_{n}} & \frac{\partial^2 h}{\partial x_{2}\partial x_n} &\ldots & \frac{\partial^2 h}{\partial x_n^2}
\end{matrix}
\right] \eqdef   \nabla^2_{\bm{x}} h,
\end{equation*}
the \textit{Hessian} which is also denoted by the square of the \textit{Nabla} operator, i.e. $\mathcal{H}(h)(\bm{x})= \nabla^2_{\bm{x}} h$. A sufficient condition for the \textit{Hessian} to be symmetric, i.e. $\mathcal{H}(h)(\bm{x}) = \mathcal{H}(h)(\bm{x})^T$ or $\nabla^2_{\bm{x}} h = (\nabla^2_{\bm{x}} h)^T$,  is that $h(\bm{x})$ must be two-times continuously differentiable (\textit{Schwarz-Clairaut} Theorem), i.e. $h \in C^{2}$. 

Now suppose that the scalar function $h$ depends on a second set of variables combined in vector $\bm{y} = [\begin{matrix}y_1 &y_2 &\ldots &y_c \end{matrix}]^T \in \mathbb{K}^{c}$, i.e. $h=h(\bm{x},\bm{y})$. Then, the mixed derivative yields
\begin{equation*}
\frac{\partial^2 h}{\partial \bm{x} \partial \bm{y}} = \frac{\partial }{\partial \bm{x}} \left( \frac{\partial h}{\partial \bm{y}} \right)^T = \frac{\partial }{\partial \bm{x}} \left( \nabla_{\bm{y}} h \right)    =\left[
\begin{matrix}
\frac{\partial^2 h}{ \partial x_1 \partial y_1} &\frac{\partial^2 h}{\partial x_2 y_1} &\ldots & \frac{\partial^2 h}{\partial x_n y_1}\\
\frac{\partial^2 h}{ \partial x_1 \partial y_2} &\frac{\partial^2 h}{\partial x_2 y_2} &\ldots & \frac{\partial^2 h}{\partial x_n y_2}\\
\vdots & \vdots & \ddots & \vdots\\
\frac{\partial^2 h}{ \partial x_1 \partial y_c} &\frac{\partial^2 h}{\partial x_2 y_c} &\ldots & \frac{\partial^2 h}{\partial x_n y_c}
\end{matrix}
\right],
\end{equation*}
and in case $h(\bm{x},\bm{y})$ is two-times continuously differentiable, one can write
\begin{equation*}
\frac{\partial^2 h}{ \partial \bm{y} \partial \bm{x}} = \left(  \frac{\partial^2 h}{\partial \bm{x} \partial \bm{y}} \right)^T = \frac{\partial }{\partial \bm{y}} \left( \frac{\partial h}{\partial \bm{x}} \right)^T = \frac{\partial }{\partial \bm{x}} \left( \frac{\partial h}{\partial \bm{y}} \right)^T.
\end{equation*}
The second-order derivatives of a vector-field such as $\bm{f}(\bm{x})$ can no longer be represented compactly by a single matrix and will be discussed in detail when needed.


\end{appendices}



\bibliography{example}

\end{document}